\pdfoutput=1
\RequirePackage{./StyleFilesMacros/lhchiggs}
\documentclass{cernyrep}
\usepackage{etoolbox}
\usepackage[stable]{footmisc}
\usepackage{adforn}

\newtoggle{paper}
\togglefalse{paper}

\usepackage[T1]{fontenc}
\usepackage{esvect}

\usepackage[dvipsnames]{xcolor}
\usepackage{graphicx}
\usepackage{tikz}
\usetikzlibrary{arrows,decorations.pathmorphing,backgrounds,positioning,fit,petri,automata,shadows,calendar,mindmap,decorations.markings,calc}
\tikzset{->-/.style={decoration={markings, mark=at position .5 with {\arrow{>}}},postaction={decorate}}}
\tikzset{-<-/.style={decoration={markings, mark=at position .5 with {\arrow{<}}},postaction={decorate}}}
\usepackage{cancel}
\usepackage{pdflscape}
\usepackage{float}
\usepackage{./StyleFilesMacros/axodraw}
\usepackage{lineno}
\usepackage{enumitem}
\setlist{nolistsep} 
\usepackage{./StyleFilesMacros/lhchiggs}
\usepackage{heppennames2}
\usepackage{cernunits}

\usepackage{./StyleFilesMacros/hepparticles}

\usepackage{fancyhdr}
\usepackage{tikz}

\usepackage{braket}
\usepackage{authblk} 
\usepackage{makecell}

\newcommand{\cref}[1]{Chapter~\ref{c.#1}}

\newcommand{\nc}{\newcommand}
\newcommand{\dimE}{dimension-eight}
\newcommand{\dimS}{dimension-six}
\nc{\gev}{\;\mathrm{GeV}}

\usepackage{hypernat}
\RequirePackage{./StyleFilesMacros/relsize}
\usepackage{units}
\usepackage[toc,page]{appendix}
\usepackage{booktabs}
\usepackage{bigdelim}
\def\reffi#1{\mbox{Figure~\ref{#1}}}

\def\be{\begin{equation}}
\def\ee{\end{equation}}

\newcommand{\GeV}{\ensuremath{\,\text{GeV}}\xspace}
\newcommand{\TeV}{\ensuremath{\,\text{TeV}}\xspace}

\newcommand{\pt}{\ensuremath{p_\text{T}}\xspace}

\newcommand{\madgraph}{{\sc\small MadGraph5\_aMC@NLO}\xspace}
\newcolumntype{.}{D{.}{.}{-1}}
\newcolumntype{d}[1]{D{.}{.}{#1}}

\renewcommand{\vec}[1]{\mathbf{#1}}

\newlength{\width}
\newlength{\height}

\def\draftdate{\relax}
\def\mda{\relax}
\def\mua{\relax}
\def\mla{\relax}
\def\draft{
\def\thtystars{******************************}
\def\sixtystars{\thtystars\thtystars}
\typeout{}
\typeout{\sixtystars**}
\typeout{* Draft mode!
         For final version remove \protect\draft\space in source file *}
\typeout{\sixtystars**}
\typeout{}
\def\draftdate{\today}
\def\mua{\marginpar[\boldmath\hfil$\uparrow$]%
                   {\boldmath$\uparrow$\hfil}\color{black}%
                    \typeout{marginpar: $\uparrow$}\ignorespaces}
\def\mda{\color{red}\marginpar[\boldmath\hfil$\downarrow$]%
                   {\boldmath$\downarrow$\hfil}%
                    \typeout{marginpar: $\downarrow$}\ignorespaces}
\def\mla{\marginpar[\boldmath\hfil$\rightarrow$]%
                   {\boldmath$\leftarrow $\hfil}%
                    \typeout{marginpar: $\leftrightarrow$}\ignorespaces}
\def\Mua{\marginpar[\boldmath\hfil$\Uparrow$]%
                   {\boldmath$\Uparrow$\hfil}\color{black}%
                    \typeout{marginpar: $\uparrow$}\ignorespaces}
\def\Mda{\color{red}\marginpar[\boldmath\hfil$\Downarrow$]%
                   {\boldmath$\Downarrow$\hfil}%
                    \typeout{marginpar: $\downarrow$}\ignorespaces}
\def\Mla{\marginpar[\boldmath\hfil\textcolor{red}{$\Rightarrow$}]%
                   {\boldmath\textcolor{red}{$\Leftarrow $}\hfil}%
                    \typeout{marginpar: $\leftrightarrow$}\ignorespaces}
\overfullrule 5pt
\oddsidemargin 15mm
\marginparwidth 29mm
}

\newcolumntype{C}[1]{>{\centering\arraybackslash}p{#1}}

\usepackage[hyperfootnotes=true,bookmarks=false,hypertexnames=true,pagebackref=true,linktocpage=false]{hyperref}
\iftoggle{paper}{%
	\definecolor{white}{rgb}{1.0, 1.0, 1.0}
	
   \hypersetup{draft}
}{%
\hypersetup{
   unicode=true,          	
   pdftoolbar=true,        	
   pdfmenubar=true,        	
   pdffitwindow=false,     	
   pdfstartview={FitH},    	
   pdftitle={My title},    	
   pdfauthor={Author},     	
   pdfsubject={Subject},   	
   pdfcreator={Creator},   	
   pdfproducer={Producer}, 
   pdfkeywords={keyword1} {key2} {key3}, 
   pdfnewwindow=true,   	
   colorlinks=true,       	
   linkcolor=MidnightBlue,	
   citecolor=OliveGreen,	
   filecolor=Cyan,          	
   urlcolor=RedViolet       	
}
}

\renewcommand*\backref[1]{\ifx#1\relax \else (#1) \fi}

\let\counterwithout\relax
\let\counterwithin\relax

\usepackage{chngcntr}
\usepackage[subfigure]{tocloft} 
\usepackage{subfigure} 
\usepackage{wrapfig}
\usepackage{afterpage}

\usepackage[export]{adjustbox} 

	\counterwithin{chapter}{part}	
	\counterwithin{section}{chapter}
	\counterwithin{equation}{chapter}
	\counterwithout{figure}{chapter}
	\counterwithout{table}{chapter}
  \counterwithout{equation}{chapter}
	\counterwithout{footnote}{chapter}
\makeatletter
	\@addtoreset{chapter}{part}
	\@addtoreset{footnote}{part}
\makeatother
\usepackage[export]{adjustbox}

\title{VBSCan Thessaloniki 2018 Workshop Summary}

\newcommand{\myabstract}
{
This document reports the first year of activity of the VBSCan COST Action network,
as summarised by the talks and discussions 
happened during the VBSCan Thessaloniki 2018 workshop.
The VBSCan COST action is aiming at a consistent and coordinated study of vector-boson scattering
from the phenomenological and experimental point of view, 
for the best exploitation of the data that will be delivered by 
existing and future particle colliders.
}

\newcommand{\editors}
{
\begin{flushright}
L.~S.~Bruni, R.~Covarelli, \\
P.~Govoni, P.~Lenzi, \\ 
N.~Lorenzo-Martinez, 
J.~Manjarres, M.~U.~Mozer, \\
G.~Ortona, M.~Pellen, \\
D.~Rebuzzi, M.~Slawinska, \\
M.~Zaro \\
\end{flushright}
}

\newcommand{\docID}{VBSCAN-PUB-05-19\\
                    DESY 19-108 \\
                    Nikhef/2019-025\\
                    UWThPh 2019-20}
\usepackage{layout}
\newcommand{\mydate}{\today}

\begin{document}


\pagestyle{plain}

  

\thispagestyle{empty}
{

\begin{center}

{\fontfamily{qhv}\selectfont

\begin{minipage}[t]{0.25\textwidth}
  \mydate
\end{minipage}
\begin{minipage}[t]{0.4\textwidth}
  \begin{center}
  \textit{VBSCan COST Action report}
  \end{center}
\end{minipage}
\begin{minipage}[t]{0.25\textwidth}
  \begin{flushright}
  \docID
  \end{flushright}
\end{minipage}


\vspace{5cm}

\makeatletter

{\LARGE\textbf{\@title}}

\makeatother

\vspace {1cm}
\adforn{50}~~\adforn{10}~~\adforn{22}
\vspace {1cm}

\begin{minipage}{0.7\textwidth}

\myabstract
\end{minipage}

\vfill


\begin{minipage}[b]{0.35\textwidth}
  \includegraphics[height=1.1cm,valign=c]{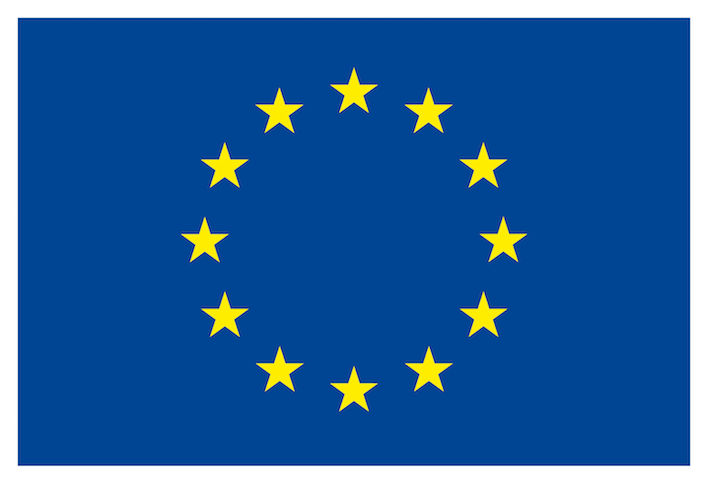}
  \hspace{0.1cm}
  \includegraphics[height=1cm,valign=c]{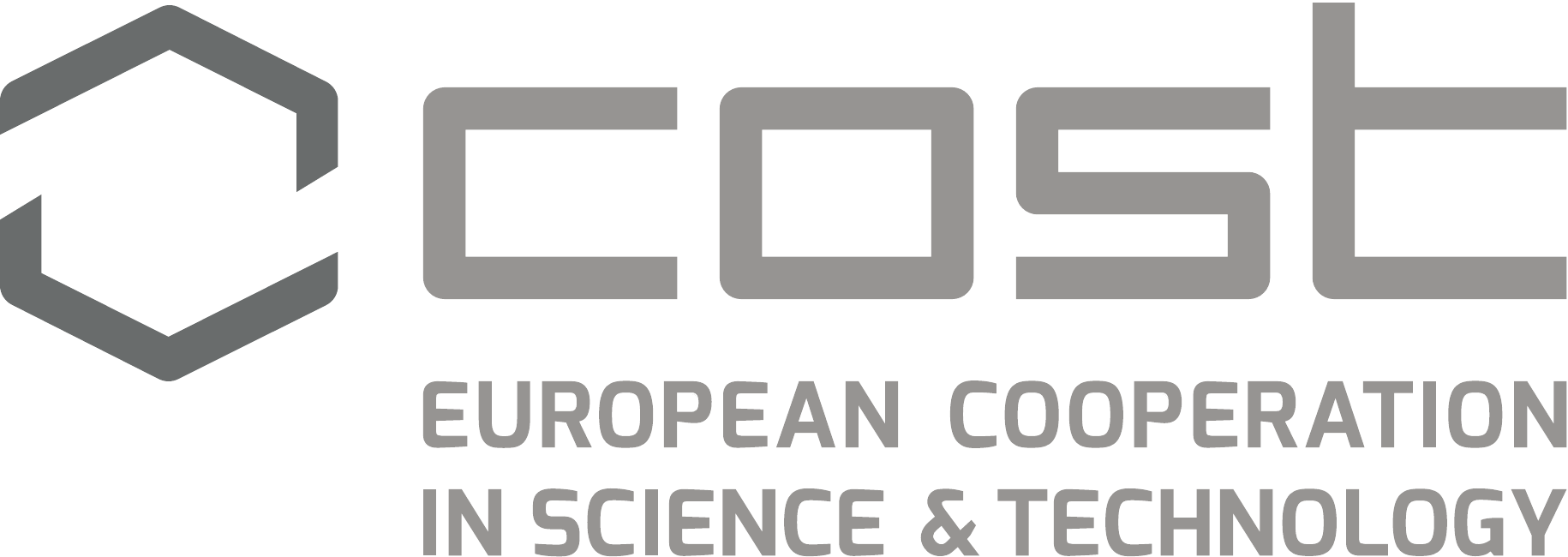}\\
\end{minipage}
\begin{minipage}[t]{0.2\textwidth}
  \begin{center}
  \includegraphics[width=0.8\textwidth]{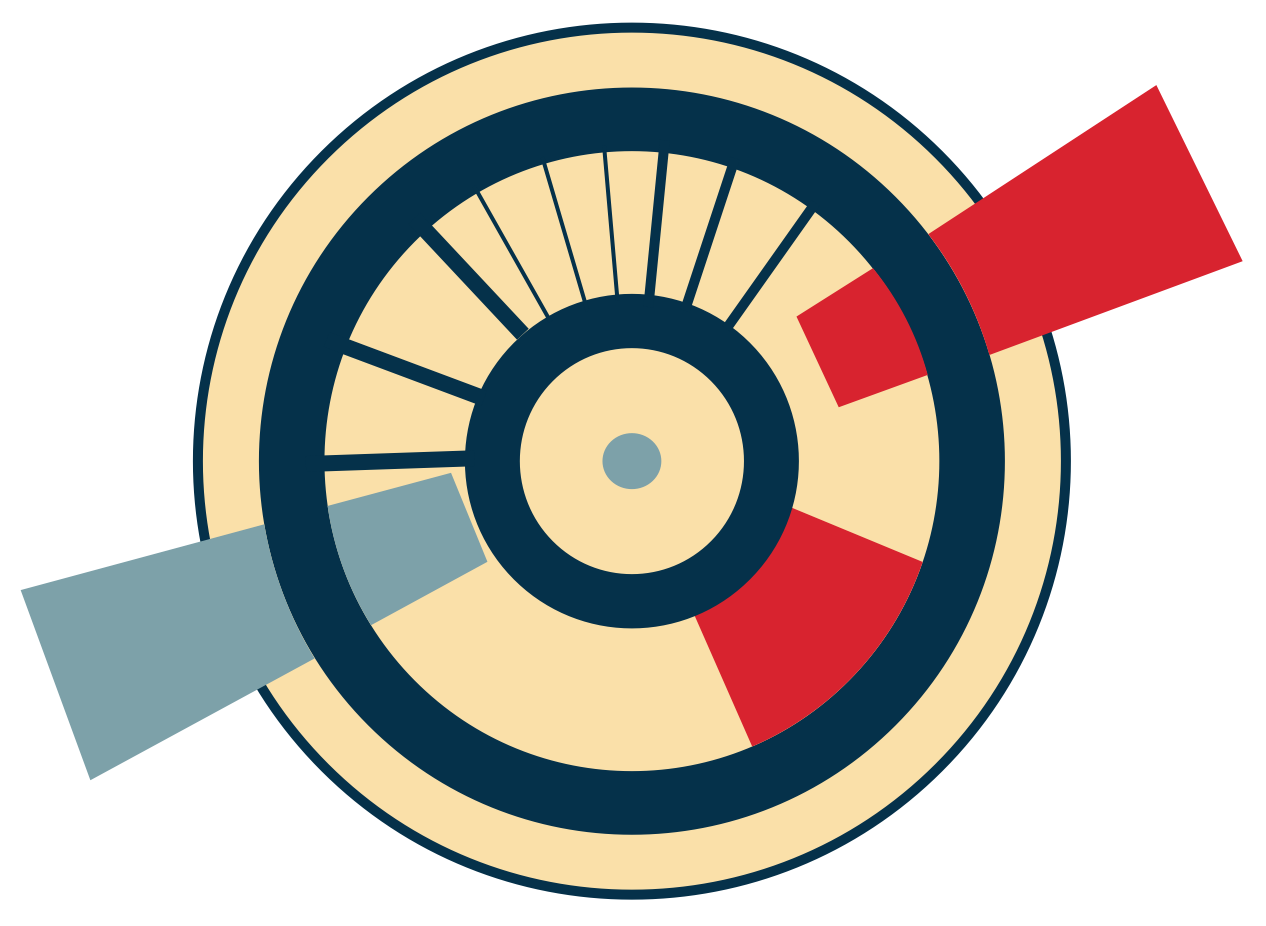}
  \end{center}
\end{minipage}
\begin{minipage}[b]{0.35\textwidth}
  \begin{flushright}
  \textit{Editors:}
  \editors
  \end{flushright}
\end{minipage}

} 
\end{center}

\newpage{}


}


\pagenumbering{roman}


{\pagestyle{plain}
\thispagestyle{plain}
{
\leftline{\textbf{Authors}}
\title{} 
\vspace{-1.5cm}
\author[TOR]{Riccardo Bellan}
\author[DES,TUD]{Jakob Beyer}
\author[TUD]{Carsten Bittrich}        
\author[MIB]{Giacomo Boldrini}            
\author[NIE,HEI]{Ilaria Brivio}            
\author[NIK]{Lucrezia Stella Bruni}    
\author[CHA]{Diogo Buarque Franzosi}    
\author[LLR]{Claude Charlot}     
\author[FIR]{Vitaliano Ciulli}     
\author[TOR]{Roberto Covarelli}
\author[SPL]{Duje Giljanovic}           
\author[FRE]{Giulia Gonella}           
\author[MIB]{Pietro Govoni}            
\author[CEA]{Philippe Gras}            
\author[PAV,IBM]{Michele Grossi}
\author[TUD]{Tim Herrmann}        
\author[WA2]{Jan Kalinowski}    
\author[ZUR]{Alexander Karlberg}    
\author[HEL]{Kimmo Kallonen}        
\author[THE]{Eirini Kasimi}       
\author[ADA]{Aysel Kayis Topaksu}        
\author[LJU]{Borut Kersevan}
\author[HEL]{Henning Kirschenmann}        
\author[TUD]{Michael Kobel}        
\author[THE]{Konstantinos Kordas}       
\author[HOU]{Antonios Leisos}     
\author[SPL]{Damir Lelas}           
\author[FIR]{Piergiulio Lenzi}     
\author[LLR]{Ang Li}     
\author[SHE]{Kristin Lohwasser}        
\author[LAP]{Narei Lorenzo-Martinez}   
\author[TUD]{Joany Manjarres}          
\author[MIB]{Dario Mapelli}
\author[HOU]{Alexandros Marantis}     
\author[MIB]{Matteo Marchegiani}            
\author[MIB]{Anna Mascellani}            
\author[THE]{Ioannis Maznas}       
\author[SHE]{Hannes Mildner}    
\author[KIT]{Matthias Ulrich Mozer}    
\author[KIT]{Max Neukum}    
\author[LJU]{Jakob Novak}                 
\author[TOR]{Giacomo Ortona}                 
\author[IST]{Kadri Ozdemir}                 
\author[CAM]{Mathieu Pellen}           
\author[TOR]{Giovanni Pelliccioli}
\author[THE]{Chariclia Petridou}       
\author[VIE]{Simon Pl\"atzer}       
\author[MIB]{Christian Quaggio}            
\author[KIT]{Michael Rauch}            
\author[PAV]{Daniela Rebuzzi}
\author[DES]{J\"urgen Reuter}
\author[THE]{Despoina Sampsonidou}     
\author[GVA]{Steven Schramm}      
\author[PAS,PRA]{Andrzej Siodmok}      
\author[PAS]{Magdalena Slawinska}      
\author[WAR]{Micha\l{} Szleper}    
\author[MIB]{Alessandro Tarabini}            
\author[TOR]{Connor Innes Thorburn}
\author[TUD]{Stefanie Todt}
\author[THE]{Spyridon E. Tzamarias}       
\author[HOU]{Apostolos Tsirigotis}     
\author[MIB]{Davide Valsecchi}
\author[TUD]{Lilly Wuest}
\author[NIK]{Marco Zaro}               
\affil[ADA]{Cukurova University, Adana, Turkey (TR)}
\affil[CAM]{University of Cambridge (UK)}
\affil[CEA]{IRFU, CEA, Universit\'e Paris-Saclay, Gif-sur-Yvette (FR)}
\affil[CHA]{Department of Physics, Chalmers University of Technology, Gotenburg (SE)}
\affil[DES]{Deutsches Elektronen-Synchrotron, Hamburg (DE)}
\affil[FIR]{University and INFN, Firenze (IT)}
\affil[FRE]{Albert-Ludwigs-Universit\"at Freiburg (DE)}
\affil[GVA]{Section de Physique, Universit\'e de Gen\`eve, Geneva, (CH)}
\affil[HEI]{Institut f\"{u}r Theoretische Physik, Universit\"{a}t Heidelberg (DE)}
\affil[HEL]{University of Helsinki and HIP (FI)}
\affil[HOU]{Hellenic Open University (GR)}
\affil[IBM]{IBM Italia (IT)}
\affil[IST]{Piri Reis University, Istanbul, Turkey (TR)}
\affil[KIT]{KIT - Karlsruhe Institute of Technology (DE)}
\affil[LAP]{LAPP, Universit\'e Grenoble Alpes, Universit\'e Savoie Mont Blanc, CNRS/IN2P3, Annecy (FR)}
\affil[LJU]{Department of Experimental Particle Physics, Jo\v{z}ef Stefan Institute and Department of Physics, University of Ljubljana (SI)}
\affil[LLR]{Laboratoire Leprince-Ringuet, CNRS/IN2P3, \'Ecole polytechnique, Institut Polytechnique de Paris, (FR)}
\affil[MIB]{University and INFN, Milano-Bicocca (IT)}
\affil[NIE]{Niels Bohr International Academy and Discovery Center, Niels Bohr Institute, Copenhagen University (DK)}
\affil[NIK]{Nikhef National institute for subatomic physics (NL)}
\affil[PAS]{Polish Academy of Sciences (PL)}
\affil[PRA]{Czech Technical University in Prague, Brehova 7, 115 19 Prague, Czech Republic (CZ)}
\affil[PAV]{University and INFN, Pavia (IT)}
\affil[SHE]{University of Sheffield (GB)}
\affil[SPL]{University of Split, FESB (HR)}
\affil[THE]{Aristotle University of Thessalon\'iki (GR)}
\affil[TOR]{University and INFN Torino (IT)}
\affil[TUD]{Technische Universit\"at Dresden (DE)}
\affil[VIE]{Particle Physics, Faculty of Physics, University of Vienna, 1090 Wien (AT)}
\affil[WA2]{Faculty of Physics, University of Warsaw, Warsaw (PL)}
\affil[WAR]{National Center for Nuclear Research, Warsaw (PL)}
\affil[ZUR]{Physics Institute, University of Z\"urich, Winterthurerstrasse 190, 8057 Z\"urich (CH)}

\maketitle
}

}
\newpage
\tableofcontents
%


\renewcommand{\thechapter}{\arabic{chapter}}
\renewcommand{\thesection}{\arabic{chapter}.\arabic{section}}
\renewcommand{\thesubsection}{\arabic{chapter}.\arabic{section}.\alph{subsection}}
\renewcommand{\thesubsubsection}{\arabic{chapter}.\arabic{section}.\alph{subsection}.\roman{subsubsection}}

\pagenumbering{arabic}
\setcounter{chapter}{0}
\setcounter{section}{0}
\setcounter{subsection}{0}
\setcounter{subsubsection}{0}
\setcounter{footnote}{0}

\let\cleardoublepage\clearpage

\cleardoublepage
\phantomsection
\addcontentsline{toc}{part}{Introduction}

\chapter*{Introduction\markboth{Introduction}{Introduction}} 
\label{chapter:intro}

The VBSCan COST Action is a four-year project,
funded by the Horizon 2020 Framework Programme of the European Union,
aiming at a consistent and coordinated study of VBS
from the phenomenological and experimental points of view, 
gathering all the interested parties in the high-energy physics community,
together with experts of data mining techniques.

After one year of collaboration,
the network developed its work in both the experimental and phenomenological directions,
to provide a state-of-the-art theoretical framework
and data analysis techniques
to exploit at best the data collected at the CERN Large Hadron Collider
in terms of precision measurements of the Stanard Model properties
and of new physics constraints.

This document, 
which follows the report of the kick-off meeting of the action 
in which the ground for scientific work was settled~\cite{Anders:2018gfr}, 
showcases the results achieved during the first year of activities. 
In particular it summarises the outcome
of the second annual meeting of the action\footnote{\url{https://indico.cern.ch/event/706178}}, held in Thessaloniki (Greece), in June 2018. 

The manuscript is structured in three chapters,
corresponding to the three working groups 
which focus on the scientific aspects of the collaboration.
Chapter~\ref{sec:wg1} is dedicated to the theoretical understanding,
targeting a detailed description of the signal 
and relative backgrounds in the SM, 
as well as effective field theory (EFT) and UV-complete modelling of BSM effects.
In this report,
particular focus is dedicated to the EFT studies,
where a detailed comparison of existing tools is performed
in order to lay the groundings for joint fits of experimental results,
and to phenomenological studies of the polarised component of VBS.
Quark-gluon discrimination,
an important ingredient to separate signal from backgrounds,
and theoretical uncertainties due to parton distribution functions
are discussed as well. 

Chapter~\ref{sec:wg2} focuses on analysis techniques,
defining data analysis protocols and performances 
to maximise the significance of the VBS analyses at hadron colliders, 
promoting the communication between theory and experiments.
Reconstruction techniques applied to leptonically-decaying W bosons
designed for the VBS case are presented, 
as well as long-term studies aimed at understanding VBS at future linear colliders.

Chapter~\ref{sec:wg3} promotes the optimal deployment of the studies 
in the experimental data analyses.
After a summary of existing experimental results, 
the studies for measurements combinations are presented,
followed by the application of machine learning techniques
to the reconstruction and identification of jets.


\renewcommand*{\thefootnote}{\fnsymbol{footnote}}


\chapter{Theoretical Understanding}
\label{sec:wg1}
\newcommand{\MP}[1]{{ {\color{blue}{ [MP: #1]}} }}
\newcommand{\GP}[1]{{ {\color{green}{ [GP: #1]}} }}

\section{Quark - gluon discrimination\footnote{speaker: A. Siodmok}}

Experimentally partons (quarks and gluons) can be studied by analysing so-called 
jets (collimated spray of particles and energy) whose kinematic properties reflect 
those of an initiating (unmeasurable) parton.
With a suitable jet definition, one can connect jet measurements made on clusters of hadrons 
to perturbative calculations made on clusters of partons.  
More ambitiously, one can try to 
tag jets with a suitably-defined flavor label\cite{Banfi:2006hf,Komiske:2018vkc}, thereby enhancing the fraction of, say, 
quark-tagged jets over gluon-tagged jets. 
Being able to distinguish quark jets from gluon jets on an
event-by-event basis could significantly enhance the reach for many new physics searches
at the Large Hadron Collider. This is because beyond the Standard Model (BSM)  signals are often dominated by quarks 
(see for example a typical gluino-pair production
topology in~\cite{Gallicchio:2011xq}) while the corresponding Standard Model (SM)
backgrounds  are dominated by gluons. Quark tagging could also help to improve our 
understanding of SM physics, this is because many interesting SM processes lead to production of quarks.  
Let me mention just a few examples such as hadronic decays of $W$-boson ($W\rightarrow u\bar{d}\textrm{ or } c\bar{s}$),
top quark physics ($t\bar{t}\rightarrow b\bar{b} + 0,2 \textrm{ or } 4$ light quarks)
and finally, the most interesting for the VBScan project, the fact that 
in Vector Boson Fusion the two forward ``tag'' jets are quark jets.
It has also been noticed recently that quark/gluon tagging is crucial for 
the extraction of the strong coupling constant from jet substructure at the LHC~\cite{Bendavid:2018nar}.
That is why it is not a surprise that a wide variety of quark/gluon discriminants have been proposed 
\cite{Gallicchio:2011xq,Gallicchio:2012ez,Krohn:2012fg,Pandolfi:1480598,
Chatrchyan:2012sn,Larkoski:2013eya,Larkoski:2014pca,Bhattacherjee:2015psa,Bhattacherjee:2016bpy}, 
and there is a growing list of quark/gluon studies at the Large Hadron Collider (LHC) 
\cite{Aad:2014gea,Aad:2014bia,Khachatryan:2014dea,Aad:2015owa,Khachatryan:2015bnx,Aad:2016oit}.

Unfortunately, the LHC measurements revealed that quark- and gluon-jets look different in
the data than in the Monte Carlo Generators, which are key tools in High Energy Physics.
To be more precise LHC measurements show that quark- and gluon-jets look more similar to each other in
the data than in the Pythia~\cite{Sjostrand:2006za, Sjostrand:2014zea} simulation and less similar than in the 
Herwig~\cite{Bahr:2008pv,Bellm:2015jjp} simulation. 
As a result, the ability of the tagger to reject gluons at a 
fixed quark efficiency is up to a factor of two better in Pythia and
up to 50\% worse in Herwig than in data\cite{Aad:2014gea}. 
We tried to understand this problem better, therefore 
we decided to study a simpler situation of electron-positron collisions. 
In this case one can define a proxy for quark 
and gluon jets based on the Lorentz structure of the production vertex. Our study revealed 
a fascinating interplay between perturbative shower effects and non-perturbative  colour reconnection effects.
It was a big surprise since the non-perturbative colour reconnection models 
were introduced in order to describe hadronic collisions. Of course 
in principle universality requires that the colour reconnection model
is also used to describe leptonic collisions. In practice however 
colour reconnection so far has little effect on the LEP distributions which 
have been used to develop and tune the models.

\begin{figure}
\centering
{
\includegraphics[width = 0.50\columnwidth]{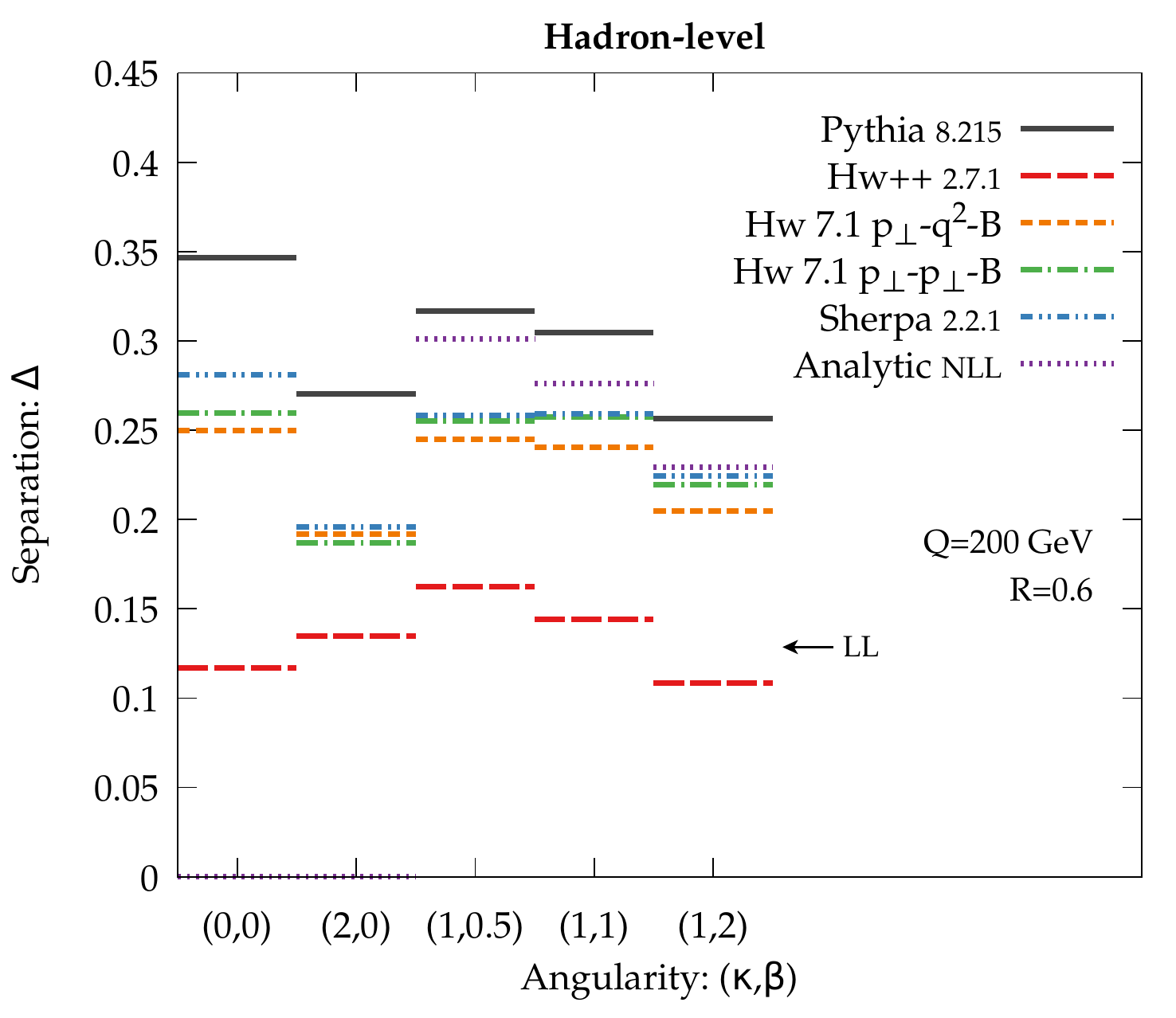}
}
\caption{Classifier separation $\Delta$ for the five angularities, determined from the various generators at hadron level
for an idealized case of $e^+ e^-$ collisions.  
The first two columns correspond to IRC-unsafe distributions (multiplicity and $p_T^D$), while the last three columns are the IRC-safe angularities.  
}
\label{fig:ee}
\end{figure}

These results triggered new developments in
the simulation of quark and gluon jets in parton-shower generator
Herwig. In~\cite{Reichelt:2017hts}, we provided deeper understanding of how the colour reconnection affects 
quark/gluon jets tagging and improve existing colour reconnection model in  Herwig 7~\cite{Gieseke:2012ft}. 
As a results we  provided more robust predictions for gluon jets which were for example appreciated in 
a recent resummation calculation by J. Mo, F.Tackmann, W. Waalewijn~\cite{Mo:2017gzp} who stated that their 
result {\it{``highlights the substantial improvement in the description of gluon jets
in the latest version of Herwig''}}. To quantify the improvements let us show
results from~\cite{Reichelt:2017hts} for five quark/gluon discriminants so-called generalized angularities $\lambda^{\kappa}_{\beta}$~\cite{Larkoski:2014pca}:
\begin{equation*}
\arraycolsep=5pt
\begin{array}{cccccc}
\label{eq:ang}
 (\kappa,\beta)&(0,0) & (2,0) & (1,0.5) & (1,1) & (1,2) \\
\lambda^{\kappa}_{\beta}: & \text{multiplicity} &  p_T^D &  \text{LHA} & \text{width} & \text{mass}
\end{array}
\end{equation*}
where 
$\lambda^{\kappa}_{\beta} = \sum_{i \in \text{jet}} z_i^\kappa \theta_i^\beta,$
 $i$ runs over the jet constituents, $z_i \in [0,1]$ is a momentum fraction, 
and $\theta_i \in [0,1]$ is an angle to the jet axis.
To quantify discrimination performance, we use classifier separation:
\begin{equation*}
\Delta =  \frac{1}{2} \int \text{d} \lambda \, \frac{\bigl(p_q(\lambda) - p_g(\lambda)\bigr)^2}{p_q(\lambda) + p_g(\lambda)},
\end{equation*}
where $p_q$ ($p_g$) is the probability distribution for $\lambda$ in a generated quark jet (gluon jet) 
sample. $\Delta = 0$ corresponds to no discrimination power and $\Delta = 1$ corresponds to perfect 
discrimination power.
In Figure~\ref{fig:ee} we show the discrimination power as a function of an angularity predicted by 
\textsc{Pythia 8.215} \cite{Sjostrand:2014zea},
\textsc{Herwig++ 2.7.1} \cite{Bahr:2008pv}, \textsc{Sherpa 2.2.1} \cite{Gleisberg:2008ta}
, the NNL analytical calculation from~\cite{Gras:2017jty} and two tunes of improved version of Herwig (denoted by $p_\perp$-$q^2$-B and 
$p_\perp$-$p_\perp$-B). Firstly, we see that the both \textsf{Herwig~7.1} 
tunes give significantly different results 
compared to \textsc{Herwig++ 2.7.1}.
Secondly, the results of bht the tunes of improved version of \textsf{Herwig~7.1} are quite similar and closer to the other predictions
giving more constrained prediction on the quark/gluon jet discrimination power in $e^+ e^-$ collisions. 
It would be interesting to estimate the parton-shower uncertainties~\cite{Bellm:2016rhh,Bellm:2016voq,Mrenna:2016sih,Bothmann:2016nao}
in the context of the quark and gluon jet discrimination observables to see whether the remaining 
discrepancy in the predictions is covered by the uncertainty band.
Finally, it would be very interesting to apply the new tools and methods in the context of VBS which we plan to do in the future.
We would also like to validate the models further and provide even better tools for SM measurements and BSM searches 
based on quark and gluon jets. Therefore, we plan to propose novel experimental strategies to measure quark/gluon jets
which then could be used to constrain models even more.

\section{EFT applicability study in the same-sign $WW$ process with leptonic decays\footnote{speaker: M. Szleper}}

There are two possible approaches to use the Effective Field Theory (EFT) to describe 
data.  One of them is by varying all the Wilson coefficients of higher-dimension 
operators simultaneously, in a global fit to all the relevant
physical processes, including VBS.  While this approach is the most correct one from
the formal point of view, there is a practical problem with it.
Apart from the technical complexity of the procedure, one needs first of
all a complete basis of operators implemented in an event generator.  Such basis exists
so far for \dimS{} operators only and its extension to \dimE{} cannot be expected
soon.  It is known that all \dimS{} operators can be probed to a better
precision in other processes than VBS, so this approach does not guarantee the optimal 
exploitation of VBS data.

The second approach is trying to fit the data from a single physical process
with one higher-dimension operator at a time (or a few, but for practical reasons
usually not more than two).  The advantage is probing the quartic
$VVVV$ couplings with \dimE{} operators, for which VBS processes are known to be
the best laboratory.  This approach has nonetheless severe limitations by construction.
Choosing a particular operator to vary, while setting all the remaining Wilson
coefficients to zero, effectively means restricting oneself to a
(rather narrow) class of BSM theories in which this choice is indeed a good approximation
for the studied process in the available energy range.
This breaks model-independence of the EFT.  

The purpose of this study is to verify the physics potential of the second approach.
Correct treatment of the EFT requires strictly watching the 
EFT cutoff parameter $\Lambda$.  The cutoff defines the maximum value of the $VV$
invariant mass for which the approach is valid.  It is unknown a priori, except that
it cannot be higher than the lowest unitarity bound.  Unfortunately, in purely leptonic 
decays of the $WW$ process, the $WW$ mass is not an observable due to two escaping
neutrinos, hence any measured distribution will in the general case be a sum of 
the respective contributions from below and above $\Lambda$.
This has important implications for the preferred data analysis strategy, as well as
for the practical usefulness of the entire approach in the case of BSM observation.  
The EFT-controlled signal is the ``clipped" one, which is calculated by applying
a sharp cutoff at $\Lambda$, and assuming only the Standard Model above it.
The total BSM signal will contain an additional contribution from above $\Lambda$,
which can be estimated within some limits from the expected asymptotic behaviour of
the regularised amplitudes.
Moreover, the value of $\Lambda$ must be treated as a free 
parameter and varied between twice the $W$ mass and the unitarity limit.  Successful 
description of the data in terms of a chosen EFT operator with Wilson coefficient $c$
is only possible provided the part above $\Lambda$ is small enough so it does not 
significantly distort the measured distributions.  This implies an effective 
upper bound on $\Lambda$ versus $c$.  On the other hand, BSM signal observability 
imposes a lower bound.

\begin{figure}[hbtp]
\centering
\includegraphics[width=0.32\linewidth]{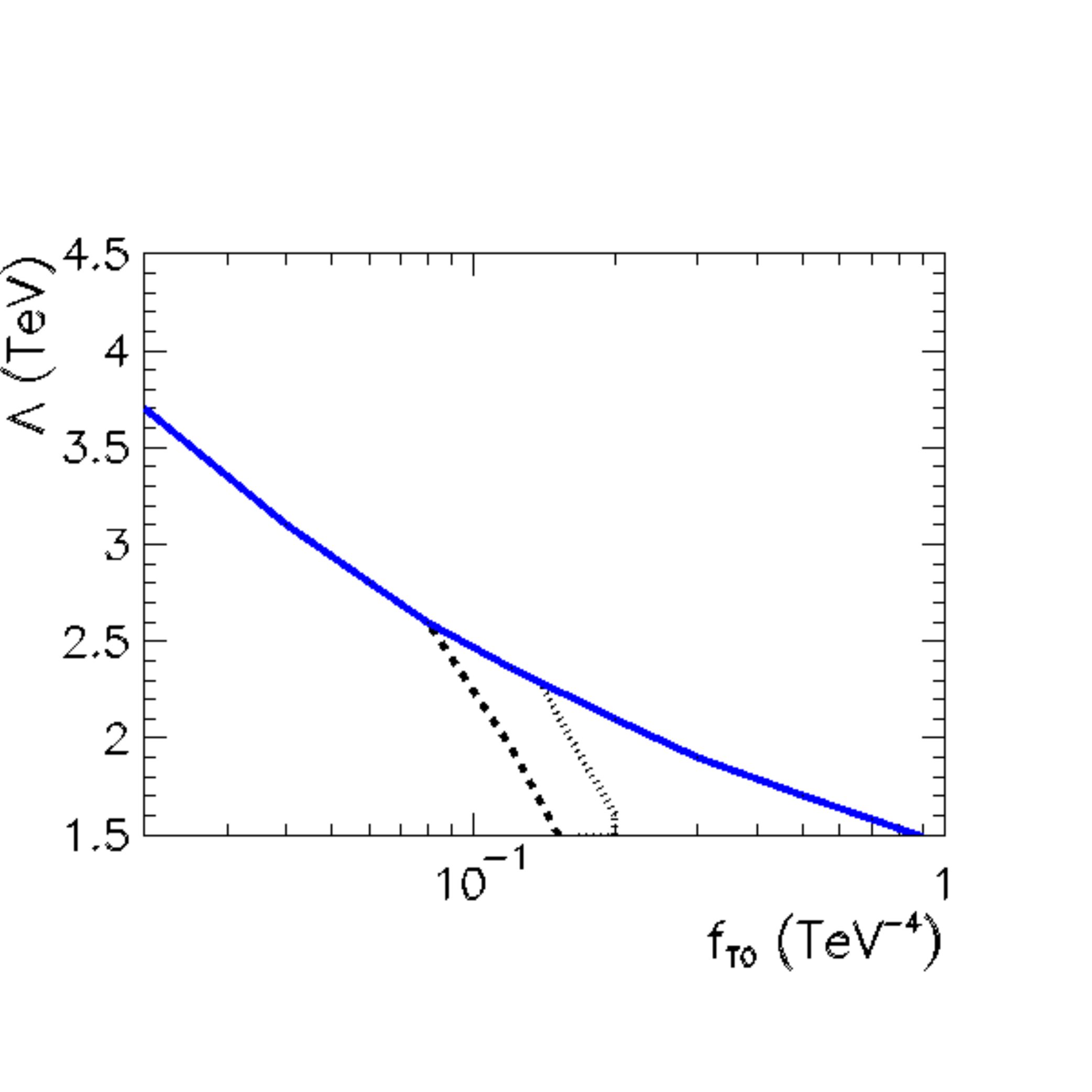}
\includegraphics[width=0.32\linewidth]{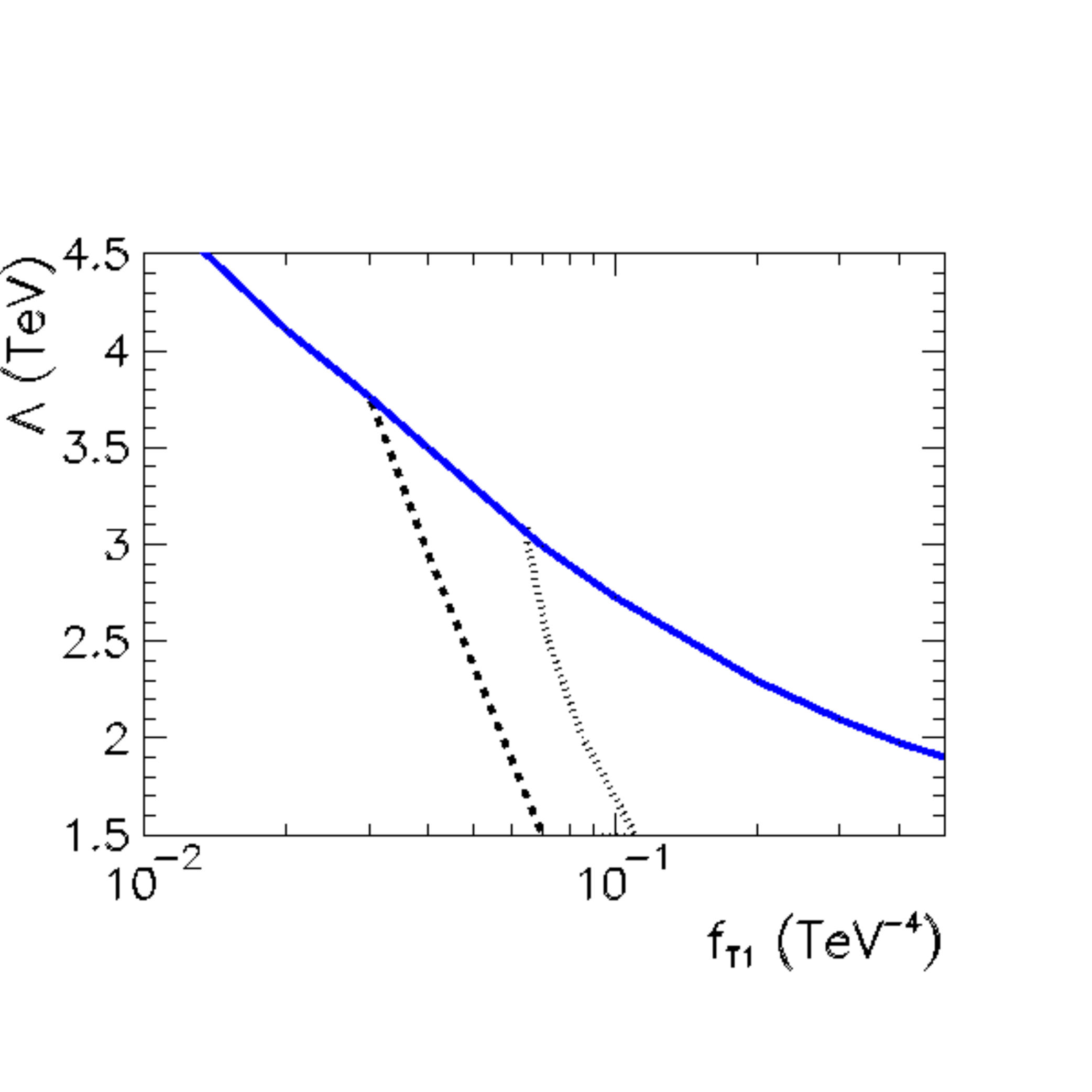}
\includegraphics[width=0.32\linewidth]{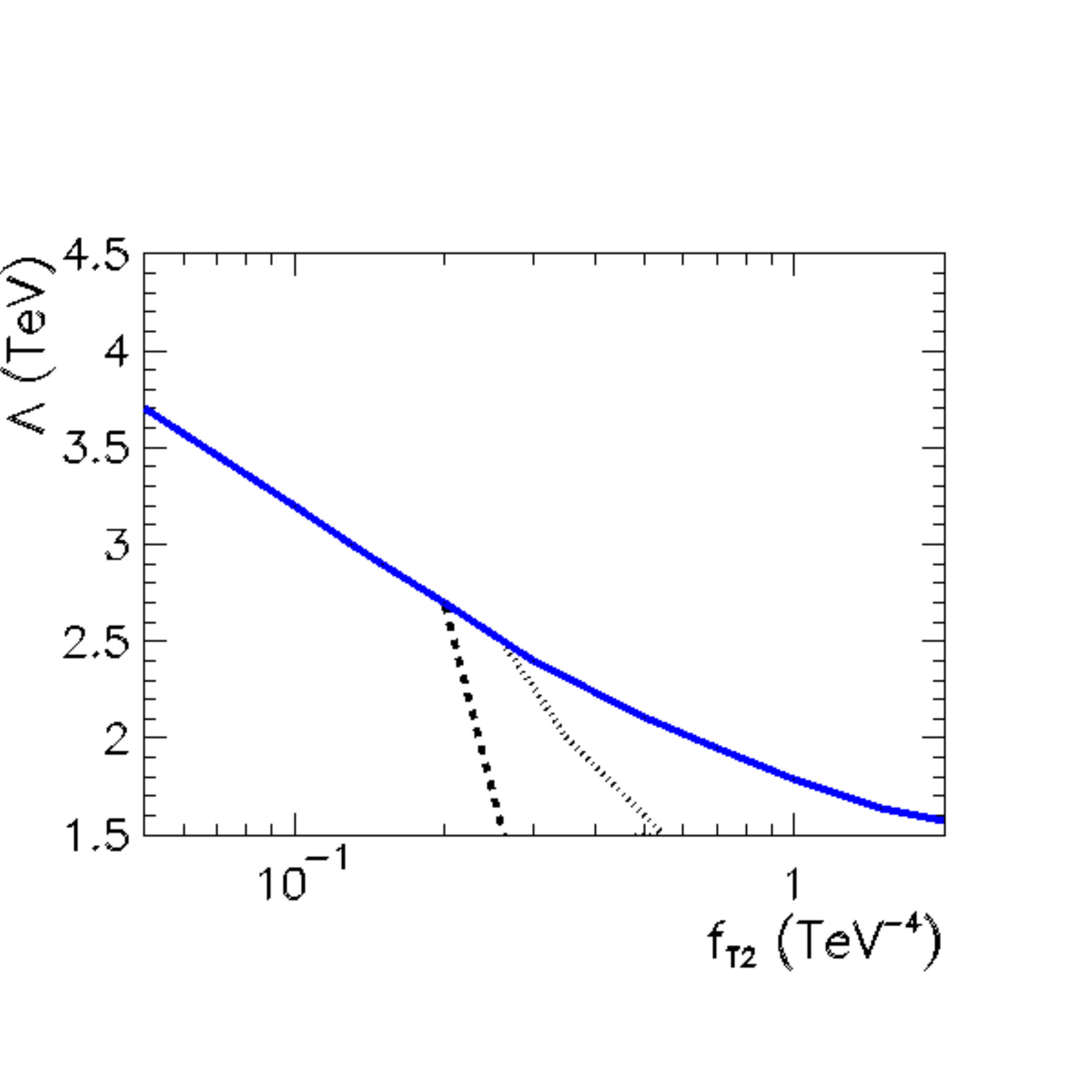}
\caption{``EFT triangles" for $f_{T0}$, $f_{T1}$ and $f_{T2}$:
regions in the $\Lambda$ versus $f$ space
where a 5$\sigma$ BSM signal can be observed and the EFT description is applicable.
Unitarity limits are shown in blue, black dashed lines denote the lower limits on
5$\sigma$ signal significance, black dotted lines denote the upper limits on
2$\sigma$ statistical consistency between the EFT-controlled signal and the total
measured signal.
Assumed is $\sqrt{s}$ = 14 TeV and an integrated luminosity of 3 ab$^{-1}$.
There is no detector simulation in this study.
}
\label{fig:EFTTriangles}
\end{figure}

Dedicated simulation work has been done to find the remaining parameter space
for each  \dimE{} operator separately, assuming proton-proton collisions at 14 TeV
and an integrated luminosity of 3 ab$^{-1}$.
It was required a BSM signal observability at a
5$\sigma$ or larger level and statistical consistency between the EFT-controlled part
of the signal and the total measured signal within 2$\sigma$.
The latter was estimated
under the working assumption of frozen amplitudes above the cutoff value.  Details 
of the study can be found in Ref.~\cite{Kalinowski:2018oxd}.  After imposing the two requirements, 
narrow ranges were found left for the $T$ operators, around 
$f_{T0} \sim 0.1$ TeV$^{-4}$, $f_{T1} \sim 0.05$ TeV$^{-4}$ and
$f_{T2} \sim 0.3$ TeV$^{-4}$, see Figure~\ref{fig:EFTTriangles}.
More problematic were $M$ and $S$ operators, for which usually only small parameter 
spaces close to the strong interaction limit ($c \to 4\pi$)
were found left.  It can be expected that some of these spaces will close up once
full detector simulation is included in the analysis. A recent study carried for the High Energy LHC option confirmed the same conclusions hold regardless of the actual proton beam energy~\cite{KozowEFT}.

\vspace{5mm}

These parameter space restrictions do not apply in the case of no BSM observation, 
and therefore setting limits on individual Wilson coefficients.  However, ``clipped"
templates from simulation should be used to fit the data in order to determine 
physically interpretable limits and a feature
of the method is that only two-dimensional limits, $c$ versus $\Lambda$, can be placed.

Given relatively little room for describing potential BSM physics using the approach
of varying one \dimE{} operator at a time, it could be worthwhile to attempt
limited global fits, where data from all $VV$ processes would be fit simultaneously,
including semi-leptonic decay modes, and many (ideally, all that affect $VVVV$ quartic
couplings) \dimE{} operators varied
at a time.  The potential impact of non-zero \dimS{} operators in such
analysis is still to be determined.

\section{EFT - codes comparison\footnote{Author: I. Brivio. Speaker: M. Rauch.}}

The presence of new physics contributions in VBS processes can be probed in a model-independent way using the language of the SM Effective Field Theory\footnote{This approach is model-independent up to the assumptions that new physics, if present, is nearly decoupled and that the SM symmetries and field content provide a correct and complete description of physics at the EW scale.}. The latter expands the SM Lagrangian with the inclusion of higher dimensional ($d>4$) operators multiplied by unknown parameters called Wilson coefficients. In practice, the impact of new physics with typical scale $\Lambda$ on accessible processes is 
organized in a Taylor expansion in $E/\Lambda$, with the largest ($L$ and $B$ conserving) BSM effects stemming from the \dimS{} terms.

A plan for the theoretical EFT analysis of VBS processes has been defined in a series of meetings that took place prior to this workshop, and the main goal is the production of accurate predictions for the VBS signal in the presence of selected \dimS{} operators.
Several Monte Carlo codes can be employed to this aim, that differ in the implemented algorithms and operator sets.
A useful preliminary step is then to compare the performance and characteristics of these codes.
This enables one to identify strengths and weaknesses of each software and to cross-check their predictions.

At present, the codes considered are: {\tt MadGraph5\_aMC@NLO} ({\tt MG5\_aMC} for short)~\cite{Alwall:2014hca} with the {\tt SMEFTsim} package~\cite{Brivio:2017btx}, {\tt VBFNLO}~\cite{Arnold:2008rz,Arnold:2011wj,Baglio:2014uba}, and {\tt Whizard}~\cite{Kilian:2007gr}.
As a starting point, we focus here on {\tt SMEFTsim} vs {\tt VBFNLO}. The main features of these codes are compared in Table~\ref{tab:SMEFTsim_vs_VBFNLO}.

The {\tt SMEFTsim} package is a FeynRules~\cite{Alloul:2013bka} and UFO~\cite{Degrande:2011ua} model that implements the complete $B$-conserving Warsaw basis of \dimS{} operators~\cite{Grzadkowski:2010es} for three generations. Field redefinitions to make the kinetic terms canonical  and parameter shifts due to fixing an input parameter set are automatically performed on the Lagrangian (see~\cite{Brivio:2017btx} for details).
The model is available in six different frameworks, namely for three flavour structures (general, $U(3)^5$ symmetric and MFV with linear expansion in the flavour spurions) and two input scheme choices for the EW sector ($\{\alpha_{em},m_Z,G_F\}$ or $\{m_W,m_Z,G_F\}$).
In what follows we consider only the $U(3)^5$ symmetric model, with $m_W$ input scheme.
The model is optimised for the estimation of LO EFT contributions in unitary gauge and it is not equipped for NLO calculations.
Although it can in principle be interfaced to any Monte Carlo that uses the UFO format, it is customarily used with {\tt MG5\_aMC}.
Interestingly, the latter allows to estimate independently contributions of different order in the anomalous couplings (i.e. separating EFT-SM interference vs. quadratic EFT term and to control the number of EFT insertions in a given diagram).
Unitarisation procedures have not been embedded in this framework.

{\tt VBFNLO} implements the $d=6$ operators of the HISZ basis~\cite{Hagiwara:1993ck} and the $d=8$ operators of the \'Eboli basis~\cite{Eboli:2006wa}. Both contain only purely bosonic interactions (CP even and CP odd). The dependence on the EFT terms is hard-coded for \emph{all} VBS processes, including diagrams with up to two EFT insertions.
As a consequence, only the full matrix element can be computed, that contains EFT contributions up to order $(c/\Lambda)^4$.
Unitarisation methods are available in {\tt VBFNLO}. An example is the dipole form factor
\begin{equation}
F=\left(1+\frac{m_{{\rm inv},\sum \ell}^2}{M^2} \right)^{-p}\,,
\end{equation}
where $m_{{\rm inv},\sum \ell}$ is the total invariant mass of the leptons, $M$ is the characteristic scale where the form factor effects become relevant and $p$ is an exponent controlling the damping. This feature can be advantageous for studies on the validity range of the EFT.


\begin{table}[t]\centering
\renewcommand{\arraystretch}{1.7}
\begin{tabular}{p{6.5cm}@{\hspace*{5mm}}|@{\hspace*{5mm}}p{6.5cm}}
\hline
\centering {\tt MG5\_aMC} + {\tt SMEFTsim} & 
\centering {\tt VBFNLO}
\tabularnewline\hline
complete $d=6$ Warsaw basis
&
$d=6$ HISZ basis + $d=8$ \'Eboli basis
\\
tree-level EFT
&
tree-level EFT
\\
$\{m_W,m_Z,G_F\}$ or $\{\alpha_{em},m_Z,G_F\}$ input schemes&
$\{m_W,m_Z,G_F\}$ input scheme
\\
can compute contributions of different order in anomalous couplings&
computes squared matrix element with all powers of operator insertions 

(with up to 2 insertions per  diagram)
\\
computes matrix elements using pure EFT expansion&
allows unitarisation of cross section by different methods
\\\hline
\end{tabular}
\caption{Comparison of the main characteristics of {\tt MG5\_aMC} with {\tt SMEFTsim} and {\tt VBFNLO}. See the text for further details.}\label{tab:SMEFTsim_vs_VBFNLO}
\end{table}

We first test the two codes on a simple process, namely $W^+ Z$ diboson production at LHC, and subsequently on same-sign WW ($W^+ W^+ j j$) production. The comparison is carried out in the SM limit and in the presence of one EFT operator at a time. A direct comparison is only possible for operators that are implemented both in {\tt SMEFTsim} and in {\tt VBFNLO}, which restricts the study to \dimS{}, purely bosonic invariants.
For consistency between the two codes, we consider all the EW Feynman diagrams with \emph{up to two insertions} of a given EFT operator and all the terms in the squared amplitude are retained.
A subtlety is the treatment of the width of intermediate particles, that in general does receive EFT corrections. Both codes calculate the widths first, and then make use of the result in the event generation. For the outputs of {\tt MG5\_aMC} and {\tt VBFNLO} to be consistent it is necessary to set the gauge bosons' widths to {\tt Auto} in {\tt MG5\_aMC} and to manually switch off the NLO QCD corrections to $W,Z \rightarrow q\bar q$ in {\tt VBFNLO}.

\subsection*{Results for diboson production}

\begin{figure}[t]\centering
\hspace*{-10mm}
\includegraphics[width=.55\textwidth]{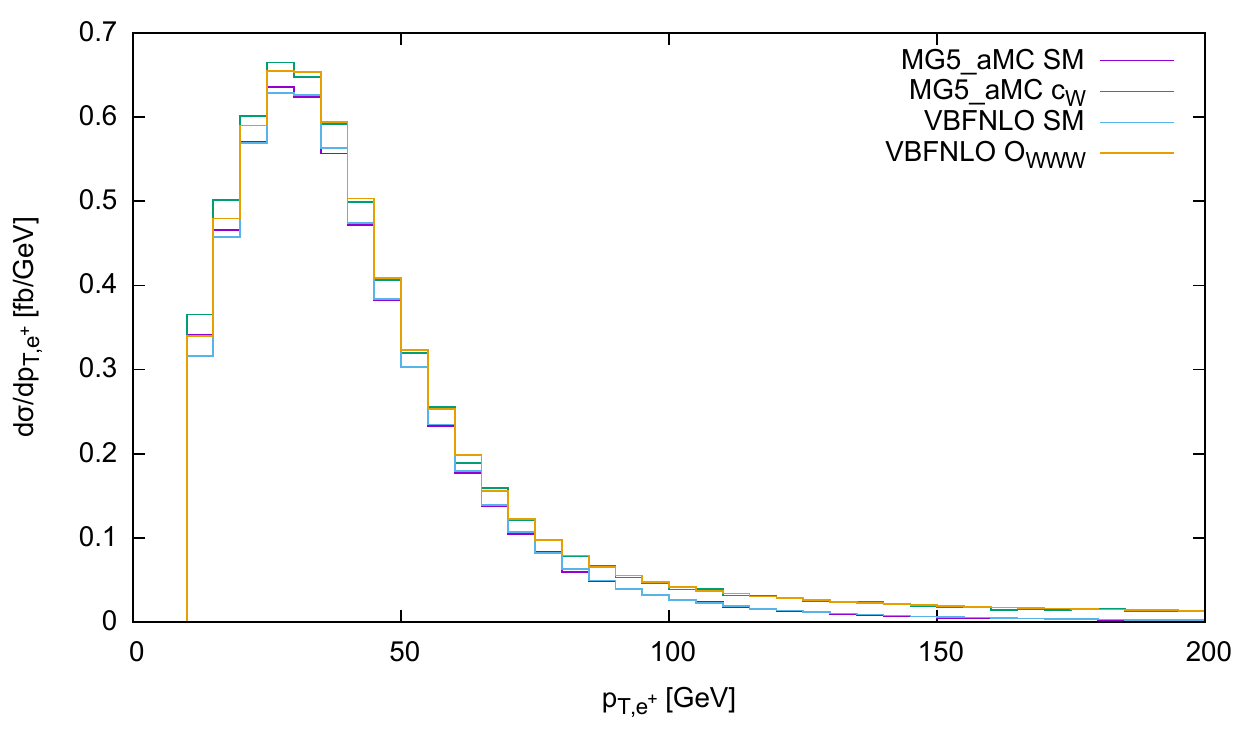}~
\includegraphics[width=.55\textwidth]{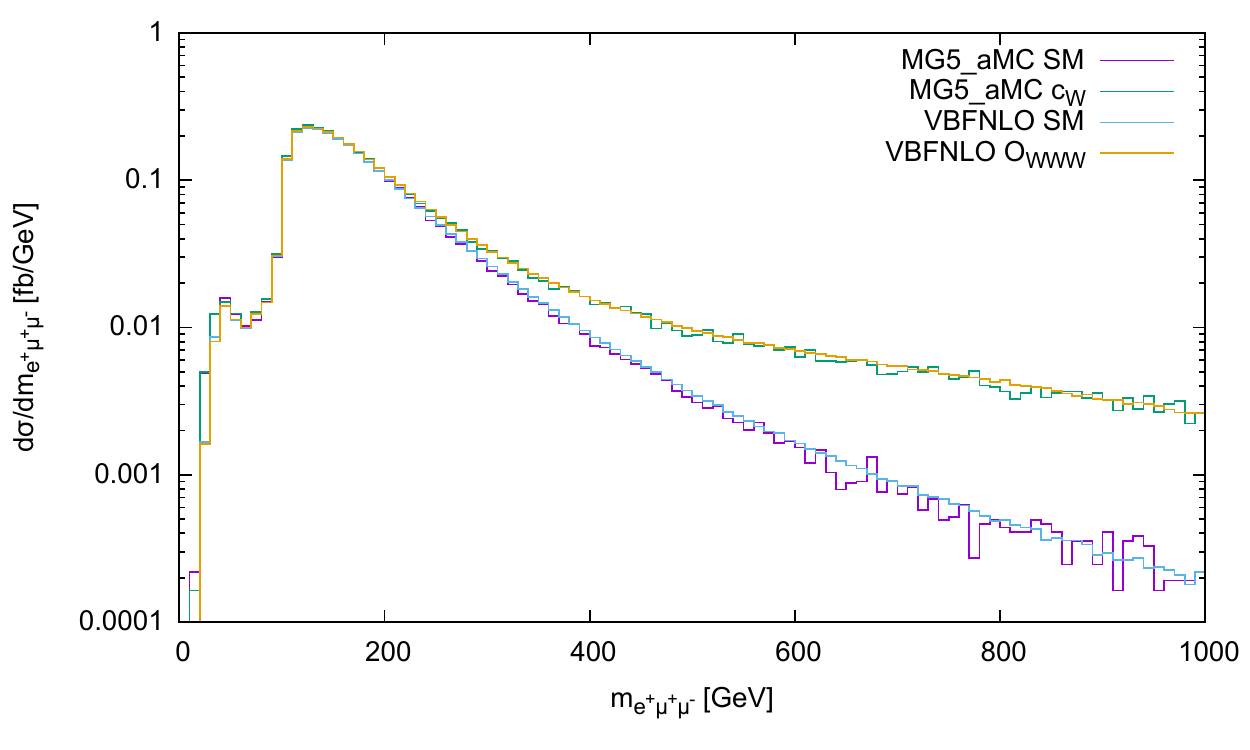}
\caption{Differential distributions $d\sigma/dp_{T,e^+}$ (left) and $d\sigma/dm_{e \mu\mu}$ (right) for $W^+Z \rightarrow e^+\nu_e\mu^+\mu^-$ production obtained with {\tt MG5\_aMC} + {\tt SMEFTsim} and {\tt VBFNLO} for two benchmark setups: $c_W = 0$ (SM) and $c_W = (g^3/4) f_{WWW}/\Lambda^2= \unit[1]{TeV^{-2}}$.}\label{plot:EFTcfr_WZ_pTe}
\end{figure}

As a preliminary test, we generate events for the process $p p \rightarrow e^+ \nu_e \mu^+ \mu^-$. The technical specifications and generation cuts are summarised in Table~\ref{tab:EFTcfr_WZ_specs}.
\begin{table}[t]\centering
\renewcommand{\arraystretch}{1.5}
\begin{tabular}{l|ll}
\hline
process & $pp\rightarrow e^+ \nu_e \mu^+ \mu^-$~~QCD=0,~~LO\\
center of mass energy& $\sqrt s = 13$ TeV\\
PDF and factor. scale& {\tt PDF4LHC15\_nlo\_mc\_pdfas}, &$\mu_F = 91.188$ GeV\\
input parameters&  $m_W = \unit[80.387]{GeV}$&
                    $G_F = \unit[1.166379\cdot 10^{-5}]{GeV^{-2}}$\\[-2mm]
                &  $m_Z = \unit[91.1876]{GeV}$&
                   $m_t = \unit[173.2]{GeV}$\\[-2mm]
                &   $m_{u,d,s,c,e,\mu} =0$&
                   $m_b = \unit[4.18]{GeV}$
                   \\
statistics& 
{\tt MG5\_aMC} + {\tt SMEFTsim}:& $10^5$ events\\[-2mm]
& {\tt VBFNLO}:& $2^{26}$ points, 6 iterations\\
generation cuts&
$p_{T,{\rm miss}}>\unit[20]{GeV}$&
$p_{T,\ell}>\unit[20]{GeV}$\\[-2mm]
& $R_{\ell\ell}>0.4$& 
$m_{\ell\ell}>\unit[15]{GeV}$
\\\hline
\end{tabular}
\caption{Technical specifications for the comparison of {\tt MG5\_aMC} + {\tt SMEFTsim} and {\tt VBFNLO} in the generation of $W^+Z$ diboson production.}\label{tab:EFTcfr_WZ_specs}
\end{table}
We look only at the impact of the operator
\begin{equation}
\mathcal{Q}_W = \epsilon_{ijk} W^{i\mu}_{\nu}W^{j\nu}_\rho W^{k\rho}_\mu\,.
\end{equation}
In the notation of Ref.~\cite{Grzadkowski:2010es}, we consider the Lagrangian $\mathcal{L}_{SM} + c_W \mathcal{Q}_W$ with the two benchmark values $c_W = 0$ and $c_W=\unit[1]{TeV^{-2}}$. Note that in the notation of Ref.~\cite{Hagiwara:1993ck} the Lagrangian reads
\begin{align}
\mathcal{L} &= \mathcal{L}_{SM} + \frac{f_{WWW}}{\Lambda^2} \mathcal{O}_{WWW}\,,\\
\mathcal{O}_{WWW} &= {\rm Tr}\left[ \hat W^{\mu}_{\nu}\hat W^{\nu}_\rho \hat W^{\rho}_\mu\right]\,,\quad \hat W^\mu_\nu = \frac{ig}{2}W^{i\mu}_\nu \sigma^i.
\end{align}
Therefore $c_W = 1$ is equivalent to
\begin{equation}
\frac{f_{WWW}}{\Lambda^2} =\frac{4}{g^3} c_W \simeq \unit[14.4]{TeV^{-2}}\,. 
\end{equation}
We compare the integrated cross section as well as the differential distributions in $p_T$ and $\eta$ of each lepton in the final state, and in the invariant masses $m_{\mu\mu}$, $m_{e \mu\mu}$, $m_\text{ all leptons}$. 
We find good agreement for all the observables.
The total cross sections are
\begin{center}
\renewcommand{\arraystretch}{1.3}
\begin{tabular}{l|cc}
\hline
 $pp\rightarrow e^+ \nu_e \mu^+ \mu^-$& $c_W=0$ (fb)& $c_W=1$ (fb)\\\hline
{\tt MG5\_aMC} + {\tt SMEFTsim}& $27.30 \pm 0.03$ & $32.80 \pm 0.03$\\
{\tt VBFNLO}& $27.24 \pm 0.01$ & $32.78 \pm 0.01$\\\hline
\end{tabular}
\end{center}
and, as illustration of the differential distributions, $d\sigma/dp_{T,e}$ and  $d\sigma/dm_{e \mu\mu}$ are shown in Figure~\ref{plot:EFTcfr_WZ_pTe}. 

\subsection*{Results for same-sign WW production}
Given the excellent agreement found for diboson production, we move on to VBS in the same-sign $WW$ channel. The technical specifications for this process  are reported in Table~\ref{tab:EFTcfr_ssWW_specs}.
\begin{table}[b]\centering
\renewcommand{\arraystretch}{1.5}
\begin{tabular}{l|ll}
\hline
process & $pp\rightarrow e^+ \nu_e \mu^+ \nu_\mu j j$~~QCD=0,~~LO\\
center of mass energy& $\sqrt s = 13$ TeV\\
statistics& 
{\tt MG5\_aMC} + {\tt SMEFTsim}:& $10^4$ events\\[-2mm]
& {\tt VBFNLO}:& $2^{26}$ points, 6 iterations\\
generation cuts&
$p_{T,\ell}>\unit[20]{GeV}$&
$p_{T,j}>\unit[20]{GeV}$\\[-2mm]
&$p_{T,{\rm miss}}>\unit[40]{GeV}$&
$|\Delta\eta_{jj}|>2.5$\\[-2mm]
& $|\eta_\ell|<2.5$&
$|\eta_j|<4.5$\\[-2mm]
& $R_{\ell\ell}>0.3$&
$R_{\ell j}>0.3$\\[-2mm]
&
$m_{\ell\ell}>\unit[15]{GeV}$&
$m_{jj}>\unit[500]{GeV}$
\\\hline
\end{tabular}
\caption{Technical specifications for the comparison of {\tt MG5\_aMC} + {\tt SMEFTsim} and {\tt VBFNLO} in the generation of $W^+W^+jj$ production. The input parameters, PDF set and factorization scale are the same as in Table~\ref{tab:EFTcfr_WZ_specs}.}\label{tab:EFTcfr_ssWW_specs}
\end{table}
We consider the same operator and benchmark values for $c_W\sim f_{WWW}$ as in the diboson case. For the integrated cross section we find
\begin{center}
\renewcommand{\arraystretch}{1.3}
\begin{tabular}{l|cc}
\hline
 $pp\rightarrow e^+ \nu_e \mu^+ \nu_\mu j j$& $c_W=0$ (fb)& $c_W=1$ (fb)\\\hline
{\tt MG5\_aMC} + {\tt SMEFTsim}& $1.602 \pm 0.003$ & $40.36 \pm 0.01$\\
{\tt VBFNLO}&  $1.5928 \pm 0.0005$ & $36.89 \pm 0.02$\\\hline
\end{tabular}
\end{center}
These results show good agreement in the SM case, while there is a small tension between the cross sections computed for $c_W=\unit[1]{TeV^{-2}}$, which is probably due to insufficient statistics in the {\tt MG5\_aMC} generation. 
A large number of differential distributions is also considered: the $p_T$ and $\eta$ of each particle in the final state (including leading and secondary lepton and the two jets sorted by their $p_T$), the total missing momentum $p_T^{\rm miss}$, the rapidity distance of the two jets $\Delta \eta_{jj}$, the invariant masses $m_{e\mu}$, $m_{4\ell}$, $m_{jj}$, $m_\text{all fermions}$, and the variable $m_T^{WWZ}\equiv m_{T,e\mu}\equiv \sqrt{(p_e+p_\mu+p_{T}^{\rm miss})^2}$~\cite{Aaboud:2016ffv}.

\begin{figure}[t]\centering
\hspace*{-1cm}
\includegraphics[width=.55\textwidth]{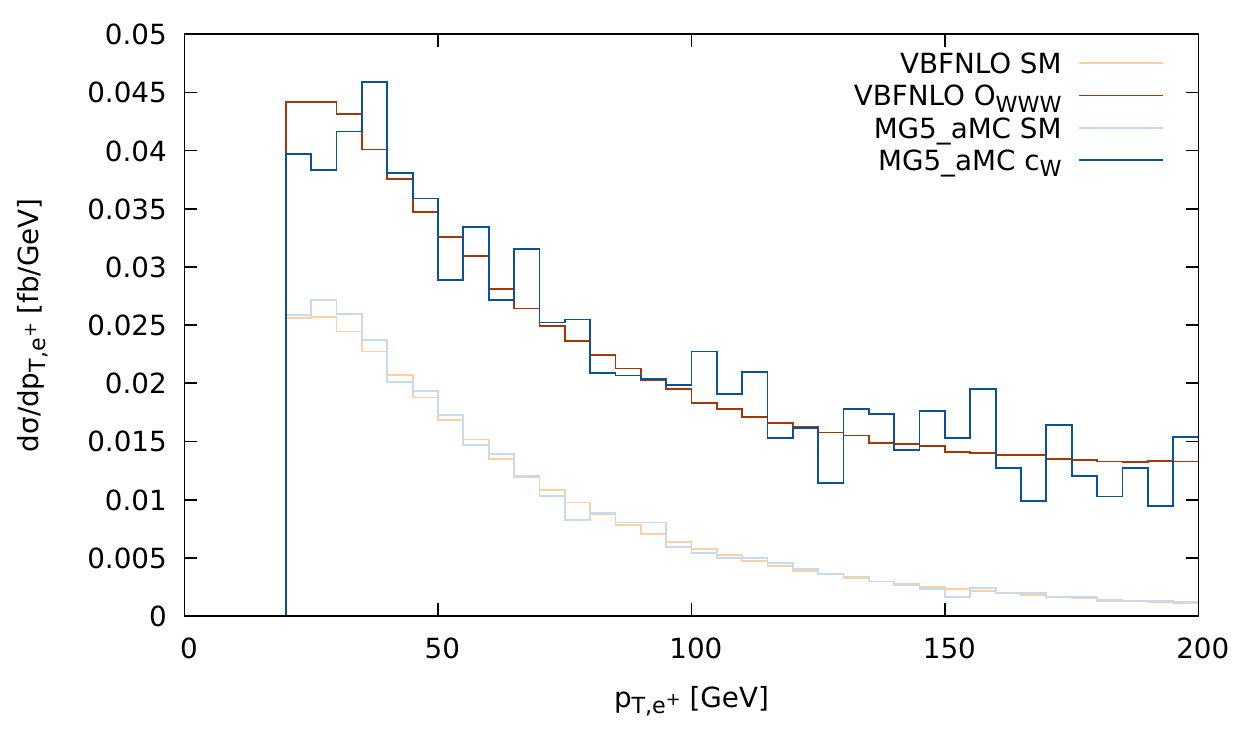}~
\includegraphics[width=.55\textwidth]{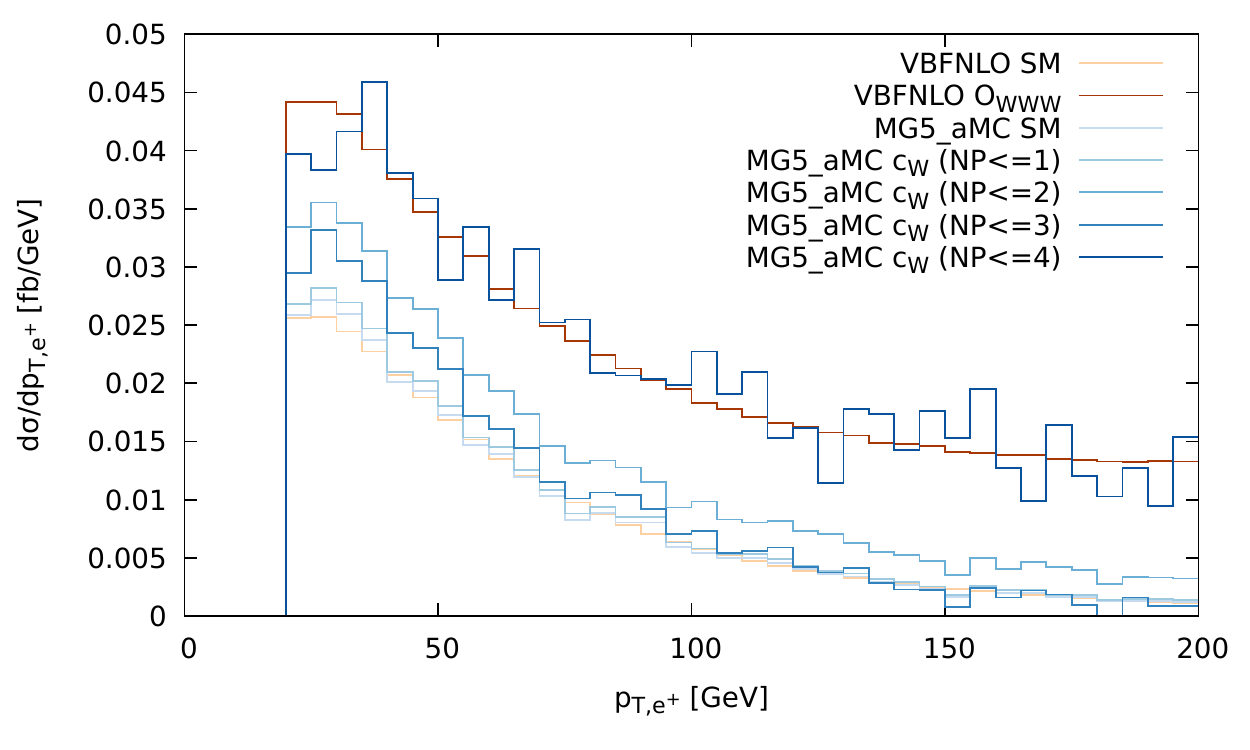}
\caption{Left: Differential distribution $d\sigma/dp_{T,e^+}$ for $W^+W^+jj \rightarrow e^+\nu_e\mu^+\nu_\mu jj$ production obtained with {\tt MG5\_aMC} + {\tt SMEFTsim} and {\tt VBFNLO} for $c_W = 0$ (SM) and $c_W = (g^3/4) f_{WWW}/\Lambda^2= \unit[1]{TeV^{-2}}.\qquad\qquad\quad$
Right: same as left, with the {\tt MG5\_aMC} distribution for $c_W=\unit[1]{TeV^{-2}}$  decomposed into contributions of different EFT order. }\label{plot:EFTcfr_ssWW_pTe}
\end{figure}
As an example, the results for $d\sigma/dp_{T,e}$ are shown in Figure~\ref{plot:EFTcfr_ssWW_pTe} (left). 
We find that all the distributions generated with the two codes are in agreement within the statistical uncertainty.
However, the estimates obtained with the specifications in Table~\ref{tab:EFTcfr_ssWW_specs} are much more stable for {\tt VBFNLO} than for {\tt MG5\_aMC} + {\tt SMEFTsim} (see Figure~\ref{plot:EFTcfr_ssWW_pTe}, left).
It is necessary to generate a larger number of events with the latter in order to verify the compatibility to higher accuracy. Because this test is computationally demanding, it is left for the future.

\begin{figure}[t]\centering
\includegraphics[width=.7\textwidth]{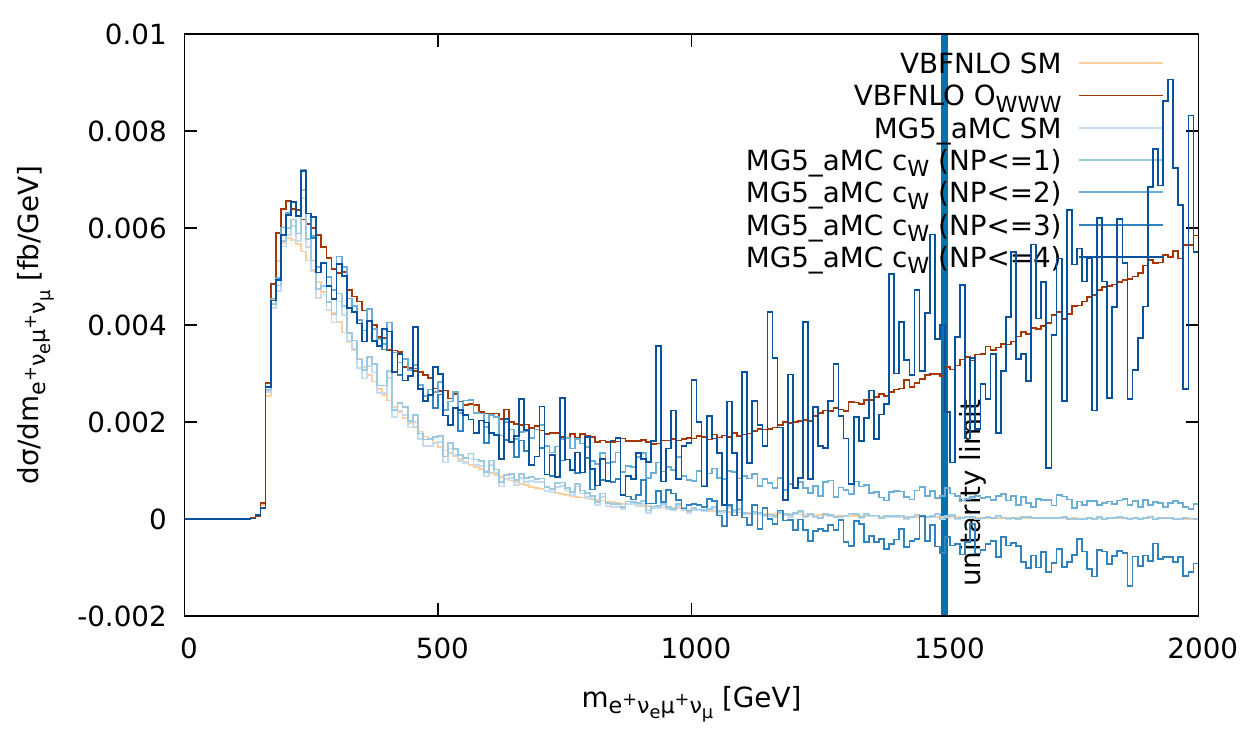}
\caption{Differential distribution $d\sigma/dm_{e\nu_e\mu\nu_\mu}$ for $W^+W^+jj \rightarrow e^+\nu_e\mu^+\nu_\mu jj$ production obtained with {\tt MG5\_aMC} + {\tt SMEFTsim} and {\tt VBFNLO} for $c_W = 0$ (SM) and $c_W = (g^3/4) f_{WWW}/\Lambda^2= \unit[1]{TeV^{-2}}$.}\label{plot:EFTcfr_ssWW_m4l}
\end{figure}

An interesting feature of {\tt MG5\_aMC} is the possibility of using interaction order specifications to estimate separately the contributions of an EFT operator at each order in the expansion.
In Figure~\ref{plot:EFTcfr_ssWW_pTe} (right), we show the decomposition for the $d\sigma/dp_{T,e}$ distribution.
In the figure the notation NP<=N denotes the inclusion of terms of order $\left(c_W\right)^{n\leq N}$ in the squared matrix element. The curve  NP<=1 includes the SM contribution and the SM-EFT linear interference pieces, the curve NP<=2 adds the interference between diagrams with two and zero $c_W$ insertions and between two diagrams with one $c_W$ insertion, and so on.
It should be noted that other effects, that are not included here, such as contributions from $d\geq 8$ operators, are also expected to contribute at order $(c_W)^{n\geq2}$.
Therefore the distributions in Figure~\ref{plot:EFTcfr_ssWW_pTe} (right) should not be interpreted as complete order-by-order estimates in the EFT.
Nonetheless, this visualisation provides a qualitative check of the behaviour of the EFT expansion. For instance, Figure~\ref{plot:EFTcfr_ssWW_pTe}, right, shows that the $p_{T,e}$ distribution is dominated by the $(c_W)^4$ contributions in the squared amplitude. This signals that the EFT expansion is breaking down, i.e. the value $c_W=\unit[1]{TeV^{-2}}$ is too large for the EFT formalism to be valid over the whole kinematic region considered. 
This is confirmed by the analysis of the distribution of the invariant mass of the four leptons in the final state (Figure~\ref{plot:EFTcfr_ssWW_m4l}).
The Lorentz structure of the operator $\mathcal{Q}_W$ enhances the cross section at large $m_{e\nu_e\mu\nu_\mu}$. Since $c_W=\unit[1]{TeV^{-2}}$ is quite large, this effect causes the violation of perturbative unitarity (and therefore of the EFT validity) at relatively low energies.
The formfactor tool of {\tt VBFNLO} estimates this to happen at $m_{e\nu_e\mu\nu_\mu}\sim \unit[1.5]{TeV}$, as indicated in Figure~\ref{plot:EFTcfr_ssWW_m4l}.  

\subsection*{Summary and Outlook}
We have performed a preliminary comparison of {\tt VBFNLO} and {\tt MadGraph5\_aMC@NLO} with the {\tt SMEFTsim} package.
We have looked at two processes, $W^+Z$ and $W^+W^+jj$ production, considering two benchmark cases, namely the SM limit and the SM + the operator $\mathcal{Q}_{WWW}$ with $c_W=\unit[1]{TeV^{-2}}$.
We found good agreement both in the total cross sections and in the differential distributions considered, although higher statistics would be required in the generation with {\tt MG5\_aMC} in order match the accuracy of the results obtained with {\tt VBFNLO}.
Notably, the two codes offer complementary tools for the control of the EFT validity over the kinematic region considered: the convergence of the EFT expansion can be directly probed via interaction order specifications in {\tt MG5\_aMC}, while the violation of perturbative unitarity can be estimated with dedicated algorithms in {\tt VBFNLO}.

The next steps will be the extension to further operator structures and other codes, such as {\tt Whizard} or {\tt Sherpa} + {\tt SMEFTsim}. 
In general, future plans for the EFT analysis include studying the behaviour of the relevant kinematic observables in the simultaneous presence of several operators, identifying optimal selection cuts to maximise the sensitivity to given operators and extending the technique to further VBS channels.

\section{Polarisation in VBS at the LHC: $WW$, $ZZ$, and $WZ$ with \texttt{PHANTOM}\footnote{speaker: G. Pelliccioli}}
The definition of polarised cross-sections in VBS represents a crucial issue at the LHC.
If physics beyond-the-Standard-Model (BSM) is present, it would interfere in the very delicate cancellation of large contributions in the high energy regime of VBS, mainly when vector bosons are longitudinal.
Thus, choosing a proper definition of polarised processes, implementing it numerically, and performing phenomenological studies constitute significant developments to the theoretical status of VBS and contribute to upcoming analyses of LHC data.

A new method to isolate $W$ bosons with definite polarisation has been proposed recently \cite{Ballestrero:2017bxn,Anders:2018gfr,Maina:2017eig}. 
This study has been performed at the leading electroweak order $\mathcal{O}(\alpha^6)$ with the \texttt{PHANTOM} Monte Carlo \cite{Ballestrero:2007xq} and provides reliable predictions for polarised cross-sections in $W^+W^-$ scattering, in the fully leptonic decay channel.
Strong evidences of the importance of the longitudinal polarisation has been pointed out in Ref.~\cite{Ballestrero:2017bxn}: the comparison of kinematic distributions obtained with underlying Standard Model (SM) dynamics with those obtained in the absence of a Higgs boson ($M_{h}\rightarrow \infty$) shows large discrepancies only in the longitudinal scattering, while transverse modes are almost insensitive to the underlying dynamics.
One of the most extreme BSM scenarios being the SM with no Higgs boson, these results suggest that if new physics is present, one has to search for it in the longitudinal scattering, as the differences between SM and other BSM models are essentially encoded in it.

The study performed in Ref.~\cite{Ballestrero:2017bxn} for opposite sign $W$'s can be extended to the same-sign case, with no further theoretical complications.
This has been done by analysing the process $p\,p \rightarrow j\,j\, e^+ \mu^+ \nu_e \nu_{\mu}$ at the LHC at $13$~TeV, which includes $W^+W^+$ scattering contributions.
In order to make the analysis as realistic as possible, we imposed a complete set of lepton cuts ($p_t^\ell>20$ GeV, $|\eta_\ell|<2.5$ and $p_t^{\rm miss} > 40 $ GeV), together with standard VBS cuts ($M_{jj}>500$ GeV, $|\Delta\eta_{jj}|>2.5$).
We investigated the behaviour of the peculiar kinematic distributions for both $W$ bosons with definite polarisation: longitudinal-longitudinal, transverse-longitudinal, longitudinal-transverse, and transverse-transverse.

\begin{figure}[h!]
  \centering
  \subfigure[$m_{WW}$]{\includegraphics[scale=0.35]{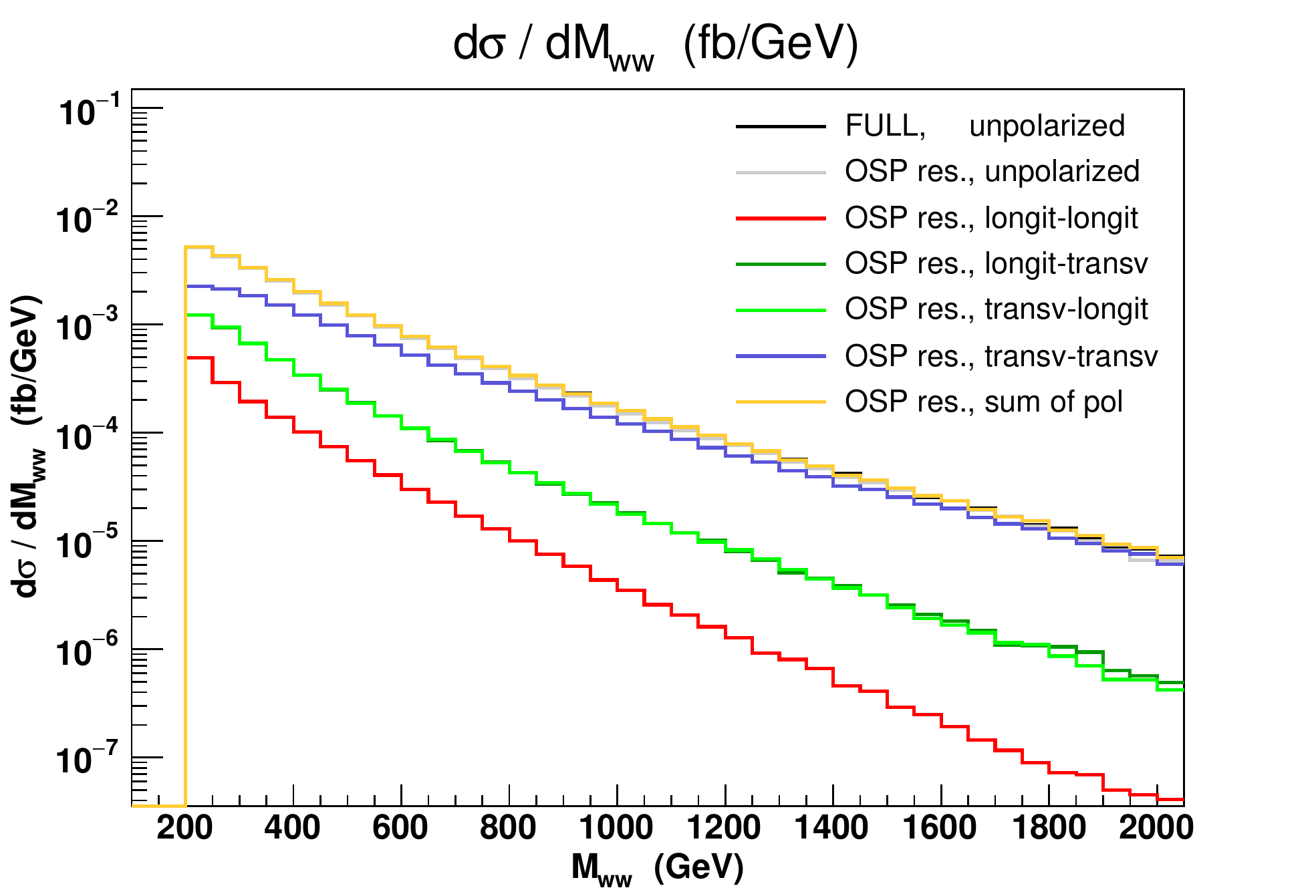}}
  \subfigure[$p_t^{e^+}$]{\includegraphics[scale=0.35]{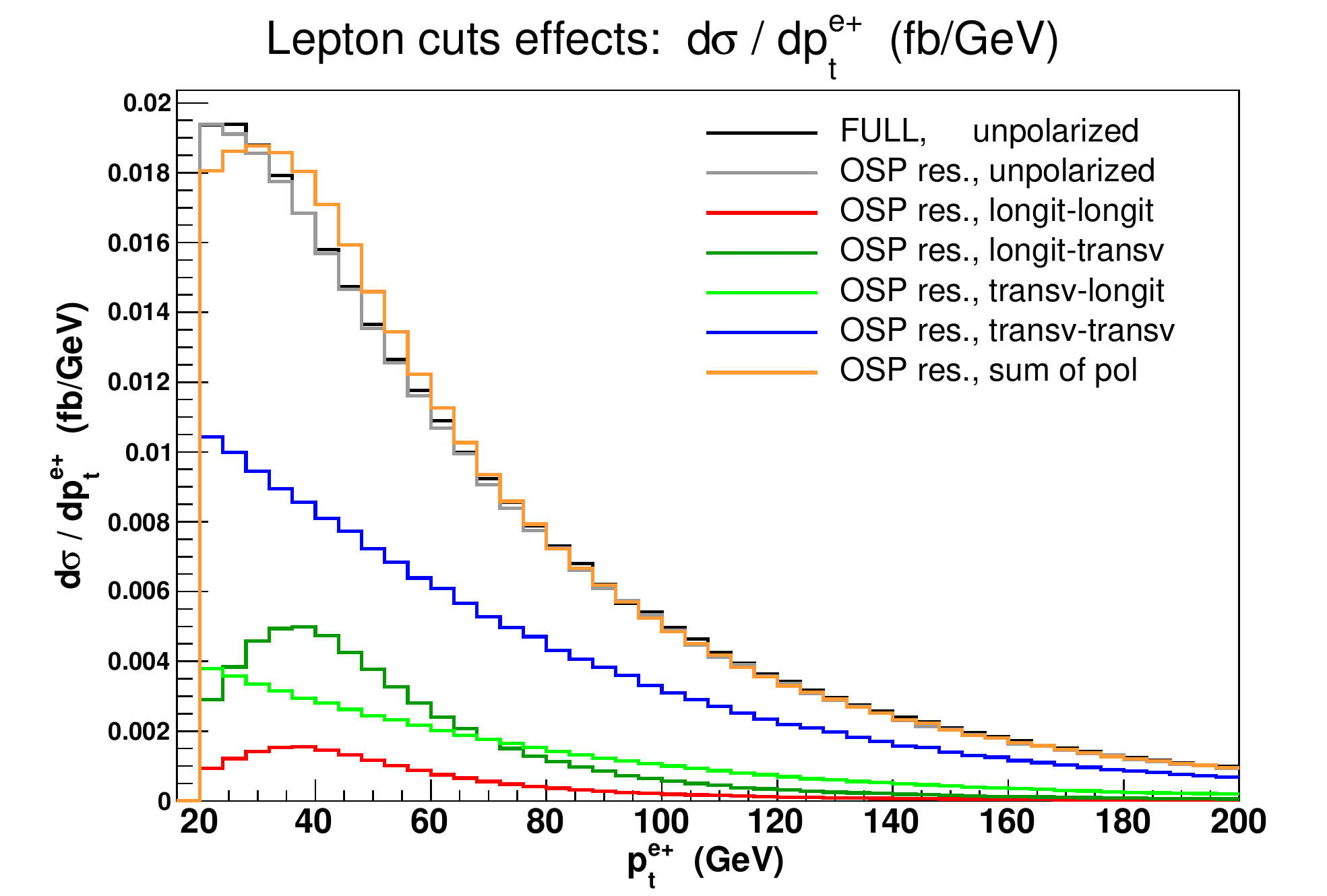}}
  \caption{
      $M_{WW}$ and $p_t^{e^-}$ distributions in $W^+W^+$ scattering, in the presence of leptonic     cuts: unpolarised full (black), unpolarised OSP (grey), longitudinal-longitudinal (red),     longitudinal-transverse (light green), transverse-longitudinal (dark green),     transverse-transverse (blue).
      The sum of the doubly-polarised distributions is shown in orange.
      In both figures the black and grey curves perfectly overlap. 
      On the left the dark green and light green curves are superposed and the orange one     coincides with the black and grey ones. 
      \label{fig:wpwp}
  }
\end{figure}

The agreement between the full and the on-shell projected (see for details Sec.~3 of Ref.~\cite{Ballestrero:2017bxn}) unpolarised calculations is very good: total cross-sections differ by $1.3\%$ and distributions agree within a few percent in all kinematic regions where statistical fluctuations are under control. This is evident in the $W^+W^+$ invariant mass and $p_t^{e^-}$ distributions, shown respectively in Figure~\ref{fig:wpwp}, left and Figure~\ref{fig:wpwp}, right (black vs. grey curve).
The sum of doubly-polarised distributions (orange curve in Figure~\ref{fig:wpwp}) is in satisfactory agreement with the full one in most kinematic distributions (see \emph{e.g.}\ the $WW$ invariant mass in Figure~\ref{fig:wpwp}, left), with discrepancies which amounts at most to $5-7\%$, mainly for leptonic kinematic variables (see \emph{e.g.} the charged leptons $p_t$ in Figure~\ref{fig:wpwp}, right).
This is due to leptonic cuts, which induce non-negligible interferences among different polarisations.

The study performed in Ref.~\cite{Ballestrero:2017bxn} for $W^+W^-$ scattering has been extended to $W^+W^+$, obtaining interesting results for the doubly-polarised electroweak $W^+W^+$ scattering, in the presence of a realistic set of cuts.


So far we have discussed results for $W$ bosons only. With the new (beta) version of \texttt{PHANTOM} it is now possible to compute VBS cross-sections also for $Z$ bosons with definite polarisation, at the perturbative order $\alpha^6$.
This gives access to polarised $ZZ$ and $WZ$ scattering, both in the fully leptonic and in the semi-leptonic channel.\\

Let us consider the process $p\,p\rightarrow j\,j\,e^+e^-\mu^+\mu^-$ at the LHC at $13$~TeV.
This embeds two different scattering sub-processes, $ZZ\rightarrow ZZ$ and $W^+W^-\rightarrow ZZ$.
In the SM, the former involves only Higgs-mediated diagrams, while the latter contains both gauge and Higgs contributions, whose interplay cancels out the bad high energy behaviour in the longitudinal scattering, restoring unitarity.

\begin{figure}[htb!]
  \centering
  \includegraphics[scale=0.45]{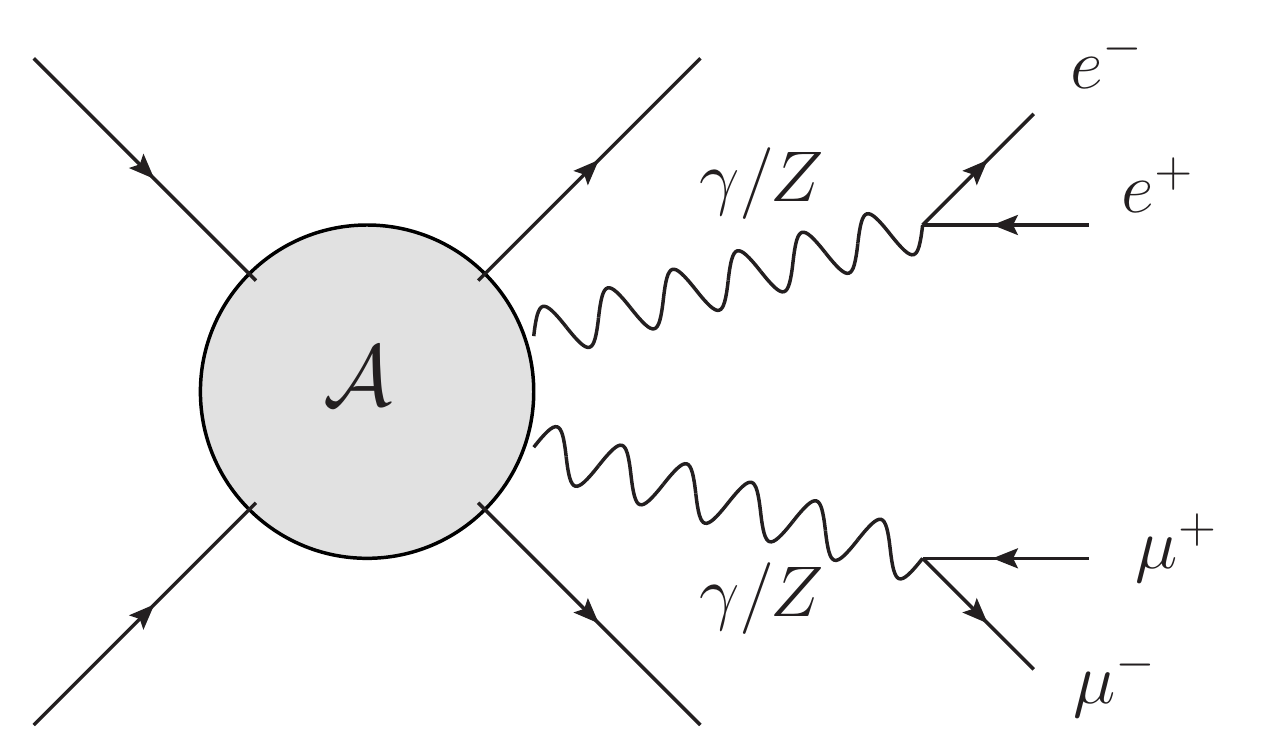}
  \caption{Doubly-resonant diagrams contributing to the VBS production of four charged leptons.}\label{fig:ampZZ}
\end{figure}

In order to separate $Z$ polarisations, we need to select only (doubly) resonant $ZZ$ diagrams and manipulate them in a gauge-invariant manner.
This turns out to be much more involved with respect to the $W$ case, due to the $\gamma/Z$ mixing in the SM.
In fact, when selecting $ZZ$ resonant diagrams (which is a gauge violating procedure), also $\gamma^* Z$ and $\gamma^* \gamma^*$ diagrams are discarded (see Figure~\ref{fig:ampZZ}), in addition to non-resonant ones.

Let us consider for simplicity the partonic sub-process $u\,u \rightarrow u\,u \,e^-e^+\mu^-\mu^+ $.
In Table~\ref{table:sigma} we provide the total leading order cross-sections obtained with three different computations: the FULL one, which is gauge invariant as it takes into account all the contributions, the RES NO OSP one, which is gauge-violating as it considers resonant diagrams only, and the RES OSP one, which considers resonant diagrams only and treats them by means of on-shell projections (OSP), as it has been done for $W$'s in Ref.~\cite{Ballestrero:2017bxn}. Note that this last procedure provides gauge invariant predictions.
In each of the three cases, we have performed the calculation both with and without a cut on the $\ell^+\ell^-$ invariant mass around the $Z$ pole mass ($|M_{\ell\ell}-M_Z| <5$ GeV).
If no $M_{\ell\ell}$ cut is imposed, resonant diagrams do not reproduce the full result ($\sim 70\%$ discrepancy in the total cross-section). 
Moreover, the employment of on-shell projections does not have any effect on the resonant description. Even in the presence of a cut on $M_{\ell\ell}$, the resonant contributions do not reproduce the full result and on-shell projections do not reduce the discrepancies.
These are clear hints of the $\gamma/Z$ mixing in the SM, which results in large discrepancies when discarding $\gamma$ contributions, mainly when small $M_{\ell\ell}$ values are allowed.
The considered sub-process is characterised by a $4Z$ topology: this means that the external particles (all of them considered outgoing) can reconstruct four $Z$ bosons. 
For sub-processes which can reconstruct two $Z$ and two $W$ bosons ($2W2Z$ topology), the discrepancies between the resonant and full description are much smaller.

\begin{table}[h!]
\begin{center}
\begin{tabular}{|C{1.5cm}||C{2.5cm}|C{2.5cm}|C{2.7cm}|}
\hline
\bf cut      & \bf FULL     & \bf RES OSP   & \bf RES NO OSP\\
\hline
\hline
\multicolumn{4}{|c|}{ $u\ u \rightarrow u\ u \ e^-\ e^+\ \mu^-\ \mu^+ $  (4Z amplitude)}\\
\hline
no cut    &  44.79   &  13.02 (-71\%)    &   13.18 (-70\%)  \\
\hline
5  \GeV   &  10.09  &   9.55 (-5\%)    &    9.53 (-5\%) \\
\hline
\end{tabular}\caption{Cross sections ($10^{-8}$ pb) with different cuts around the $Z$ pole mass ($|M_{\ell\ell}-M_Z| <$ cut) for the full  calculation (FULL), resonant diagrams only with (RES OSP) and without (RES NO OSP) on-shell projections. Final state: 4 charged leptons + 2 jets.}
\label{table:sigma}
\end{center}
\end{table}

The situation slightly improves when switching off the final state $\gamma$ decaying into $\ell^+\ell^-$, by asking for 4 final state neutrinos ($4Z$ process, $u\,u \rightarrow u\,u\,\nu_e \bar{\nu}_{e} \nu_{\mu} \bar{\nu}_{\mu}$), but also in this case the resonant calculation differs from the full one by the $7\%$ and the on-shell projections do not cure such discrepancy (see Table~\ref{table:sigma2}).

\begin{table}[h!]
\begin{center}
\begin{tabular}{|C{1.5cm}||C{2.5cm}|C{2.5cm}|C{2.7cm}|}
\hline
\bf cut      & \bf FULL     & \bf RES OSP   & \bf RES NO OSP\\
\hline
\hline
 \multicolumn{4}{|c|}{$ u\ u \rightarrow u\ u \ \nu_e\ \bar{\nu}_e\ \nu_{\mu}\ \bar{\nu}_{\mu}$  (4Z amplitude)}\\ 
\hline
no cut    &   55.80   &   51.13 (-8\%)   &  51.65 (-7\%)   \\
\hline
\end{tabular}\caption{Cross sections ($10^{-8}$ pb) with different cuts around the $Z$ pole mass ($|M_{\ell\ell}-M_Z| <$ cut) for the full  calculation (FULL), resonant diagrams only with (RES OSP) and without (RES NO OSP) on-shell projections. Final state: 4 neutrinos + 2 jets.}
\label{table:sigma2}
\end{center}
\end{table}
Summarising the previous comments, neither with nor without on-shell projections resonant diagrams describe correctly scattering processes involving $Z$ bosons, if the $\ell^+\ell^-$ pair invariant mass is allowed to be sufficiently off $Z$-mass-shell. A good description of the full computation can be obtained only by imposing a sharp cut on $M_{\ell\ell}$ around the $Z$ pole mass.

\begin{figure}[h!]
  \centering
 \hspace*{2.5cm} \includegraphics[scale=0.6]{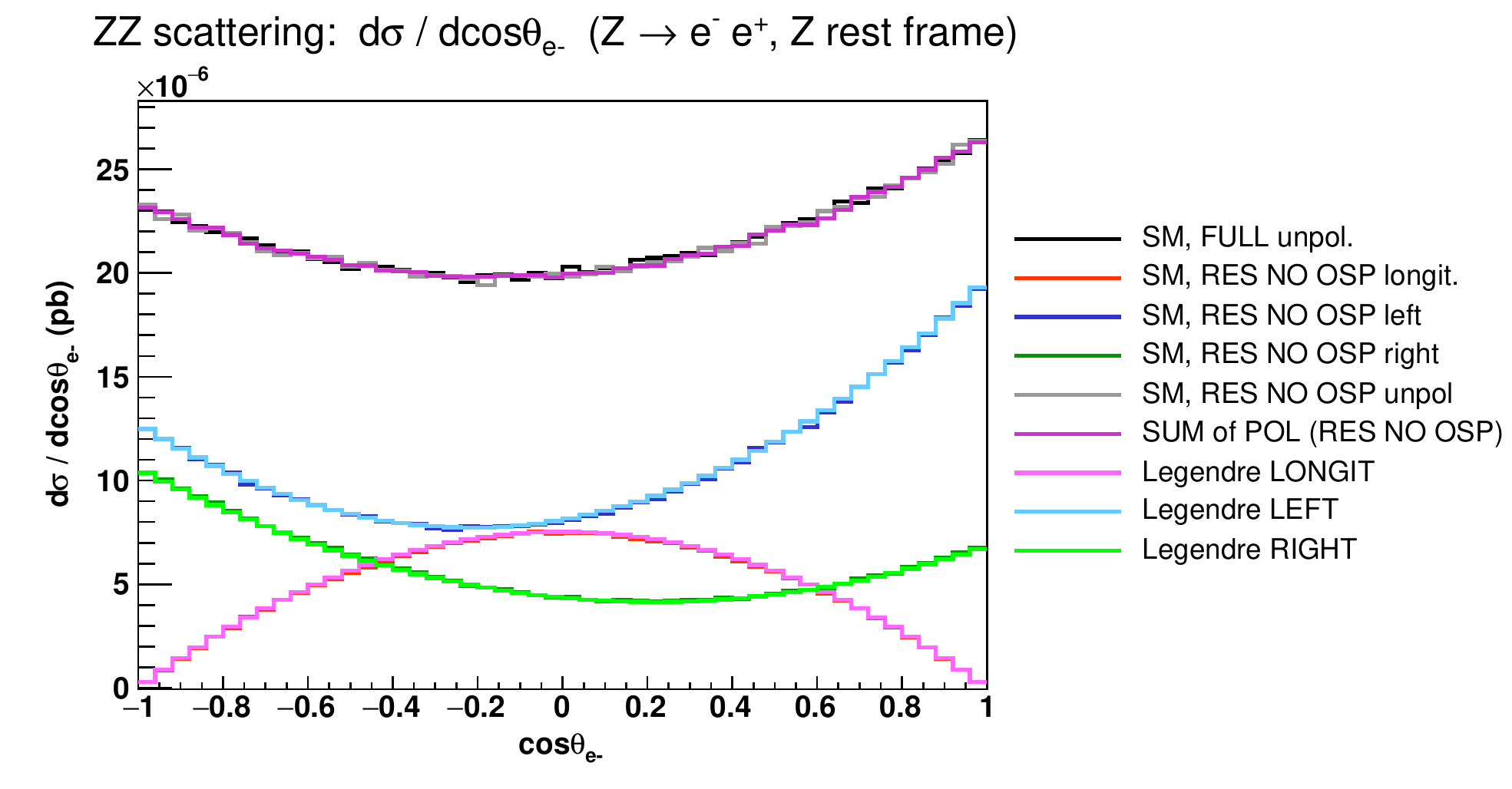}  
  \caption{Distributions in $\cos\theta_{e}$ in $ZZ$ scattering, without lepton cuts: unpolarised full (black), unpolarised OSP projected (grey), sum of polarised (violet), longitudinal (red), left (dark blue) and right (dark green), obtained with \texttt{PHANTOM}. The Legendre analysis results of the full distribution are shown in pink for the longitudinal, light blue and light green for left and right respectively. polarisation modes are separated only for the $Z$ boson decaying into $e^+e^-$.}
  \label{fig:zz_nolepcut}
\end{figure}

\begin{figure}[h!]
  \centering
 \hspace*{2.5cm} \includegraphics[scale=0.6]{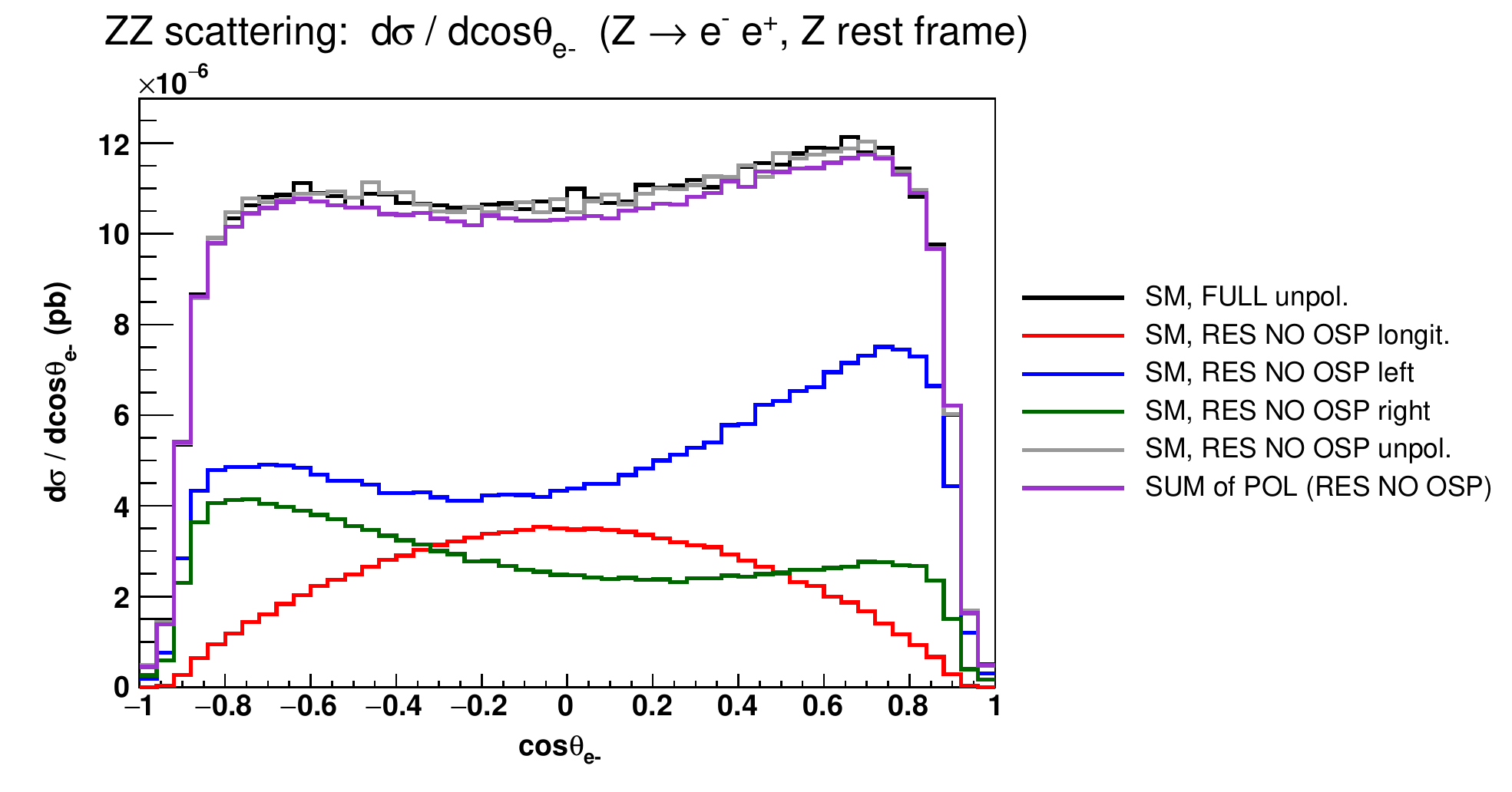}
  \caption{Distributions in $\cos\theta_{e}$ in $ZZ$ scattering, in the presence of lepton cuts: unpolarised full (black), unpolarised OSP projected (grey), sum of polarised (violet), longitudinal (red), left (dark blue) and right (dark green), obtained with \texttt{PHANTOM}. polarisation modes are separated only for the $Z$ boson decaying into $e^+e^-$.}\label{fig:zz_lepcut}
\end{figure}

Taking into account the previous reasoning, we move to LHC phenomenology.
Let us consider the process $p\,p\rightarrow j\,j\,e^+e^-\mu^+\mu^-$, including all possible partonic channels.
We note that $4Z$ processes are sub-dominant ($0.5\%$ of the total cross-section) w.r.t. the others, thus, given a sharp cut on $M_{\ell\ell}$, the resonant approximation is expected to reproduce the full computation with satisfactory precision. We choose not to employ on-shell projections, as they don't improve the approximation.
Moreover, we neglect $b$-quarks contributions, as they account for $0.04\%$ of the total cross-section.
The imposed cuts are $p_t^j>20$ GeV, $|\eta_j|<5$, $M_{jj}>600$ GeV, $|\Delta\eta_{jj}|>3.6$, $M_{4\ell}>300$ GeV and $|M_{\ell\ell}-M_Z|< 5 \,\GeV$.
The results obtained selecting $ZZ$ resonant diagrams reproduce those obtained with full matrix-element within less than a percent, at the level of the total cross section and of the main differential distributions.
The separation of polarisation modes of one $Z$ boson (the one decaying into $e^+e^-$) in the absence of leptonic cuts has been validated with a Legendre analysis (for details see Sect.~5 of Ref.~\cite{Ballestrero:2017bxn}), relying on the analytic form of the $\cos\theta_{\ell^-}$ distribution in the $Z$ center-of-mass reference frame,
 \hspace*{-0.5cm}\begin{eqnarray}\label{eq:zff}
        \hspace*{-0.5cm}\frac{d\sigma_{Z~\to~\ell^+\ell^-}}{d\cos\theta}\! &\propto& \!\frac{3}{4} \left({\sin\theta}^2\right)\,f_{\rm longit}\,+\,\frac{3}{8} \left(1+{\cos\theta}^2-2\,A_{Z\ell\ell}\,\cos\theta\right)\,f_{\rm left}\,+\,\nonumber\\[-0.1cm]
        &&
        \hspace*{-0.4cm}+\,\frac{3}{8} \left(1+{\cos\theta}^2
        +2\,A_{Z\ell\ell}\,\cos\theta\right)\,f_{\rm right}\,\,,\hspace*{0.2cm} A_{Z\ell\ell}=\frac{|c_L|^2-|c_R|^2}{|c_L|^2+|c_R|^2}\,\,,
      \end{eqnarray}
where $f_{\rm longit},\, f_{\rm left},\, f_{\rm right}$ are the polarisation fractions and $c_L,\,c_R$ represent the SM couplings of the $Z$ boson to the left- and right-handed fermions.
The polarisation fractions extracted with Eq.~(\ref{eq:zff}) from the full results are consistent with those obtained from the polarised computations performed with \texttt{PHANTOM} ($<1\%$ discrepancy).
The unpolarised and polarised $\cos\theta_{e^-}$ distributions are shown in Figure~\ref{fig:zz_nolepcut} together with the curves obtained analytically from the Legendre analysis of the full result.
In the experimentally accessible situation in which lepton cuts are imposed ($p_t^\ell>20\,\GeV$, $|\eta_{\ell}|<2.5$), some interesting results can be drawn (see Figure~\ref{fig:zz_lepcut}).
\begin{itemize}
\item The unpolarised resonant and the full results agree rather well.
The total cross-sections are
\[\sigma_{\rm full}\,=\,1.982(8) \cdot 10^{-5}\,\rm pb,\qquad\sigma^{\rm res}_{\rm unpol}\,=\,1.979(9)\cdot 10^{-5} \,\rm pb\,\,,\]
and the $\cos\theta_{e^-}$ shapes are in perfect agreement, as shown in Figure~\ref{fig:zz_lepcut} (black and grey curves).
\item The sum of the three (singly) polarised distributions (violet curve) agrees with the full results ($\sim 2\%$ discrepancy). The polarised total cross-sections are:
\[
\sigma_{\rm longit}\,=\,0.443(5)\cdot 10^{-5}\,\rm pb\,,\qquad
\sigma_{\rm left}\,=\,0.957(0)\cdot 10^{-5} \,\rm pb\,,\qquad
\sigma_{\rm right}\,=\,0.537(1)\cdot 10^{-5}\,\rm pb
\]
We do not have the equivalent of Eq.~(\ref{eq:zff}) in the presence of lepton cuts, but this is irrelevant.
In fact, interferences amongst different polarisations are well under control, though not negligible. 
\item When considering the coherent sum of the transverse modes (left and right), rather than the incoherent one, the discrepancy between the sum of polarised processes and the full one decreases to less than $0.5\%$.
\end{itemize}
The presented results are very promising and suggest further interesting phenomenological developments.
In this paragraph we considered only one of the two $Z$ bosons with definite polarisation, but we are confident that the study of doubly-polarised scattering will give satisfactory results as well.

\begin{figure}[h!]
  \centering
   \includegraphics[scale=0.43]{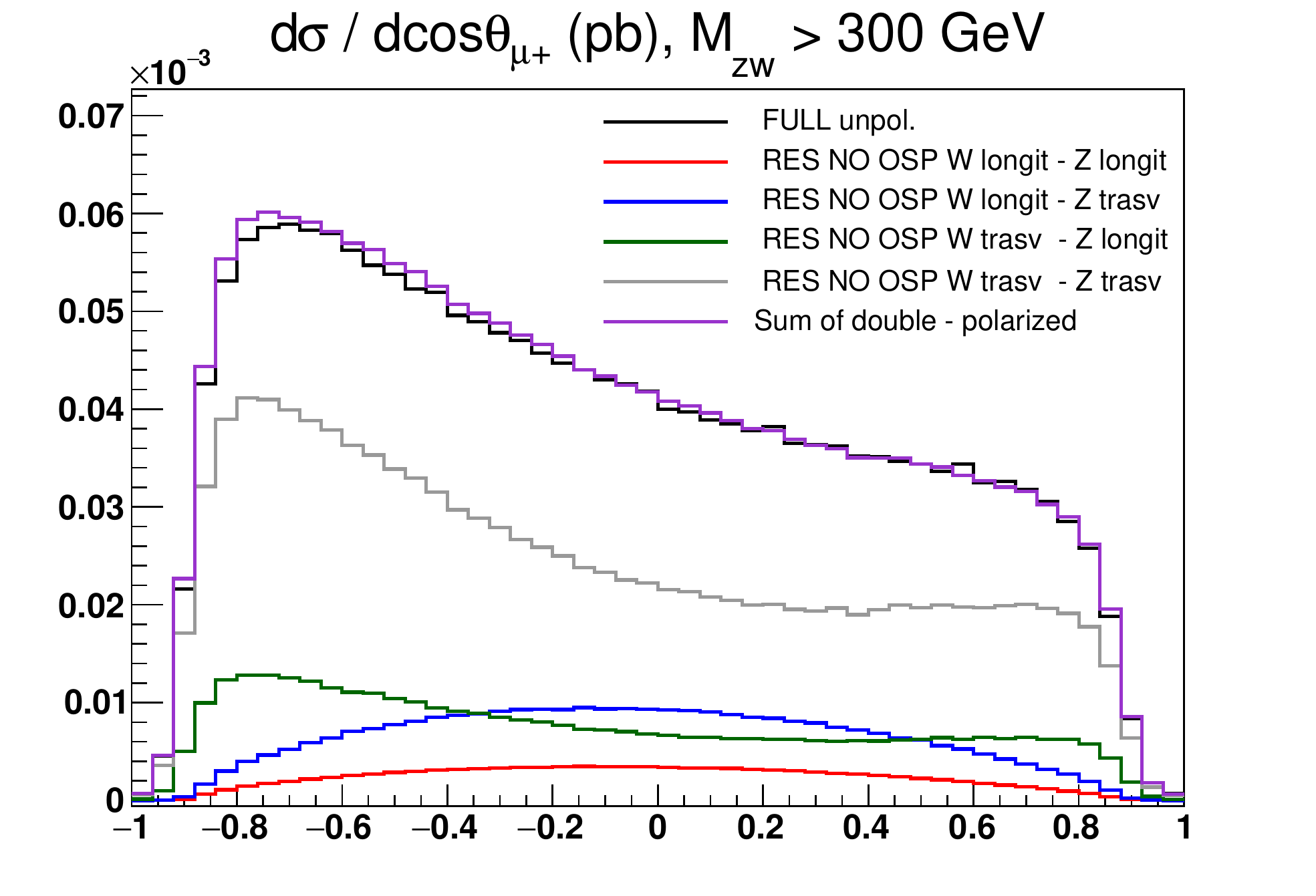}
  \caption{Distributions in $\cos\theta_{\mu^+}$ in $W^+Z$ scattering, in the presence of lepton cuts: unpolarised full (black), sum of polarised (violet), $W_{\rm longit}Z_{\rm longit}$ (red), $W_{\rm longit}Z_{\rm transv}$ (blue), $W_{\rm transv}Z_{\rm longit}$ (green) and $W_{\rm transv}Z_{\rm transv}$ (grey), obtained with \texttt{PHANTOM}.}\label{fig:wpz}
\end{figure}

In the lights of the $ZZ$ conclusions, we also performed a preliminary study of the process $p\,p\rightarrow j\,j\,e^+e^-\mu^+\nu_\mu$, which contains $W^+Z$ scattering diagrams.
We have chosen to fix the polarisation of both bosons to be either longitudinal or transverse.
Given a sufficiently sharp cut on the $e^+e^-$ invariant mass around $M_{Z}$ and on the $\mu^+\nu_{\mu}$ invariant mass around $M_{W}$ (5 GeV), we checked that resonant diagrams describe rather well the full computation in the unpolarised case.
Furthermore the separation of polarisation modes gives very good results both with and without leptonic cuts ($p_t^\ell>20\,\GeV$, $|\eta_{\ell}|<2.5$, $p_t^{\rm miss}> 20$ GeV), as shown in $\cos\theta_{\mu^+}$ distributions (Figure~\ref{fig:wpz}).
Leptonic and $p_t^{\rm miss}$ cuts induce interferences amongst different polarisation modes which are not negligible, but well under control (at most $5\%$ in some kinematic region).

This preliminary study of $W^+Z$ scattering in the fully leptonic decay channel has provided promising results.
We note that a complete comprehension of this channel would pave the way to polarisation analyses in semileptonic VBS.
The presence of only one neutrino makes $e^+e^-\mu^+\nu_{\mu} \,j\,j$ and $j j j j \mu^+\nu_{\mu}$ very similar final states and it is expected that their phenomenology is similar as well, provided that difficulties related jet structure reconstruction are overcome.

To conclude this section, we have proposed a coherent procedure to compute polarised VBS processes involving both $W$ and $Z$ bosons, at the leading electroweak order.
This has been implemented in \texttt{PHANTOM}.
In the same fashion as the recent study of $W^+W^-$ scattering Ref.~\cite{Ballestrero:2017bxn}, we have provided analogous results for the same sign channel $W^+W^+$.
Then, after a detailed explanation of the theoretical obstacles in the selection of resonant $ZZ$ diagrams, we have presented very promising results for $ZZ$ and $WZ$ polarised scattering, in the fully leptonic channel.

\section{VBS Polarization in ZZ fully leptonic channel at LHC: study of lepton cuts effect and LL separation\footnote{speaker: A. Li}}

While all the Vector Boson Scattering (VBS) processes will be observed by the Run III of the LHC, the longitudinal scattering represents only a small fraction of the VBS cross section. It is expected that the  longitudinal-longitudinal gauge boson scattering can be detected at the High Luminosity LHC (HL-LHC). 
Compared with the other final states, the $ZZ$ fully leptonic channel allows for the precise measurement of all final state leptons and therefore of the angular correlations. It also allows for the precise measurement of the center-of-mass energy of the scattering process through the invariant mass of the four final state leptons. On the other hand, the cross-section is very small.

We present here results for the polarized $ZZ$ scattering, in the $pp\rightarrow jj e^-e^+\mu^-\mu^+$ decay channel. Simulated events have been generated with \texttt{PHANTOM 1.5b} \cite{Ballestrero:2007xq}, using the NNPDF30\_lo\_as\_0130 \cite{Ball:2014uwa} with scale $Q = \frac{m_{4l}}{\sqrt{2}}$. We consider only ElectroWeak processes at $\mathcal{O}(\alpha_{EM}^6)$. 
All presented results refer to a center-f-mass energy of 13~TeV and have been obtained with the following set of cuts:

\begin{enumerate}
	\item	$\left| \eta_j \right| < 5$ 
	\item	$p_t^j > 20~\mathrm{GeV}$
	\item	$M_{jj} > 600~\mathrm{GeV}$
	\item	$\left| \Delta\eta_{jj} \right| > 3.6 $
	\item	$\eta_{j1}\cdot\eta_{j2} < 0$
	\item	$M_{ZZ} > 2M_{Z} $ 
	\item	$m_{ll} > 40~\mathrm{GeV}$
	\item	$ 86.2~\mathrm{GeV} <  m_{Z} < 96.2~\mathrm{GeV}$
\end{enumerate}

In addition the lepton $p_t$ and $\eta$ cuts are varied in order to study their effects.

Table~\ref{Table:ZZ_Generations} presents the cross sections obtained for the polarized $ZZ$ scattering in the $jj e^-e^+\mu^-\mu^+$ decay channel, when only one $Z$ boson ($Z \rightarrow e^-e^+$, denoted as $Z_e$) is polarized and when the two $Z$ bosons are polarized, for different combinations of the polarizations. The cross sections are quoted both without any $p_t$ nor $\eta$ cut on the leptons and with $p_t^{e^-} > 20 \mathrm{GeV}$ and $\left|\eta_{e^-} \right| < 2.5 $.

\begin{table}[hbt!]
	\centering	
	\begin{tabular}{lcccc}
		\toprule
		& nocut && lepcuts& \\
		\midrule
		Full. &6.37&&5.36&\\
		\midrule
		$Z_e$ Left &2.89&45.37\%&2.39&44.59\%\\
		$Z_e$ Right&1.71&26.84\%&1.42&26.49\%\\
		$Z_e$ Long&1.77&27.79\%&1.49&27.80\%\\
		\midrule
		Left-Left&1.45&22.76\%&1.22&22.76\%\\
		Right-Right&0.58&9.11\%&0.48&8.96\%\\
		Long-Long&0.53&8.32\%&0.44&8.21\%\\
		\midrule
		TransTrans&3.36&52.75\%&2.83&52.80\%\\
		TransLong&1.25&21.19\%&1.03&19.22\%\\
		LongTrans&1.25&21.19\%&1.05&19.59\%\\
		\bottomrule
	\end{tabular}
	\flushbottom
	\caption{ Cross sections in $10^{-5} pb$ for different polarization combinations: lepcuts means $p_t^{e^-} > 20 \mathrm{GeV}$ and $\left|\eta_{e^-} \right| < 2.5 $ .}
	\label{Table:ZZ_Generations}
\end{table}

The sum of Left-Left, Right-Right and Longitudinal-Longitudinal combinations contributes to $ ~ 40\%$ of the total cross section, which means that the cross polarization combinations contribute to $ ~60\%$. The Longitudinal-Longitudinal scattering contributes only to $8\%$ of the total cross section, which makes the separation of Longitudinal-Longitudinal from the other polarization components challenging.  

Experimentally, 
cuts on leptons are due either to the detector acceptance ($\eta$) 
or to the lepton identification performance ($p_t$). 
We study here the effect of a larger lepton acceptance, changing the lepton $\eta$ cut from  2.5 to 3. Since in the $ZZ$ channel there are four detectable leptons and it is possible to lower the $p_t$ requirement on the leptons, compared to the $WW$ case. The cuts have been  lowered to $p_t^{\text{lepton}} > 20, 10, 10, 10~\mathrm{GeV}$ and the cross sections compared to a cut at 20 GeV applied on all leptons. 

Lowering the $p_t$ cut on the softest leptons, it is found that
the increase of background ($41.36\%$) is much higher than that of the signal ($13.65\%$).  We conclude therefore that it is not interesting to lower the lepton $p_t$ threshold for the separation of the LL component from the other polarization combinations in VBS.
However one should keep in mind that lowering the $p_t$ thresholds on the leptons enables an increase in the overall VBS yield. 

Figure \ref{Fig:Acceptance_of_Eta} shows the ratio of cross sections for $|\eta^{\text{lepton}}| < 3$ to that for $|\eta^{\text{lepton}}| < 2.5$ for the different polarization combinations and as a function of the $ZZ$ invariant mass. Increasing the acceptance of $|\eta^{\text{lepton}}| < 2.5$ to $|\eta^{\text{lepton}}| < 3.0$ leads to a significant gain, with an increase of the LL cross section of $27.03\%$ in average and an increase in the background cross section of $14\%$.

From now on, it is implied that $p_t^{lepton} > 20\mathrm{GeV}$ and $|\eta^{\text{lepton}}| < 3.0$.
In order to discriminate the LL component from the others, we first consider the distribution of the angle between the lepton direction in the Z boson rest frame and the Z direction in the laboratory frame, $\cos\theta_{e^-}$ and $\cos\theta_{\mu^{-}}$. 

The signal (single boson longitudinal component) exhibits a parabolic shape maximum at 0 while the backgrounds (left and right components) are maximum at the edges of the $\cos\theta$ distribution ($\vert \cos\theta \vert \sim 1$), as shown in Figure~\ref{fig:zz_nolepcut}. Therefore, choosing a working area around the center of $(\cos\theta_{e^-}, \cos\theta_{\mu^{-}})$ plan can probably filter out a substantial part of the backgrounds while keeping most of the signal. Fixing the efficiency of the signal to $70\%$ and $80\%$, the corresponding efficiencies for the background (left or right polarizations) are $49.57\%$ ( $50.43\%$ bkg rejection) and $61.09\%$ ( $38.92\%$ bkg rejection). Additional discriminant variables are looked at to separate the longitudinal component.
The variables $p_t^{sum}$ and $p_t^{dif}$ were identified, defined as follow.

\begin{equation}
	p_{t}^{sum} = p_t^{\text{lepton max}} + p_t^{\text{lepton min}} \qquad \text{and} \qquad p_{t}^{dif} = p_t^{\text{lepton max}} - p_t^{\text{lepton min}} .
\end{equation}

\begin{figure}
	\begin{minipage}[t]{0.5\linewidth}
		\begin{center}
			\scalebox{0.4}{\input{WG1/Figures/h_Mzz_Eta_ZZ_transpol_stat.tex}}

			\caption{Ratio of the the differential cross section $d\sigma/dM_{ZZ}$ for $\left| \eta^{\text{lepton}}\right| < 3.0$ to that for $\left| \eta^{\text{lepton}}\right| <2.5$}
			\label{Fig:Acceptance_of_Eta}
		\end{center}
	\end{minipage}
	\begin{minipage}[t]{0.5\linewidth}
		\begin{center}
			\scalebox{0.4}{\input{WG1/Figures/h_Pt_sum_Efficiency_stat.tex}}
			\caption{Fraction of events with $p_{t}^{sum}>p_{t\quad\text{cut}}^{sum}$ as a function of $p_{t}^{sum}$, where $p_{t}^{sum}$ is defined as $p_t^{\text{lepton max}} + p_t^{\text{leptom min}}$}
			\label{Fig:Pt_sum}
		\end{center}
	\end{minipage}
	{\color{black} Black: Unpolarized ( res.)}, {\color{blue}Blue: Longit.-Longit. (res.)}, {\color{red}Red: Longit-Trans. (res.)}, {\color{green}Green: Trans.-Longit. (res.)}, {\color{violet}Violet: Trans.-Trans. (res.)} and {\color{cyan} Cyan: Bkg} ({\color{red}LT}+{\color{green}TL}+{\color{violet}TT})
\end{figure}

In order to compare their performance at separating the longitudinal component with that of $(\cos\theta_{e^-}$ and $ \cos\theta_{\mu^{-}})$, we studied their background efficiencies while signal efficiencies are kept at $70\%$ and $80\%$.
The efficiency of signal is $80\%$ when $p_t^{sum} < 124  \mathrm{GeV}$ or $p_t^{dif} < 62 \mathrm{GeV}$  and the corresponding backgrounds efficiencies are $57.03\%$ and $67.66\%$. The efficiency of signal is $70\%$ when $p_t^{sum} < 116 \mathrm{GeV}$ or $p_t^{dif} < 54 \mathrm{GeV}$ and the corresponding backgrounds efficiencies are $46.28\%$ and $55.23\%$. 

To conclude, we studied the cross sections for the different polarization combinations  for the VBS scattering in the $ZZ$ channel. The longitudinal-longitudinal component amounts to 8\% of the VBS cross section, averaged over $m_{ZZ}$ in the separation of the longitudinal-longitudinal component, lowering the lepton $p_t$ thresholds does not appear interesting as it decreases the signal-to-background ratio.
On the contrary a sizeable improvement is obtained extending the $\eta$ acceptance of leptons, the longitudinal Z bosons being produced more forward.
New discriminant variables such as $p_t^{sum}$ and $p_t^{dif}$ are seen to have performance in the separation of the longitudinal-longitudinal component from the other polarization combinations.

\section{PDF uncertainty for VBS\footnote{speaker: J. Novak}}

The uncertainty due to parton distribution functions (PDFs) is an important contribution 
to the theoretical uncertainty of the VBS predictions. Therefore, it is very 
important to evaluate such an uncertainty carefully.
There exist two leading representations of the uncertainties of PDF: 
Monte Carlo \cite{DelDebbio:2004xtd} (MC) and Hessian \cite{Pumplin:2001ct}. The MC 
representation contains an ensemble of replicas, which are the instances of 
uncertain PDF parameters, sampled according to a Gaussian distribution, 
around their central values. 
The central PDF is the average of the PDF set, while the 
PDF uncertainty is its standard deviation. Under the assumption of Gaussian 
distribution of the cross-sections obtained from the PDF set, the same 
treatment can be applied to the cross-section value: 
\begin{equation}
\delta^{\text{pdf}}\sigma = 
\sqrt{\frac{1}{N_{\text{mem}}-1}\sum_{k=1}^{N_{\text{mem}}}(\sigma^{(k)}-\braket{\sigma})^2},\qquad
 \braket{\sigma} = 
\frac{1}{N_{\text{mem}}}\sum_{k=1}^{N_{\text{mem}}}\sigma^{(k)},
\label{Eq:PDFMCGaus}
\end{equation}
where $\sigma^{(k)}$ represents the cross-section calculated from the $k$-th 
member of the set and $N_{\text{mem}}$ represents the number of members. 
On the other hand, 
when the distribution of cross-sections differs significantly from a Gaussian, 
it is better to employ a more robust estimate of the coverage interval. 
This means that $16\%$ of the largest and $16\%$ of the smallest cross-sections 
fall out of the $68\%$ C.L. interval, 
the rest of them lies within. 
The symmetric PDF uncertainty is then calculated as:
\begin{align}
\delta^{\text{pdf}}\sigma = \frac{\sigma^{(N_{\text{mem}}84/100)} - 
\sigma^{(N_{\text{mem}}16/100)}}{2};\qquad\quad \sigma^{(1)} \leq \sigma^{(2)} 
\leq \ldots \leq \sigma^{(N_{\text{mem}})}.
\label{Eq:PDFMCNonGaus}
\end{align}
This kind of treatment better accounts for the outliers. 
To roughly estimate how much does a specific distribution resemble a Gaussian, 
we can compare the values calculated from equations (\ref{Eq:PDFMCGaus}) and 
(\ref{Eq:PDFMCNonGaus}). If they give similar results, the distribution can be 
approximated by the Gaussian and uncertainty from equation 
(\ref{Eq:PDFMCGaus}) can be taken as a result. In the opposite case, it is 
better to use equation (\ref{Eq:PDFMCNonGaus}). In this study we performed the 
Shapiro-Wilk Gaussianity test.

As opposed to a MC PDF set, an Hessian PDF set is not composed of the random 
replicas, but each of its members coincides with one eigenvalue and eigenvector 
of the pdf fit covariance matrix in the parameter space. Central PDF cannot be 
calculated from its members and have to be appended separately to a set. In 
\texttt{LHAPDF} sets, this is always the first member. The equation for the 
calculation of the cross-section uncertainty is:
\begin{equation}
\delta^{\text{pdf}}\sigma = \sqrt{\sum_{k=1}^{N_{\text{mem}}}(\sigma^{(k)}-\sigma^{(0)})^2},
\end{equation}
where $\sigma^{(0)}$ represents the cross-section of the central PDF and 
$N_{\text{mem}}$ represents the number of members of the set 
(without the central one).

One of the uncertain parameters of the PDF is also $\alpha_S$. The 
experimental value of this parameter is based on several measurements, 
the current PDG average is \cite{Bethke:2012jm}:
\begin{equation}
\alpha_S(m_{\text{Z}}^2) = 0.1184 \pm 0.0007.
\label{Eq:as_measurement}
\end{equation}
The $\alpha_S$ variation is not a part of the standard PDF variation. Instead, 
it is appended to certain PDF sets in the form of two additional members. For 
the calculation of the total PDF uncertainty it is important to merge the 
$\alpha_S$ uncertainty with the uncertainty of the rest of parameters, in a way 
that accounts for the correlations among them. In Ref. \cite{Lai:2010nw} they 
propose the calculation of variation of all parameters, except $\alpha_S$, at 
the central value of the $\alpha_S$. The boundaries of the $68\%$ $\alpha_S$ 
C.L. are redefined in a way, which insures that the sum of the squares of the 
two uncertainties reproduces the total uncertainty. Therefore, the upper and 
the lower limits of the confidence interval, taken into account in PDF sets are 
slightly more conservative than in equation (\ref{Eq:as_measurement}): 
$\alpha_S(m_{\text{Z}}^2) = 0.118 \pm 0.0015$. The combined uncertainty can be 
calculated as:
\begin{equation}
\delta^{\text{pdf}+\alpha_S}\sigma = \sqrt{(\delta^{\text{pdf}}\sigma)^2 + 
(\delta^{\alpha_S}\sigma)^2},\qquad \delta^{\alpha_S} = 
\frac{1}{2}\left[\sigma(\alpha_S^+) - \sigma(\alpha_S^-)\right],
\label{Eq:Combined}
\end{equation}
where the $\sigma(\alpha_S^+)$ and $\sigma(\alpha_S^-)$ are the cross-sections 
calculated with the values $\alpha_S^+(m_{\text{Z}}^2) = 0.1195$ and 
$\alpha_S^-(m_{\text{Z}}^2) = 0.1165$ and the central values of the rest of 
parameters.

Because it is necessary to generate $N_{\text{mem}} + 1$ MC samples to evaluate 
the cross-section uncertainty due to PDF uncertainty ($N_{\text{mem}} + 3$ for 
the combined uncertainty), the reweighting of the events is very useful, 
as it can significantly reduce the computational time. For LO reweighting of 
the sample, produced with the PDF set A, to the PDF set B, the corresponding 
weights are
\begin{equation}
w_{A\to B} = \frac{f^B_1(x_1,Q)f^B_2(x_2,Q)}{f^A_1(x_1,Q)f^A_2(x_2,Q)},
\label{Eq:PDFReweigh}
\end{equation}
where the $f^A_i(x_i,Q)$ are the PDFs of the two incoming partons from the 
sample A and $f^B_i(x_i,Q)$ are the PDFs of the two incoming partons from the 
sample B. For the NLO reweighting the use of built-in generator routines is 
necessary to calculate the weights. NLO reweighting is, for example, 
implemented in the generators \texttt{MadGraph5\_aMC@NLO} \cite{Alwall:2014hca}, 
\texttt{POWHEG} \cite{Oleari:2010nx}, \texttt{Sherpa} \cite{Gleisberg:2008ta}, or 
\texttt{FEWZ} \cite{Gavin:2010az}.

The \texttt{LHAPDF} library \cite{Buckley:2014ana} offers a wide range of PDF 
sets from different groups. Because 
different PDF sets are based on different experimental data and use different 
assumptions, it is better to take into account more than just one PDF set. For 
this purpose statistical combinations can be used. This approach is competitive 
with the older PDF envelope method, by which the PDF uncertainty has to be 
calculated for each input PDF set in order to obtain uncertainty bands for the PDF.
The combined PDF uncertainty is defined as the envelope of these bands.
On the other hand, the statistical combination of the PDF sets 
already contains characteristics of several input sets. Uncertainty has to be 
calculated only on the statistical combination, while the obtained uncertainty 
takes into account the uncertainties of all the input sets. The PDF uncertainty 
from the envelope method is often overestimated, while this is not the case for 
the statistical combination method. The \texttt{LHAPDF} library currently contains 
one statistical combination: PDF4LHC15 \cite{Botje:2011sn}. This is a 
combination of CT14 \cite{Dulat:2015mca}, MMHT2014 \cite{Harland-Lang:2014zoa},
and NNPDF3.0 \cite{Ball:2008by} sets.

If the input PDF sets are represented by MC replicas, the construction of the 
statistical combination is pretty straightforward. Namely, the input PDF sets, 
with equal number of replicas, can simply be merged into one larger PDF set 
\cite{Forte:2010dt}, which is called the prior set. The only input PDF set into 
the PDF4LHC15, represented with the MC replicas, is NNPDF3.0, while the CT14 
and MMHT2014 are Hessian representations. Therefore, the first step of 
construction of the statistical combination is transformation of the Hessian 
representations into MC replicas. This is done by sampling along directions of each 
eigenvector, according to the corresponding eigenvalue, assuming Gaussian 
distributions.

The authors of the PDF4LHC15 statistical combination tested different sizes of 
the prior set: $N_{\text{rep}}=300$, $N_{\text{rep}}=900$, and 
$N_{\text{rep}}=1800$. After the comparison of the central values and the 
uncertainty for different partons, they concluded that there is a (small) 
difference between the sets with 300 and 900 replicas, while the differences 
between 900 and 1800 are completely negligible \cite{Butterworth:2015oua}. 
Therefore prior set with 900 replicas has been adopted for uncertainty estimates.

Since the prior set is too large to be handled in the analysis, reduction 
methods are applied to it. PDF4LHC15 is distributed in three options, 
which use different reduction algorithms as indicated in the brackets:
\begin{itemize}
\item Monte-Carlo (CMC-PDFs), 
\item Hessian with 30 eigenvectors (META-PDFs),
\item Hessian with 100 eigenvectors (MCH-PDFs).
\end{itemize}
In the case of Monte-Carlo compression method (CMC-PDFs) the number of MC 
replicas is reduced in a way, that the agreement between the certain statistical 
characteristics of the prior and the reduced set is the best 
\cite{Carrazza:2015hva}. For the transformation of the MC representation into a 
Hessian one, the authors of PDF4LHC15 used two different methods. META-PDFs is 
based on fitting a functional form to the set of MC replicas, while MCH-PDFs 
uses singular value decomposition, followed by principal component analysis 
\cite{Gao:2013bia, Carrazza:2015aoa}. META-PDFs offers a more accurate 
description of the prior set at smaller numbers of eigenvectors in the reduced 
set, while the opposite holds for the larger reduced sets. For this reason the 
representation with 30 eigenvectors is reduced with the META-PDFs and the 
representation with 100 eigenvectors is reduced with the MCH-PDFs. The latter 
is more appropriate, when the highest accuracy is desired, on the other hand 
the META-PDFs is useful, when simple statistical analysis is the priority. The 
main PDF4LHC15 sets, available in the \texttt{LHAPDF} library are gathered in 
Table \ref{Tab:PDF4LHC}.

\begin{figure}[htbp]
\begin{center}
\includegraphics[scale=0.35]{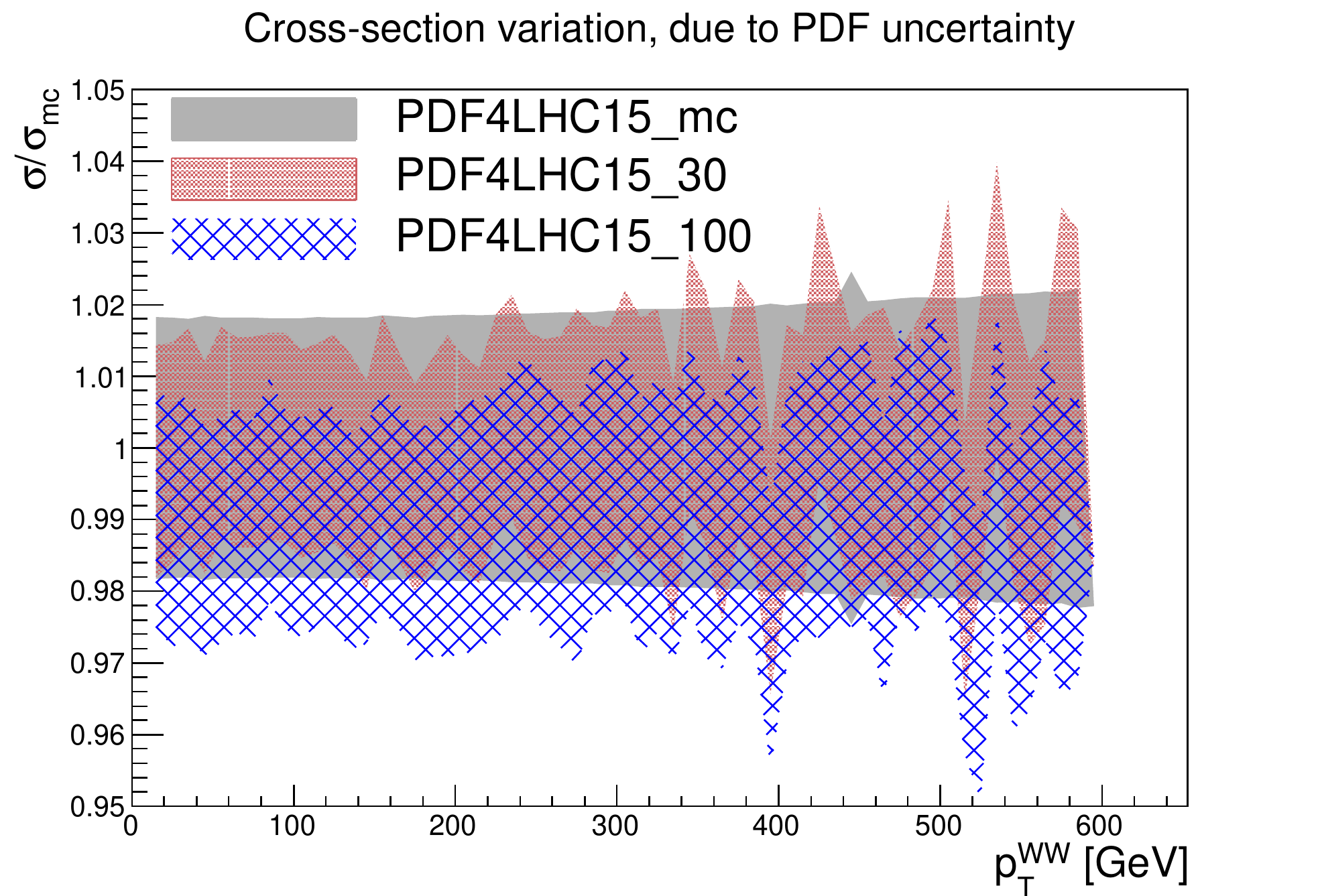}
\includegraphics[scale=0.35]{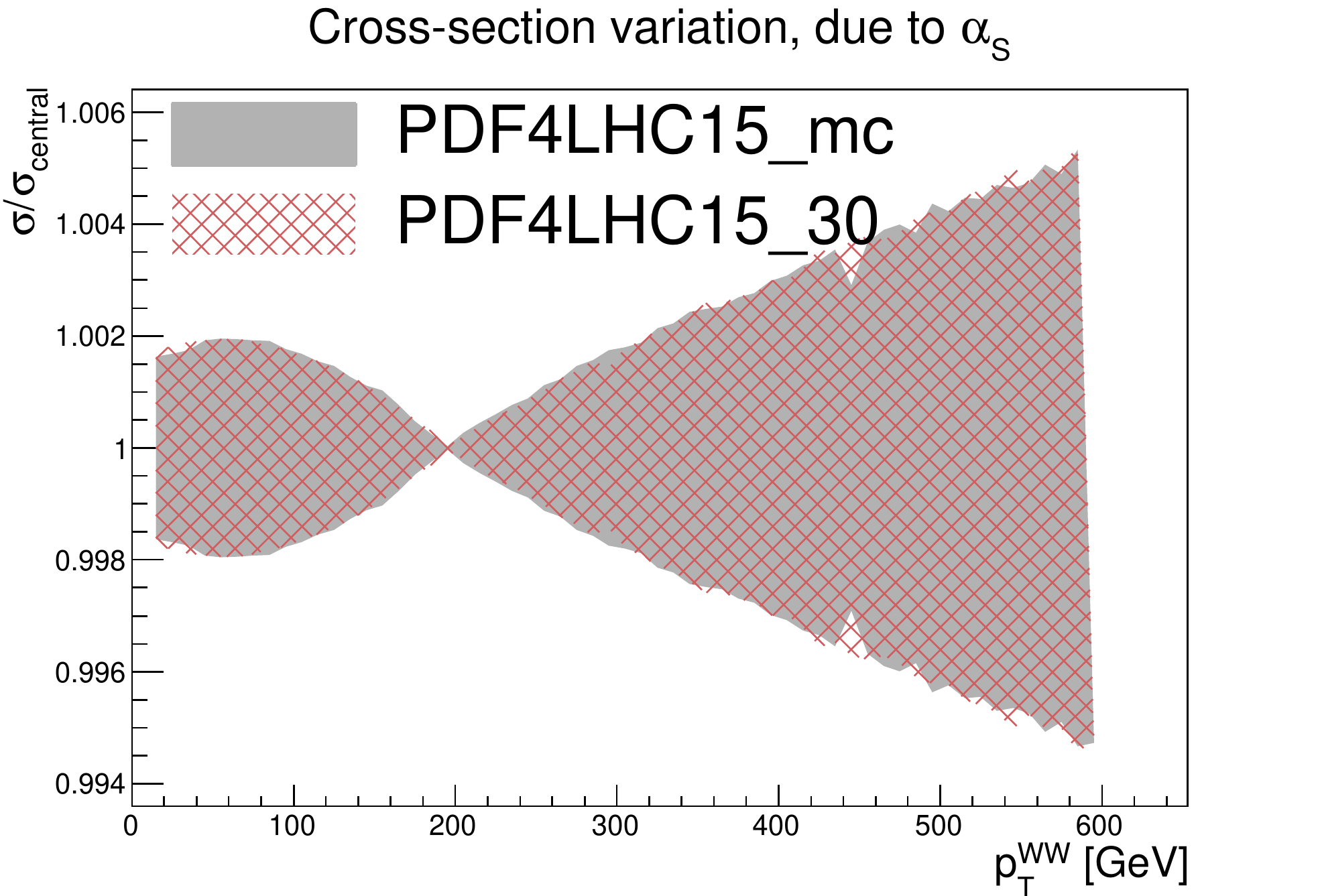}
\end{center}
\caption{PDF uncertainty bands (left) and $\alpha_S$ uncertainty bands (right), 
along the variable $p_{\text{T}}^{WW}$, calculated with the PDF sets 
\texttt{PDF4LHC15\_nlo\_mc}, \texttt{PDF4LHC15\_nlo\_30} and 
\texttt{PDF4LHC15\_nlo\_100}. The cross-sections without $\alpha_S$ variation 
are normalized to the central PDF of MC PDF set, while the PDFs from $\alpha_S$ 
variation are normalized to the central PDF of their own set. The spikes in 
the shape of the uncertainty bands are the consequence of the statistical 
error. The differences among different PDF sets are consistent with the 
calculated uncertainties.}
\label{Fig:PDFResults1}
\end{figure}

\begin{figure}
\begin{center}
\includegraphics[scale=0.35]{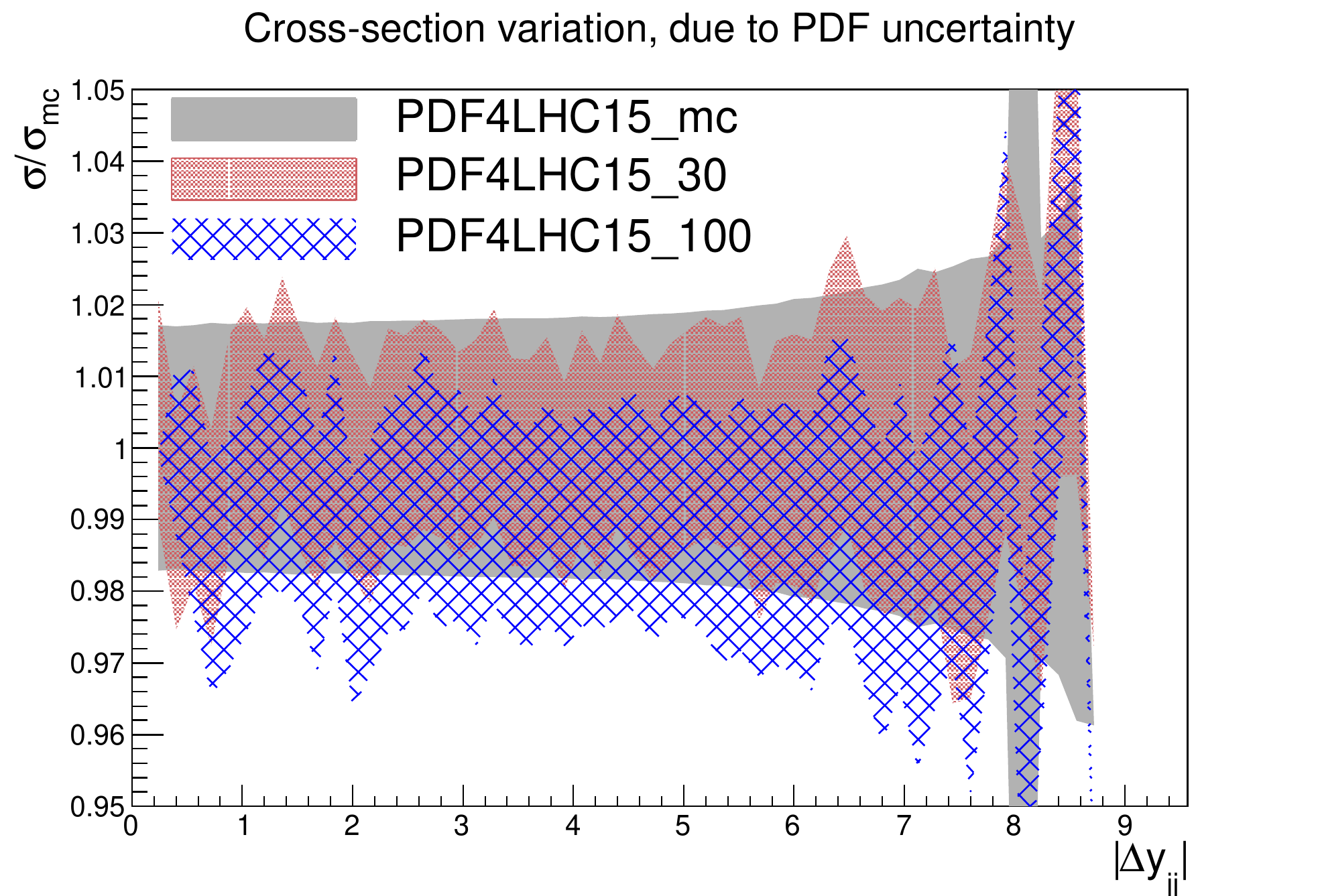}
\includegraphics[scale=0.35]{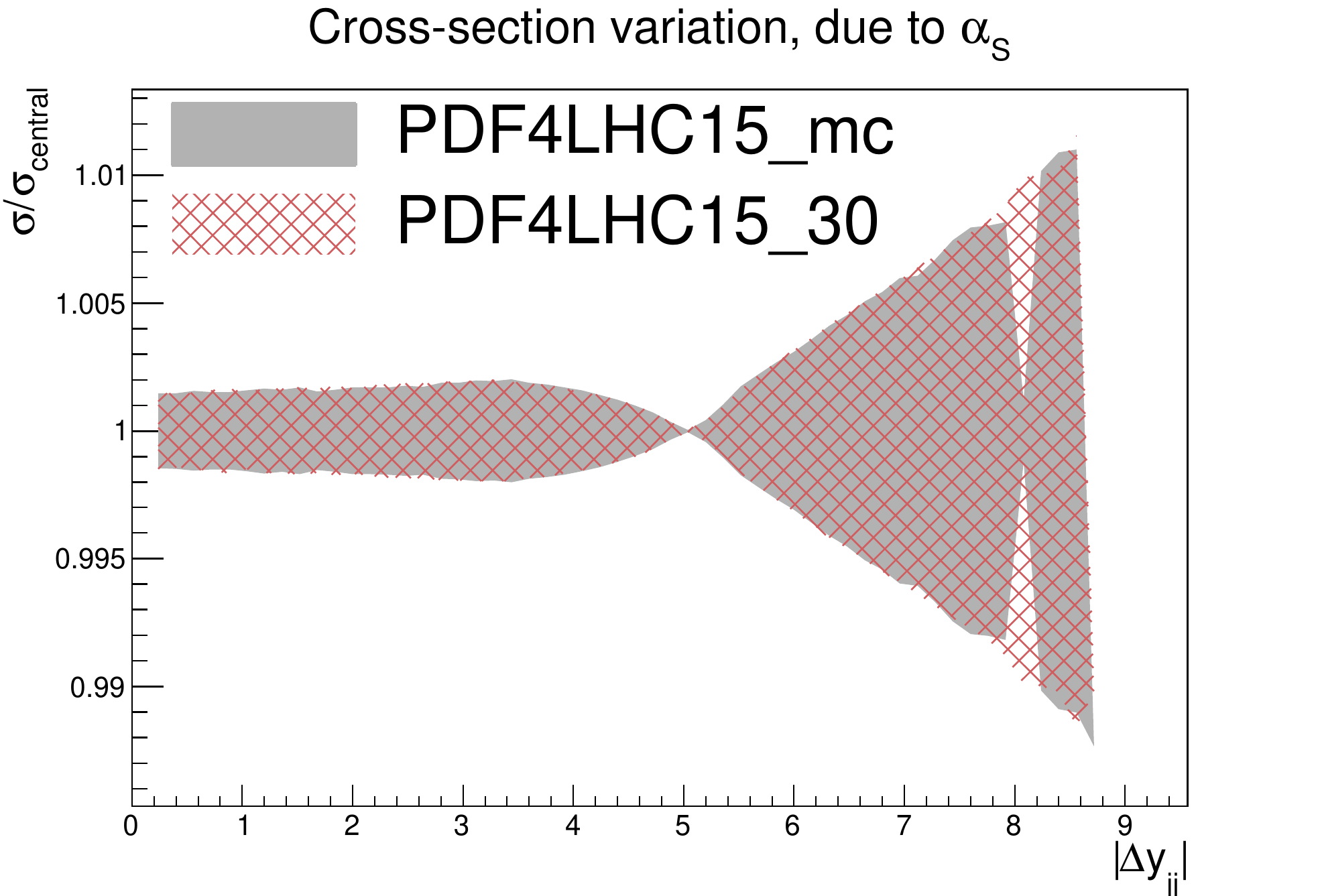}
\end{center}
\caption{PDF uncertainty bands (left) and $\alpha_S$ uncertainty bands (right), 
along the variable $\Delta y_{jj}$, calculated with the PDF sets 
\texttt{PDF4LHC15\_nlo\_mc}, \texttt{PDF4LHC15\_nlo\_30} and 
\texttt{PDF4LHC15\_nlo\_100}. The cross-sections without $\alpha_S$ variation 
are normalized to the central PDF of MC PDF set, while the PDFs from $\alpha_S$ 
variation are normalized to the central PDF of their own set. The spikes in 
the shape of the uncertainty bands are the consequence of the statistical 
error. The differences among different PDF sets are consistent with the 
calculated uncertainties.}
\label{Fig:PDFResults2}
\end{figure}

\begin{figure}
\begin{center}
\includegraphics[scale=0.35]{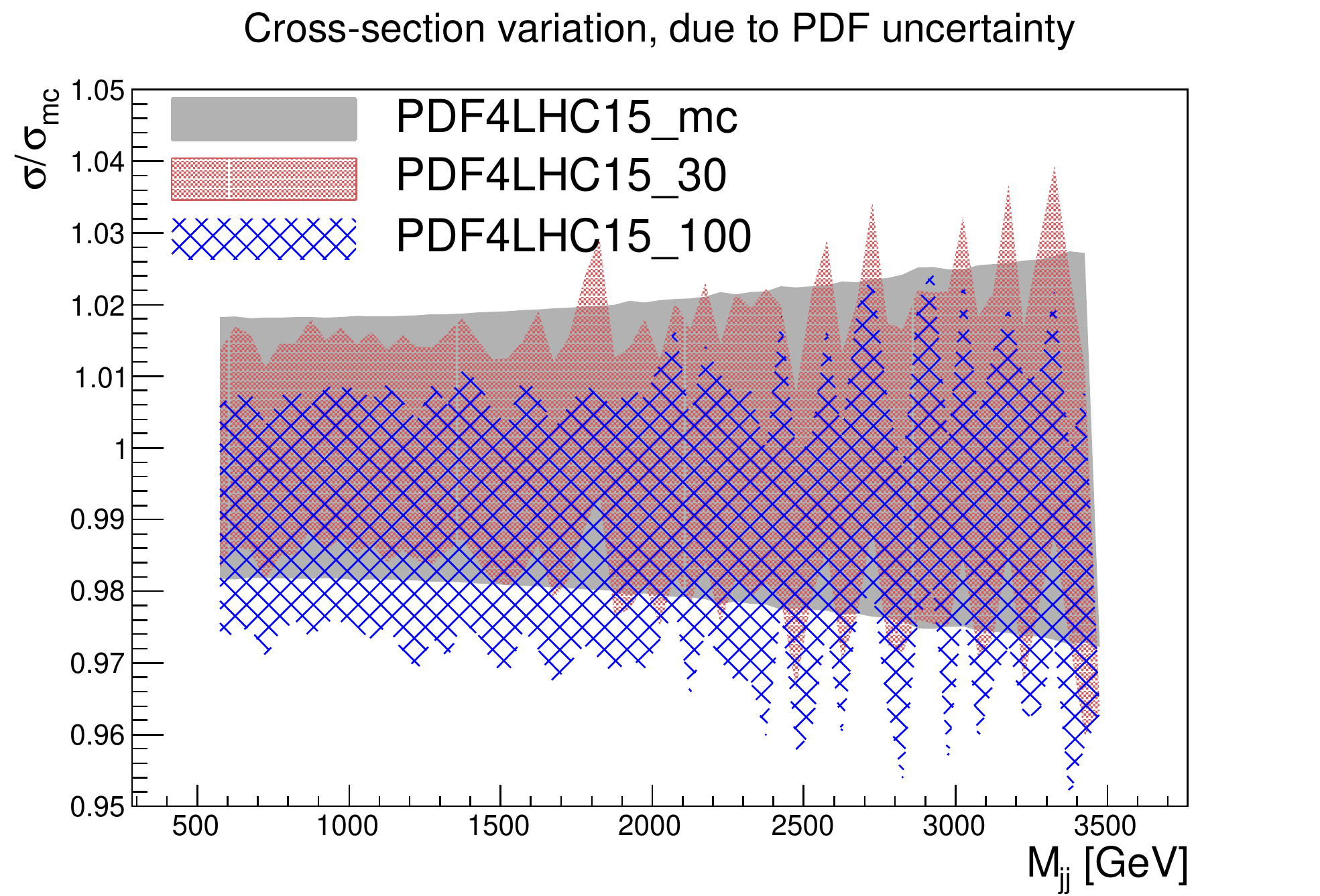}
\includegraphics[scale=0.35]{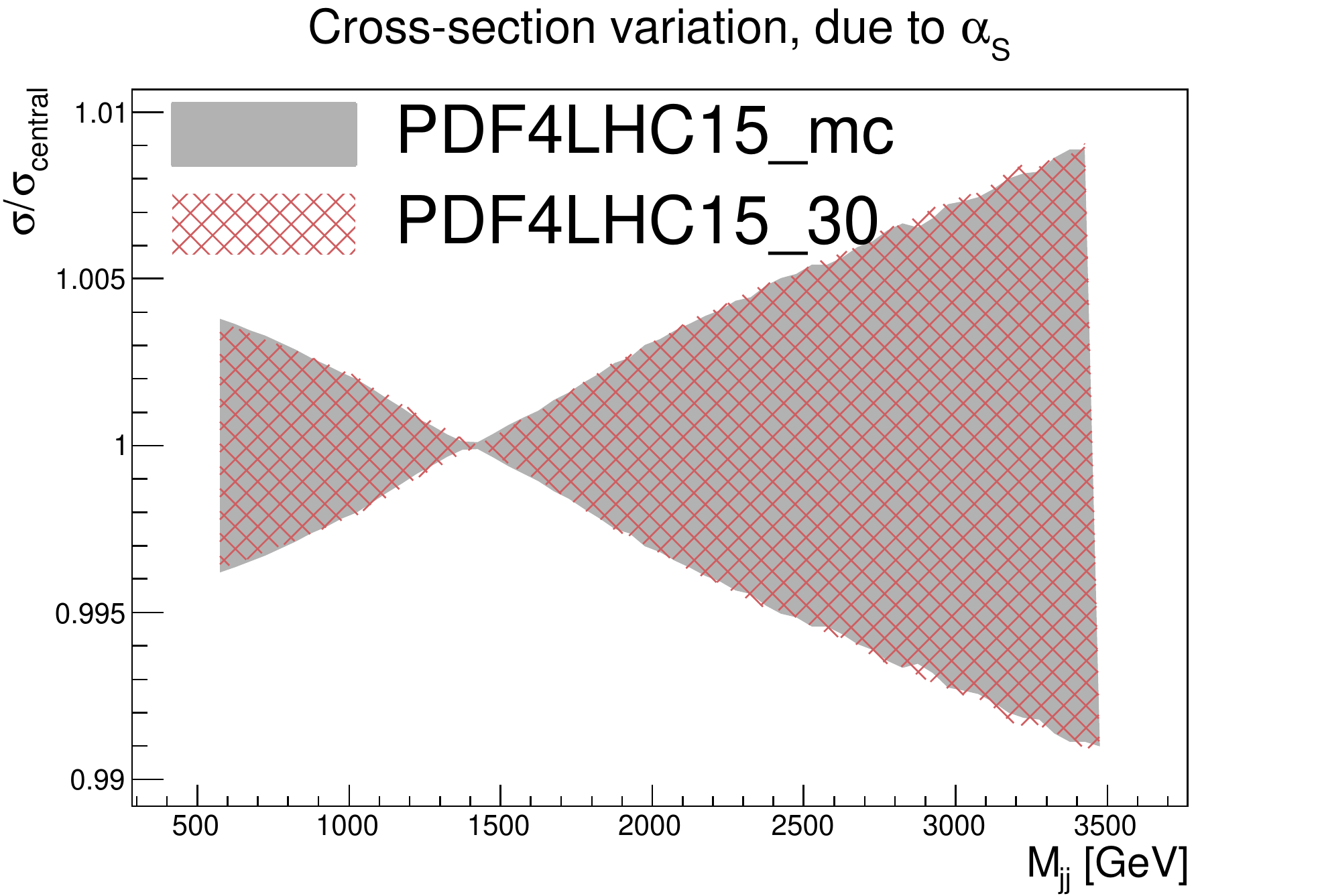}
\end{center}
\caption{PDF uncertainty bands (left) and $\alpha_S$ uncertainty bands (right), 
along the variable $m_{jj}$, calculated with the PDF sets 
\texttt{PDF4LHC15\_nlo\_mc}, \texttt{PDF4LHC15\_nlo\_30} and 
\texttt{PDF4LHC15\_nlo\_100}. The cross-sections without $\alpha_S$ variation 
are normalized to the central PDF of MC PDF set, while the PDFs from $\alpha_S$ 
variation are normalized to the central PDF of their own set. The spikes in 
the shape of the uncertainty bands are the consequence of the statistical 
error. The differences among different PDF sets are consistent with the 
calculated uncertainties.}
\label{Fig:PDFResults3}
\end{figure}

\begin{table}
\begin{center}
{\small
\begin{tabular}{|c|c|c|c|c|c|}
\hline
PDF set & Reduct. algo. & Pert. order & Uncertainty type & $N_{\text{mem}}$ & 
$\alpha_S$ var. \\ \hline
\texttt{PDF4LHC15\_nlo\_100} & MCH-PDFs & NLO & \texttt{symhessian} & 100 & No \\ \hline
\texttt{PDF4LHC15\_nlo\_30} & META-PDFs & NLO & \texttt{symhessian} & 30 & No \\ \hline
\texttt{PDF4LHC15\_nlo\_mc} & CMC-PDFs & NLO & \texttt{replicas} & 100 & No \\ \hline
\texttt{PDF4LHC15\_nlo\_30\_pdfas} & META-PDFs & NLO & \texttt{symhessian+as} & 32 & Yes \\ \hline
\texttt{PDF4LHC15\_nlo\_mc\_pdfas} & CMC-PDFs & NLO & \texttt{replicas+as} & 102 & Yes \\ \hline
\texttt{PDF4LHC15\_nnlo\_100} & MCH-PDFs & NNLO & \texttt{symhessian} & 100 & No \\ \hline
\texttt{PDF4LHC15\_nnlo\_30} & META-PDFs & NNLO & \texttt{symhessian} & 30 & No \\ \hline
\texttt{PDF4LHC15\_nnlo\_mc} & CMC-PDFs & NNLO & \texttt{replicas} & 100 & No \\ \hline
\texttt{PDF4LHC15\_nnlo\_30\_pdfas} & META-PDFs & NNLO & \texttt{symhessian+as} & 32 & Yes \\ \hline
\texttt{PDF4LHC15\_nnlo\_mc\_pdfas} & CMC-PDFs & NNLO & \texttt{replicas+as} & 102 & Yes \\ \hline
\end{tabular}
}
\end{center}
\caption{The list of the main PDF4LHC15 sets, in five quark flavour scheme, 
available in the \texttt{LHAPDF} library. Currently $\alpha_S$ variation is 
available only in the META-PDFs and the CMC-PDFs sets.}
\label{Tab:PDF4LHC}
\end{table}\begin{table}[ht]
\begin{scriptsize}
\begin{center}
\begin{tabular}{|c|c|c|c|c|c|c|c|}
\hline
PDF type & \makecell{Total LO\\ xsection [fb]} & \makecell{PDF unc\\ MC (a)  
[$\%$]} & \makecell{PDF unc\\ MC (b)  [$\%$]} & \makecell{PDF unc\\ Hess 
[$\%$]} & $\alpha_{S}$ unc [$\%$] & \makecell{Combined\\ unc [$\%$]} \\ \hline
\texttt{PDF4LHC15\_nlo\_100} & $2.15271$ & - & - & $1.76815$ & - & - \\ \hline
\texttt{PDF4LHC15\_nlo\_30\_pdfas} & $2.15298$ &  - & - & $1.6248$ & $1.3131\times 10^{-2}$ & $1.6249$ \\ \hline
\texttt{PDF4LHC15\_nlo\_mc\_pdfas} & $2.15333$ &  $1.8329$ & $1.92104$ & - & $1.1046\times 10^{-2}$ & $1.8329$\\ \hline
\end{tabular}
\end{center}
\end{scriptsize}
\caption{PDF uncertainties of the total cross-section for the process 
$pp\rightarrow\mu^+\nu_{\mu} e^+\nu_e jj$. The PDF uncertainty is observed to 
be around $2\%$, while the $\alpha_S$ uncertainty is two orders of magnitude 
lower. The difference between the uncertainties calculated from the equations 
(\ref{Eq:PDFMCGaus}) and (\ref{Eq:PDFMCNonGaus}) is small, which is consistent 
with the Shapiro-Wilk test.}
\label{Tab:PDFTotUnc}
\end{table}In the study of the PDF uncertainty for the process 
$pp\rightarrow\mu^+\nu_{\mu} e^+\nu_e jj$ with the \texttt{PHANTOM} LO 
generator \cite{Ballestrero:2007xq}, we used the reweighting from equation 
(\ref{Eq:PDFReweigh}). The kinematical cuts used in generation are:
\begin{itemize}
\item $p_{\text{T}}^{\ell} > 20$ GeV, $|\eta^{\ell}| < 2.5$, $p_{\text{T}}^{\text{miss}} > 40$ GeV;
\item $p_{\text{T}}^{j} > 30$ GeV, $|\eta^{j}| < 4.5$, $|\Delta\eta_{jj}| > 2.5$, $m_{jj} > 500$ GeV;
\item  $\Delta R_{\ell\ell} > 0.3$, $\Delta R_{j\ell} > 0.3$;
\end{itemize}
The Shapiro-Wilk normality test on the cross-section distribution of the MC PDF 
set gives $p$-value $p = 0.5928$. This value indicates that 
the distribution is Gaussian and equation (\ref{Eq:PDFMCGaus}) is adopted for 
the calculation of the PDF uncertainty for the MC PDF set. The resulting 
uncertainties of the total cross-sections, calculated with the NLO 
representations of the PDF4LHC15 set, are listed in Table~\ref{Tab:PDFTotUnc}. 
We note that the $\alpha_S$ uncertainty is significantly smaller than the PDF 
uncertainty. Uncertainty bands, calculated with the same PDF sets, are 
presented in the Figures \ref{Fig:PDFResults1}, \ref{Fig:PDFResults2}, and \ref{Fig:PDFResults3}. 
Different representations give consistent predictions. The PDF uncertainty is 
observed to be more or less uniform along the whole phase space.
[htbp]
[ntbp]



\setcounter{footnote}{0}
\chapter{Analysis Techniques}
\label{sec:wg2}
 \section{Reconstruction of the W boson rest frame in VBS \footnote{speaker: M. Grossi}}

As well known, the longitudinal \emph{WW} scattering carries the most direct information about the mechanism of electroweak symmetry breaking. 
Experimental investigation of the $W_LW_L$ scattering is now becoming feasible at the LHC and several techniques for disentangling the longitudinal components from the transwerse one have been proposed. Some of them (for instance \cite{Maina:2017eig}) are crucially connected with the experimental capability to reconstruct the \emph{W} boson rest frame, in which some of vector boson properties, like indeed the polarisation, can be studied by exploiting the angular distribution in terms of Legendre polynomials.
However, the \emph{W} rest frame reconstruction is experimentally extremely challenging, both in the case when the \emph{W} decays hadronically (final states with two hadronic jets) or when it decays leptonically (final state with one lepton and one neutrino, which is escaping the detector). 

Considering a single \emph{W} boson as a starting point, by working out the neutrino energy-momentum equation in the ultra-relativistic limit, and solving it for the longitudinal component of the neutrino, one finds:
\be
\begin{aligned}[c]
\label{pLnu}
\begin{split}
&\underbrace{\left(p^2_{\ell L} -E_\ell ^2\right)}_{a} p_{\nu L}^2 + \\
&\underbrace{\left( { m_{\textrm{W}}^2} p_{\ell L} +2 p_{\ell L} \vec{p}_{\ell T} \vec{p}_{\nu T}\right)}_{b} p_{\nu L} + \\
&\underbrace{\frac{m_{\textrm{W}}^4}{4}+ ( \vec{p}_{\ell T} \vec{p}_{\nu T})^2 + {m_{\textrm{W}}^2} \vec{p}_{\ell T} \vec{p}_{\nu T} - E_\ell ^2 \vec{p}_{\nu T}^ {\;2}) }_{c} = 0
\end{split}
\end{aligned}\quad\implies\,\,\,
\begin{aligned}[c]
p_{\nu L} = \frac{-b \pm \sqrt{b^2 - 4ac}}{2a} ,
\end{aligned}
\ee
where $p_{\ell L}$, $\vec{p}_{\ell T}$ are respectively the longitudinal and transverse component of the lepton momentum and $E_\ell$ represents its energy, while $\vec{p}_{\nu T}$ is the transverse component of the momentum of the neutrino. This represents a second order equation in $p_{\nu L}$, which has the standard solution shown on the right side of \ref{pLnu}. Solving it, means to determine the full events kinematics, and therefore to reconstruct the frame in which the \emph{W} is at rest.
It is important to note that transverse neutrino components are measurable in collider experiments, but longitudinal ones are not.

As a matter of fact, when $b^2 - 4ac$ is negative, imaginary solutions for $p_{\nu L}$ appear, which are not physically meaningful. There are some ad-hoc shortcuts adopted in literature, such as forcing the discriminant to be zero or recalculating the discriminant using the \emph{W} transverse mass. However, we do not want to discuss these cases here, rather to focus on the ambiguity of the two possible solutions ($+/-$) in the positive discriminant ($\Delta = b^2 - 4ac >0$) case.
A priori, both are physical solutions but nature chooses one of them only. 

In the following, we describe a possible approach to determine which of the two ($+/-$) solutions is chosen, based on a cut based analysis. We analyse as first the case when one \emph{W} boson decays leptonically and the other hadronically ({\em semi-leptonic} VBS process), to move afterwars to the more complitate situation whem both \emph{W}'s are decaying leptonically ({\em fully-leptonic} VBS processes).

We wrote a code which could solve sign ambiguity in longitudinal neutrino momentum reconstruction by means of different selection criteria. The reconstruction capability of the algorithm is then evaluated comparing the identified $p_{\nu L}$ against the truth one ($p_{\nu L}^{rec}$ - $p_{\nu L}^{th}$), provided by the event generator. This relative error is scanned in a 2D plot, to investigate whether there are preferred phase space regions, by each of the two ($+/-$) solution.

The code can choose among the following algorithms, to estimate the $p_{\nu L}$:
\begin{itemize}
\item Selection 0: the sign pf the solution of \ref{pLnu} is chosen randomly;
\item Selection 1: the solutions with absolute value smaller than 50 GeV are discarted;
\item Selection 2: all the solutions for which $-p_{\nu L}*a/b < 0.5$ are discarted (this option chooses the solutions which fall on the right hand side w.r.t.\, the parabola axis); 
\item Selection 3: if the scalar product of the reconstructed neutrino three-momentum (solution is taken for the longitudinal component) with the reconstructed \emph{W} three-momentum is smaller than 2500~ GeV$^2$, the solution is discarded;
\item Selection 4: if the value of the scalar product of the reconstructed neutrino three-momentum  with the reconstructed emph{W} three-momentum, multiplied by $a/b$, is smaller than 30~GeV or larger than $-$25~GeV, the solution is discarded. 
\end{itemize}
In the selections above, $a$ and $b$ represents the first and second coefficients in Equation~\ref{pLnu}.

Semi-leptonic VBS events are generated with \texttt{PHANTOM}\cite{Ballestrero:2007xq} with the following characteristics:
\begin{itemize}
\item Statistics: 1 million events;
\item PDF choice: \texttt{NNPDF30\_nnlo\_as\_0118};
\item Perturbative order: $\alpha_e^6$ at 13~TeV c.m.e.\,;
\item QCD scale choice: (invariant mass of the two central jets and of the two leptons)$/\sqrt2$.
\item kinematical cuts:
\begin{itemize}
\item $p_{\text{T}}^{\ell} > 20$ GeV, $|\eta^{\ell}| < 2.5$, $p_{\text{T}}^{\text{miss}} > 40$ GeV;
\item $p_{\text{T}}^{j} > 30$ GeV, $|\eta^{j}| < 4.5$, $|\Delta\eta_{jj}| > 2.5$, $m_{jj} > 500$ GeV;
\item  $\Delta R_{\ell\ell} > 0.3$, $\Delta R_{j\ell} > 0.3$.
\end{itemize}
\end{itemize}
The events generated have mixed polarisations.

\begin{figure}[h]
\centering
\includegraphics[width=0.75\textwidth]{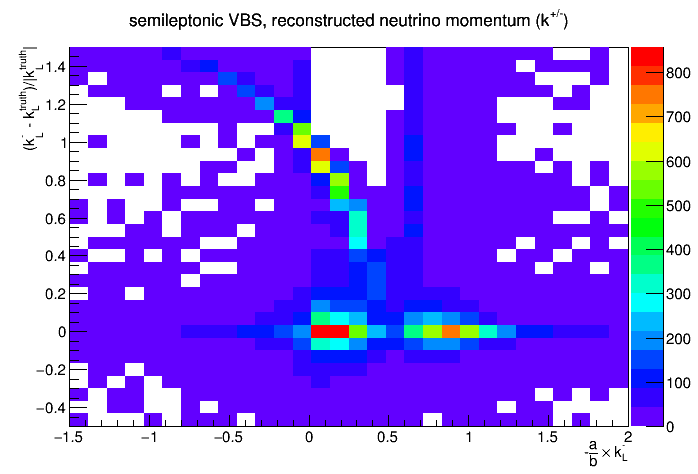}
\caption{Relative error of longitudinal neutrino ($p_{\nu L}^{rec}$ - $p_{\nu L}^{th}$) as a function of one of the event kinematic variables.} \label{2d}
\end{figure}

In the following, we show the effect of applying one of the selection criteria above, for instance Selection 2: $-p_{\nu L}(\pm)*a/b < 0.5$.
By looking at \reffi{2d}, we can clearly distinguish two peaks (in red in the 2D plot), which correspond either to the positive or to the negative solution. By applying the  $-p_{\nu }L(\pm)*a/b < 0.5$ cut, one is selecting the left peak, more populated than the right one. 

The relative distance $p_{\nu L}^{rec}$ - $p_{\nu L}^{th}$ of the reconstructed $p_{\nu L}$ has been studied both at parton level and after detector simulation effects\footnote{Detector fast simulation performed by using \texttt{Delphes}\cite{deFavereau:2013fsa} framework with ATLAS card set up.}.
The evaluation of the $p_{\nu L}$ performed by the code on events {\em\/ after} the detector smearing is crucial to evaluate the contribution of experimental effect on the performance of the reconstruction code.

\begin{figure}
\centering
\includegraphics[width=0.7\textwidth]{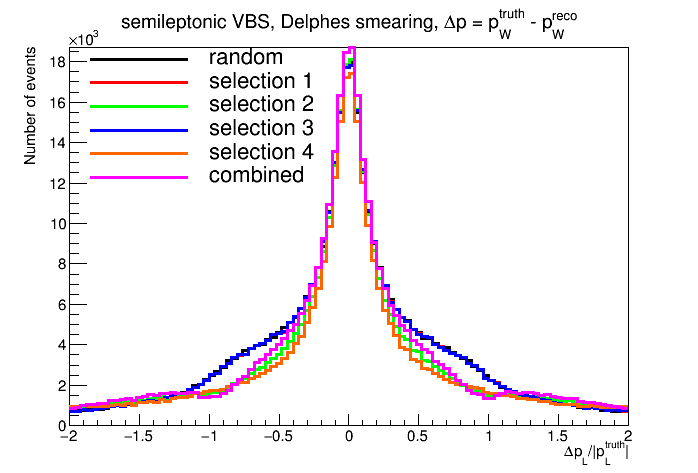}
\caption{Reconstruction efficiency of \emph{W} transverse momentum at simlation level, for the different Selections listed in the text and for their combination.}
\label{1ddetec}
\end{figure}

A qualitative analysis of plots like \reffi{2d} and similars, demonstrates that a combined selection criterion (i.e.\, a combination of all the selections listed above) performs better than any single one. This can be better appreciated at simulation level, where the separation among the different lines is more evident, as shown in \ref{1ddetec}.


The reconstruction code, which makes use of the selection criteria described above, can be applied also to fully leptonic VBS (i.e., events in which both \emph{W} decay leptonically).
In this case, the kinematics is more complicated and to reconstruct both \emph{W}'s rest frames, one has to deal with 8 parameters (2 for each neutrino four momentum) and to solve 6 equations similar to \ref{pLnu}.
To handle this complicated situation, which is not analytically solvable, we adopted the $MT2$-Assisted On-Shell (MAOS) quantity techniques \cite{Sonnenschein:2006ud,Choi:2009hn}, which performs a minimisation of the transverse masses of the lepton-neutrino pairs.
The MAOS estimations of $\vec{p}^{\,\nu_e\prime}_{\text{T}}$ and $\vec{p}^{\,\nu_{\mu}\prime}_{\text{T}}$ for neutrinos transverse momenta can be obtained by minimising the function $f(\vec{p}_1,\vec{p}_2) = \max\{M^{W_1}_\text{T},M^{W_2}_\text{T}\}$, constrained by a bond $\vec{p}_1 + \vec{p}_2 = \vec{p}_{\text{T}}^{\,\text{miss}}$, where: \\
\begin{equation*}
M_{\text{T}}^{W_1} = 2(|\vec{p}_{\text{T}}^{\,\mu}||\vec{p}_1| - \vec{p}_{\text{T}}^{\,\mu}\cdot\vec{p}_1),\qquad M_{\text{T}}^{W_2} = 2(|\vec{p}_{\text{T}}^{\,e}||\vec{p}_2| - \vec{p}_{\text{T}}^{\,e}\cdot\vec{p}_2) .
\end{equation*}
The minimum of the function $f$ defines the quantity $M_{\text{T2}}$:
  \be
M_{\text{T2}} \equiv \min_{\vec{p}_1 + \vec{p}_2 = \vec{p}_{\text{T}}^{\,\text{miss}}} f(\vec{p}_1,\vec{p}_2) = \left. f\right|_{\vec{p}^{\,\nu_e\prime}_{\text{T}},\vec{p}^{\,\nu_{\mu}\prime}_{\text{T}}} .\label{Eq:MAOS}
\ee
Without entering into mathematics details, the exact solution of this problem $\min\left[\max\{M^{W_1}_\text{T},M^{W_2}_\text{T}\}\right]$ lies always at the intersection of $M^{W_1}_\text{T}$ and $M^{W_2}_\text{T}$.
In addition, in this case, the additional bond $M_{\text{T}}^{W_1} = M_{\text{T}}^{W_2}$ holds.

\begin{figure}
\centering
\includegraphics[width=.7\textwidth]{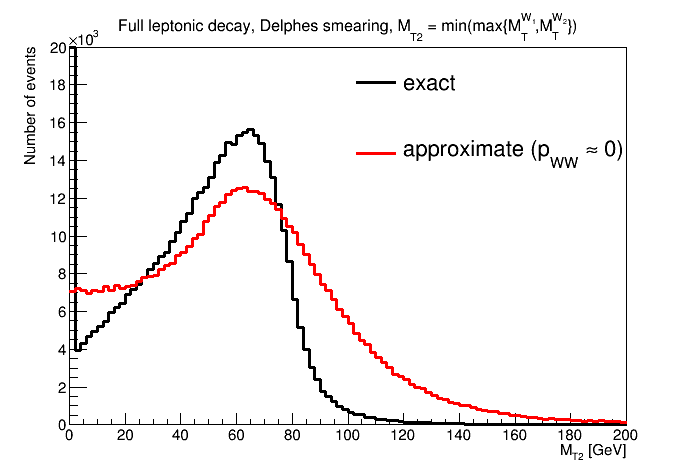}\\
\caption{Evaluation of $M_{\text{T2}}$ as a result of the maximization problem described, using data coming from detector smearin}\label{maosdetec}
\end{figure}

Introducing the angle $\varphi_0$ - between $\vec{p}_{\text{T}}^{\,\text{miss}}$ and $\vec{p}_{\text{T}}^{\,\ell\ell}$ we obtain a second order equation, in parametric form where the $x$-axis of coordinate system coincides with the $\vec{p}_{\text{T}}^{\,\ell\ell}$ direction.
\begin{equation*}
|\vec{p}_1| = \frac{-g(\varphi) \pm \sqrt{g(\varphi)^2 - 4cf(\varphi)}}{2f(\varphi)},\qquad \vec{p}_2 = \vec{p}_{\text{T}}^{\,\text{miss}} - \vec{p}_1 .
\end{equation*}
Minimum of $M_{\text{T2}}$ on the intersection curve can be found numerically.
\reffi{maosdetec} is produced by evaluating $M_{\text{T2}}$ in 2000 points.

\begin{figure}
\centering
\label{costheta}
\includegraphics[width=.7\textwidth]{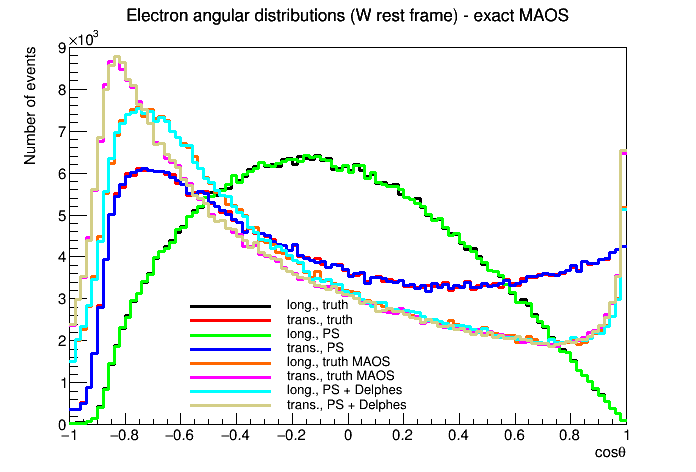} \\
\caption{Distribution of $\cos\theta$ for the longitudinal and the transverse components of the $electron$ using MAOS algorithm, and before and after detector simulation.}\label{maosfull}
\end{figure}

Fully-leptonic VBS events are generated with PHANTOM with the same configuration and set of cuts described above.
In \reffi{maosfull}, the distribution of $\cos\theta$ of the lepton (electron, in this case) for each of polarisation components of the \emph{W}, are shown, for each of the stage considered: truth coming from the event generator, events after the parton shower (labeled as 'PS', in the plot), the thruth coming from MAOS algorithm, and events after parton shower and detector effect smearing. 

Comparing the curves in \ref{costheta}, it appears that it is not the detector effect responsible for the distorsion of the curve with respect to the truth, rather the application of the MAOS technique.
The curves already deviate from the corresponding truth ones at parton level, proving that MAOS cannot be adopted to handle the complicated \emph{WW} scenario.


To summarise the status of these preliminary studies about the \emph{W} rest frame reconstruction, we can conclude that:
\begin{itemize}
\item In the semi-leptonic channel, the reconstruction of the \emph{W} reference frame can be obtained up to a sign ambiguity. The use of selection criteria represents the most promising way to select correct solution.
\item In full-leptonic case, instead, the same approach gives almost no possibility to disentangle the two \emph{W} polarisations (transverse and longitudinal) and the method should be improved by studying different selection criteria for longitudinal solution (for instance finding a variable, i.e. $p_T$ of \emph{WW}, with a larger capability of discriminating between polarisations).
It has been also verified that the effect of detector smearing in the fully-leptonic channel is sub-leading w.r.t.\, MAOS techniques adopted to overcome the complication due to the large number of parameter involved. Different approaches should be evaluated.
\end{itemize}

\section{VBS at linear colliders \footnote{speaker: J. Reuter}}

Surpassing the $WW$ and $ZZ$ threshold at LEP2 in 1996/97 proved the
non-Abelian structure of the electroweak theory and offered for the
first time the possibility to search for anomalous triple gauge
couplings. Future lepton colliders will provide an indispensable
physics program by precision consistency tests of the Standard Model
framework in the Higgs and electroweak sector. Vector boson fusion
into a Higgs boson at 350 GeV center-of-mass energy and beyond
contribute significantly to the Higgs coupling measurements. VBS
measurements will be feasible for energies of 500 GeV and with 
interesting rates at 1~TeV and beyond. This includes the 1~TeV upgrade
option of the International Linear Collider
(ILC)~\cite{Baer:2013cma,Behnke:2013lya}, where a Japanese 
proposal for hosting the project is currently under investigation, and
the high-energy 1.4/1.5 and 3~TeV stages of the Compact Linear
Collider (CLIC)~\cite{Lebrun:2012hj}, studied at CERN. Lepton collider
measurements take place in a clean environment with a well-defined
initial state and with a triggerless operation. This allows to study
fully hadronic final states in order to use the larger hadronic
branching ratios. As the largest cross section comes from the $WW \to
WW$ and $WW \to ZZ$ subprocesses, the signal consists of two very
forward neutrinos, i.e. missing energy in the detector, and four QCD
jets paired into two electroweak vector bosons. One of the largest
experimental challenges is the separation of hadronic $W$ and $Z$
bosons for energies of 1 TeV and beyond. This can be achieved with an
efficiency of close to 90 \% using tight particle-flow algorithms
(photon-induced backgrounds can deteriorate this efficiency to a bit
below 80 \%)~\cite{Marshall:2012ry}. The largest background comes from
four-jet processes (dibosons mainly) where the missing energy has been
produced by undetected photons from initial-state radiation. ILC and
CLIC detectors offer a low angle coverage with the Lumi and Beam
calorimeters down to 15 mrad. This allows to veto collinear ISR
photons very close to the beam axis. Other backgrounds are triboson
production (which in contrast to the LHC) is irreducible as it is in
the same (EW) gauge-invariance class than VBS and cannot be separated
theoretically~\cite{Boos:1997gw,Boos:1999kj,Beyer:2006hx}, top pairs,
single $W$, radiative Bhabha and $Z$ production as well as QCD di- and
multijets~\cite{Fleper:2016frz}. 

The signal cross sections rise a factor 3-4 from 1.4 to 3 TeV. The
dominant cuts are on the missing mass to suppress $Z\to\nu\nu$, $WW$
and QCD 4-jets, cuts on the $p_\perp(W/Z)$ and beam angles of the $W/Z$ to
suppress multi-peripheral diagrams, $p_\perp(WW/ZZ)$ and the beam
angle of electrons and positrons to suppress photon-induced
backgrounds and invariant mass cuts on the diboson system to suppress
massive EW radiation events. After all cuts, total cross sections are
at the level of 0.2 to 0.8~fb, so integrated luminosities of an inverse
attobarn and more are necessary for precise
measurements~\cite{Fleper:2016frz}. Further enhancement of
signal-to-background ratios can be achieved by using 80 \% electron
polarization (at CLIC), or even further with 30\% positron
polarization at ILC. This tremendously helps extraction of \dimS{} and
\dimE{} EFT operator coefficients and disentangling different operator 
coefficient contributions. Limits on new physics contributions using
an sanitized approach of \dimE{} EFT operators including a unitarization
procedure using the formalism
of~\cite{Alboteanu:2008my,Kilian:2014zja,Kilian:2015opv} have been provided in 
Ref.~\cite{Fleper:2016frz} and compared to projections for the full
LHC program. These EFT setup has been implemented and together with
the unitarization procedure made publicly available within the WHIZARD
event generator~\cite{Kilian:2007gr}. 

The largest theoretical challenges are precision predictions,
i.e. full EW next-to-leading order calculations are necessary,
particularly electroweak Sudakov logarithms from massive final-state
radiation and resummation of soft and hard-collinear photons in the
initial state to get the correct normalization of cross section and
the correct description of beam spectra.


\section{VBS at the ILC \footnote{speaker: J. Beyer}}

Precision studies of the quartic interaction between vector bosons are a crucial
step in testing the validity of the Standard Model at high energies. A lepton
collider provides an ideal environment for such measurement due to its clean and
well-known initial state. The study of vector boson scattering requires
a high center-of-mass energy. \\
Such could be provided by the proposed International Linear Collider (ILC) or
the Compact Linear Collider (CLIC) which could ultimately reach energies of
$1\,$TeV and $3\,$TeV, respectively. Proposed detectors for these experiments
are optimized for such precision measurements and utilize Particle Flow event
reconstruction. The International Large Detector (ILD) is one of the ILC
detector concepts and has been shown in simulation to achieve jet energy
resolutions down to 3\%. At this precision it is possible to separate hadronic
decays of a $Z$ from those of a $W$. It is therefore feasible to make precision
electroweak measurements in fully hadronic final states which are at the moment
technically inaccesible at hadron colliders.\\
Sensitivity studies searching for anomalous Quartic Gauge Couplings in the
$e^+e^- \rightarrow \nu \bar{\nu} q \bar{q} q \bar{q}$ channel have been
performed for the ILD Technical Design Report \cite{Baer:2013cma}. Full
simulation and a cut-based analysis were employed to set limits on anomalous
couplings.
The goal of the work is to update the study, 
taking into account advances in the detector model,
description and simulation as well as in particle physics in general. \\
First studies have been performed on the reconstruction of
$E_{\text{miss}} + 4\text{jets}$ final state. The di-boson $WW/ZZ$ mass peaks
are found to be well separated, but long tails towards low masses and a small
shift of the peaks with respect to the generator level mass can be seen.
Challenges are identified in the reconstruction of jets originating from heavy
quarks. Studies are in progress attempting to correct for effects
specific to these jets, such as leptonic decays and hadronic jet content. \\
The final goal is to study the ILCs sensitivity to anomalous coupling in a
\dimE{} SM-EFT framework.

\section{Reinterpretation studies: search for VBS(ZZ) into $4l$, $2l2q$ and $2l2\nu$ final states with the CMS experiment \footnote{speaker: C. Thorburn}}

A search for ZZ vector boson scattering into $4l$, $2l2q$ and $2l2\nu$ final
states, based on matrix element techniques (MELA\cite{PhysRevD94055023,
PhysRevD89035007,PhysRevD86095031,PhysRevD81075022} ), is presented.
2016 CMS detector proton-proton collision ($\sqrt{s}=13$~TeV) data is employed,
with an integrated luminosity of $35.9~\mathrm{fb}^{-1}$. \\
The  VBS(ZZ) 4l channel was already addressed in studies employing MVA/BDT\cite{2017682},
as well as matrix element techniques \cite{CMS-PAS-SMP-16-019}.\\
As shown in previous VBS(ZZ) jj $\rightarrow$ 4l jj channel studies, MELA and BDT efficiencies
are comparable and both better than classic cut-based methods\cite{CMS-PAS-SMP-16-019}.
Therefore, the result is a confirmation of the validity of MELA.\\
This new VBS(ZZ) analysis will follow the methodology employed in a MELA based ZZ-high mass
higgs study, with the same three final states\cite{Sirunyan:2018sfa}. 
The work was focused on a potential H(125) heavy scalar partner decaying into four fermions. Standard
$p_{T}$ and $\eta$ selection for leptons and jets was applied, in addition to low mass $m_{ZZ}$ cuts
at 130 GeV (4l), 300 GeV (2l2$\nu$) and 550 GeV (2l2q), while no $m_{jj}$ restrictions were set. \\
The main feature of MELA is the study of processes at generator level, using 
JHUGen~\cite{Gao:2010qx,Gritsan:2016hjl,Bolognesi:2012mm,Anderson:2013afp} and 
MCFM~\cite{Campbell:1999ah, Campbell:2011bn, Campbell:2015qma} matrix
elements. As a first step, discriminants are defined for event categorization and signal over background
separation. 2D templates are then created for mass and discriminant distributions and a profile likelihood
analysis is performed with the aid of a statistical tool. Lastly, the significance of signal over background
and an upper limit on the cross section are computed. \\
This VBS(ZZ) analysis, at present, has addressed the only $4l$ channel (VBS(ZZ) jj $\rightarrow$ 4l jj) with
the three final states $4e, 4\mu, 2e2\mu$. The main backgrounds are QCD(ZZ) production and single Z plus
jets. A low mass selection cut is set at 160 GeV. \\
The kinematic discriminant is redefined as $K_D = \frac{P_{VBS}}{P_{VBS} + 0.02*P_{q\bar{q}ZZ+ggZZ}}$,
where $P_{q\bar{q}ZZ+ggZZ}$ and $P_{VBS}$ are aggregated probabilities of several independent variables 
for a given 4l total mass. Finally, following the same procedure explained above, significance of signal over
background is obtained. Only an expected, Monte Carlo-based, significance is quoted: this is 1.7$\sigma$ for the 4l channel only (it is 1.6$\sigma$ for the BDT-based analysis), and goes up to approximately 2.2$\sigma$ for the combination of the three final states.

\setcounter{footnote}{0}
\chapter{Experimental Measurements}
\label{sec:wg3}
 \section{Summary on the ATLAS+CMS anomalous quartic couplings combination \footnote{speaker: S. Todt}}

\newcommand\sww{\ensuremath{W^{\pm}W^{\pm}jj}\xspace}
\newcommand\wzjj{\ensuremath{W^{\pm}Zjj}\xspace}
\newcommand\mll{\ensuremath{m_{ll}}\xspace}

One of the efforts of the VBSCan network aims for an inter-experimental combination of limits on anomalous quartic coupling (aQGC) parameters from the ATLAS and CMS experimental results. The proposed strategy suggests to use public results of both experiments. ATLAS and CMS have published results for the \sww and \wzjj final state VBS analyses at $13$ TeV center-of-mass energy \cite{ATLAS-CONF-2018-033, Sirunyan:2017ret, ATLAS-CONF-2018-033, CMS-PAS-SMP-18-001}. Hence, these final states will be included for a first iteration, others can be added once they have been published. The ingredients for this combination are a common signal modelling, the publication of analysis details in a HEPdata format, and  a tool that performs the statistical analysis of the limit setting. This contribution summarizes the studies and suggestions that have been conducted to establish such an effort. Its main focus herein is put on the signal modelling studies.

The theoretical model to be used for the description of the anomalous quartic coupling effects is the effect field theory (EFT) prescription in \cite{Eboli:2016kko} using a \dimE{} basis of operators. The mechanism employed to restore unitarity at large center-of-mass energies is determined to be the so-called clipping method \cite{Lindquist:2261444}. A scan of the clipping energy will be performed in order to set limits depending on the choice of the unitarisation clip off. The signal process for both final states is proposed to be simulated with {\tt MadGraph5\_aMC@NLO}~\cite{Alwall:2014hca} and interfaced to the {\tt Pythia8} event generator \cite{Sjostrand:2014zea} at leading order (LO) in perturbation theory. The intention is to use the LO matrix-element reweighting \cite{Mattelaer:2016gcx}. In this report the various studies using this reweighting technique in Madgraph and based on the \sww final state are presented. The aim of these studies is to propose a systematic grid of aQGC parameter sampling including the choice of baseline samples for the reweighting to various other points.

\begin{figure}
  \centering
  \includegraphics[width=.45\linewidth]{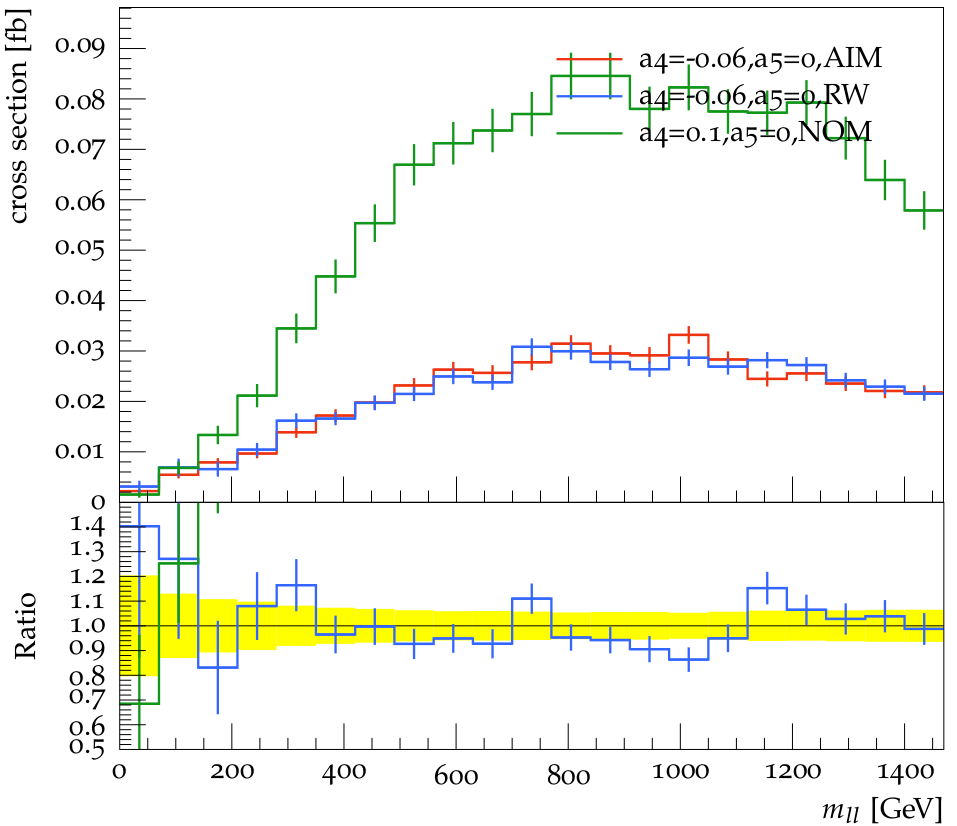}	
  \includegraphics[width=.45\linewidth]{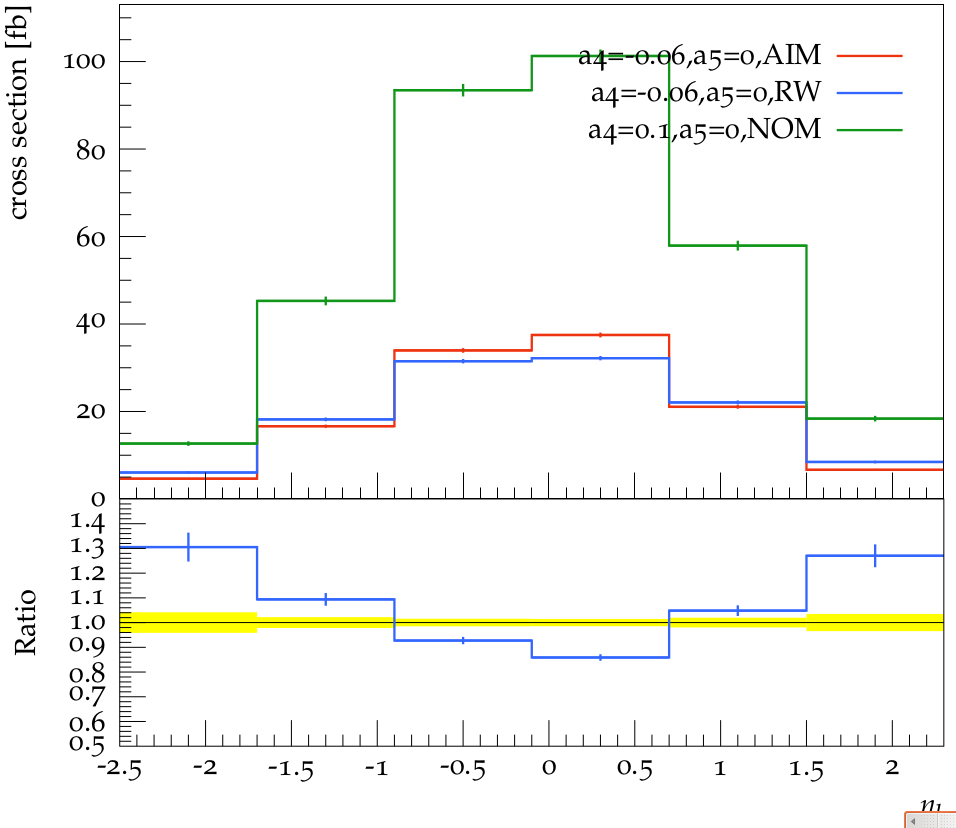}	  
  \caption{Comparison of the stand-alone (red) and the reweighted (blue) distributions in the di-lepton invariant mass (left) and lepton $\eta$ (right) distributions. The aQGC parameter points chosen are $(\alpha_4, \alpha_5) = (0.06,0.00)$ The baseline point used for the reweighting is depicted for reference: $(\alpha_4, \alpha_5) = (0.10,0.00)$ (green).}
  \label{fig:cx_comp1}
\end{figure}

One study is comparing the total cross sections and differential distribution of aQGC samples simulated stand-alone and with using the reweighting tool. Three baseline samples for the reweighting have been chosen which have been reweighted to 14 different parameter points in the $f_{S,0} = f_{S,2}, f_{S,1}$ plane based on the $8$-TeV-limits in \cite{Aad:2014zda}. It has been found that the generator cross sections of the reweighted samples agree within $5 \%$ and the extended statistical uncertainty calculated from \cite{Mattelaer:2016gcx} for most of the points. Possible bigger discrepancies can be identified by using the underlying theoretical evolution of the cross section with the aQGC parameters which follows a paraboloid in the aQGC plane. Differential distribution in this study show a generally good agreement. However, variables can be identified which are not modelled well within the statistical uncertainties of the samples (see Fig.~\ref{fig:cx_comp1}). Therefore a sorrow study of all kinematic distributions has to be made, especially those which are used for the selection of the phase space. 

\begin{figure}
  \centering
  \includegraphics[width=.45\linewidth]{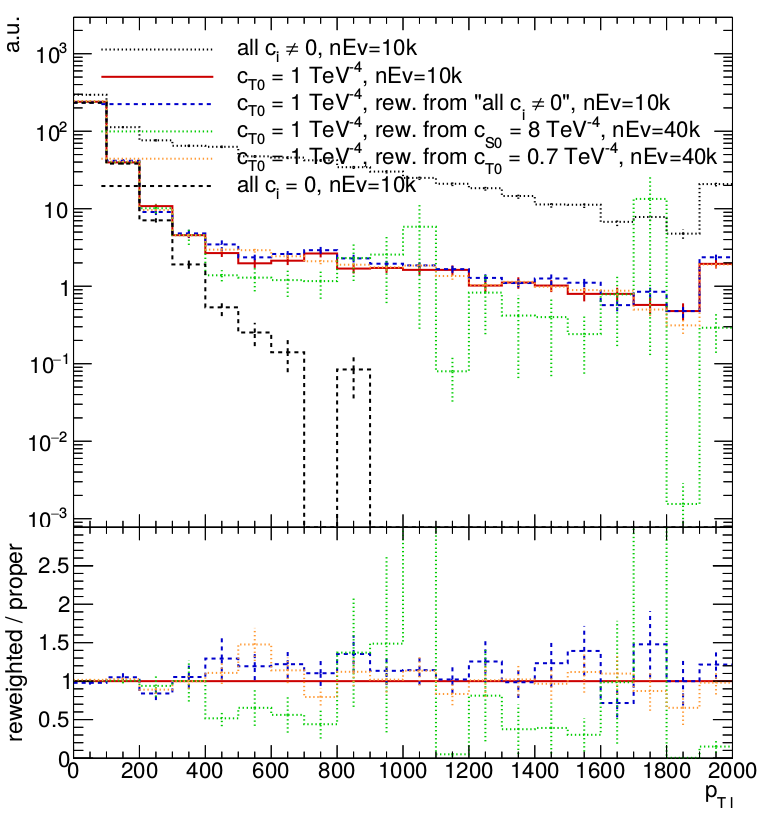}	
  \includegraphics[width=.45\linewidth]{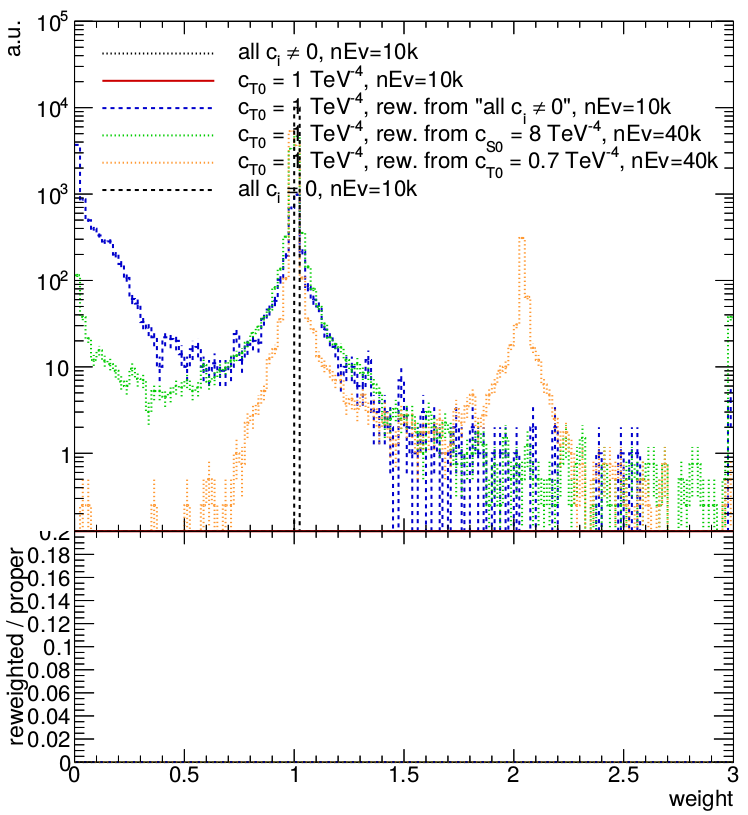}	  
  \caption{Comparison of the lepton $p_T$ (left) and MC weight (right) distribution different reweighting samples to the aQGC parameter point $f_{T,0} = c_{T0} = 1$ TeV$^{-4}$. The stand-alone prediction is shown in red. One of the baseline points used for reweighting is depicted in dotted black. The corresponding reweighting result is shown in dashed blue. In dotted green and orange, two additional reweighted samples are shown, reweighted from $f_{S,0} = 8$ TeV$^{-4}$ and $f_{T,0} = 0.7$ TeV$^{-4}$, respectively. Additionally the SM prediction is shown (dashed black).}
  \label{fig:cx_comp2}
\end{figure}

A second study compares the reweighting between different sets of aQGC parameters. It can be shown that the reweighting reveals the best statistical stability if it is performed within the same family of parameters. This is shown in Fig.~\ref{fig:cx_comp2} where also the distribution of event weights is shown.

The finding of these studies on the matrix element reweighting shows that it is necessary to choose baseline aQGC parameter points which share a similar phase space as the points which should be reweighted to in order for the reweighting to be statistically stable. In this respect also it has to be made sure that the tails of the distributions are populated since these reveal the greatest sensitivity to the anomalous coupling signal. The studies show that an accuracy of the order of $10 \%$ can be reached with the reweighting procedure. This accuracy has been decided to be acceptable also in the light of the current theoretical uncertainties of the signal process.

Based on the results of the reweighting studies a grid of aQGC parameter points can now be generated. In addition to the reweighting further aspects have to be considered. The coverage of the parameters has to include the un-unitarised one dimensional limits published by CMS at $13$ TeV but also the range of sensitivity for low values of the clipping energy. The density of the parameter points has to allow for proper interpolation. The aim is to set two-dimensional limits, therefore we propose to generate EFT signal samples in a two-dimension grid with pair-wise non-zero parameters $f_i$. The chosen grid then also has to allow for a one-dimensional limit extraction.

\section{Combination Studies \footnote{speaker: M. Neukum}}
\newcommand{\software}[1]{\textsc{#1}\xspace}
\newcommand{\MADGRAPHAMC}{\software{MadGraph5\_aMC@NLO}}
\newcommand{\PYTHIA}{\software{Pythia8}}

Combining limits on aQGC parameters makes them stricter by increasing the effective luminosity and is a step towards further studies. This Combination Study focuses on previous analyses by the ATLAS and CMS collaborations at $13~\mathrm{TeV}$ center-of-mass energy and utilizes publicly available information in the HEPData format. The goal is setting one and two dimensional limits on the coefficients of \dimE{} operators in the EFT framework.
The Signal modeling describes the dependence of the signal contribution with varying EFT parameters, which is achieved by generating a MC sample with \madgraph~\cite{Alwall:2014hca,Mattelaer:2016gcx} and \PYTHIA~\cite{Sjostrand:2006za,Sjostrand:2007gs}. Events are then reweighted to different points in EFT parameter space, selection citeria are applied, and resulting yields are fitted with a quadratic polynomial to interpolate between discrete values and describe the signal scaling. Finally, one and two dimensional limits on EFT parameters are extracted by performing a Maximum Likelihood Fit in the sensitive variable, e.g., $m_{\ell \ell}$ in the WW channel, of the signal and background contributions to the measured data. A scan of the negative log-likelihood ratio~\cite{Cowan:2010js} for different EFT parameters results in $95\, \%$ C.L. limits according to Wilks' theorem~\cite{Wilks:1938dza}. \\

\begin{figure}[h]
\centering
\subfigure{\includegraphics[width=.45\textwidth]{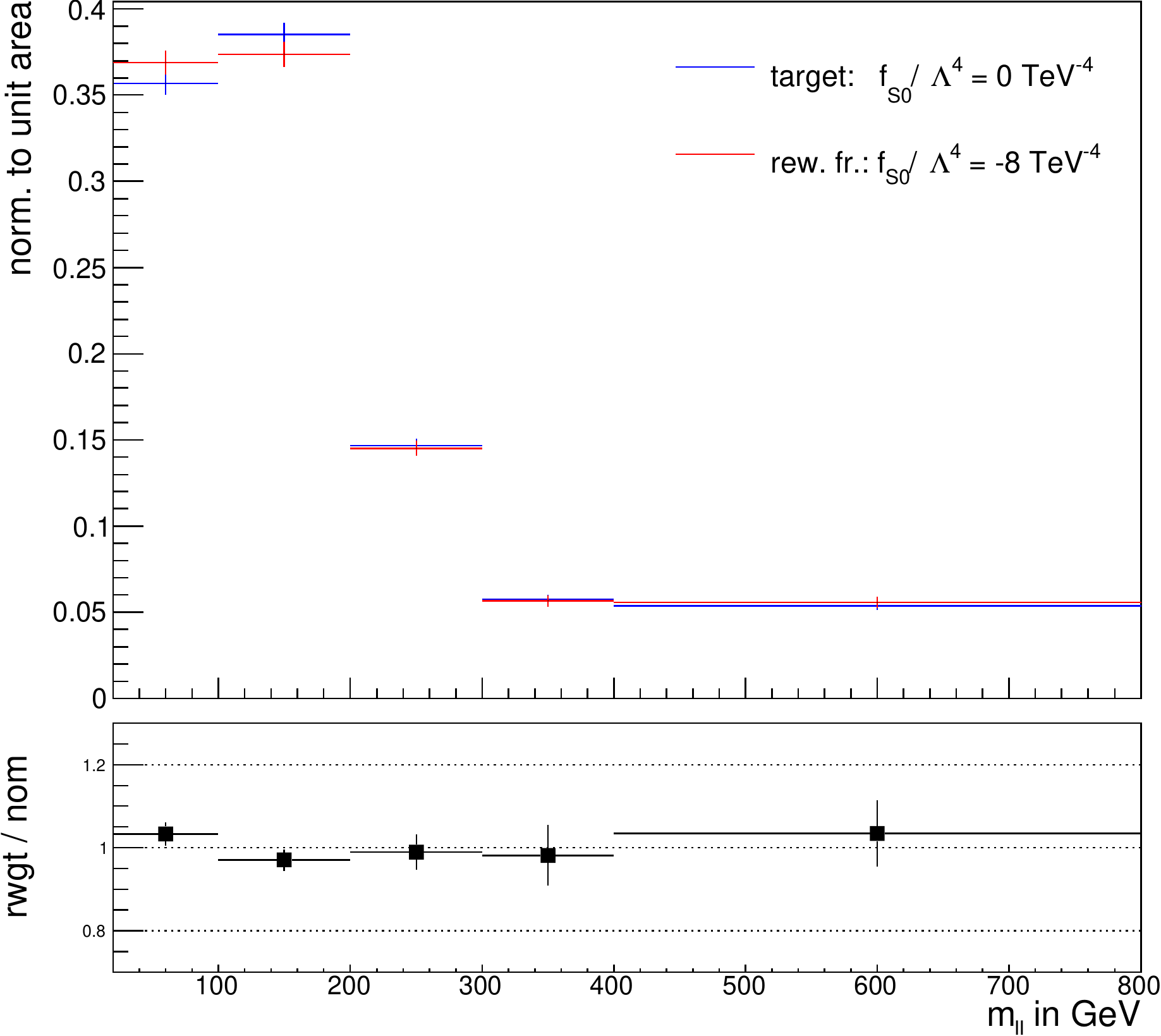}}\hspace*{5mm}
\subfigure{\includegraphics[width=.45\textwidth]{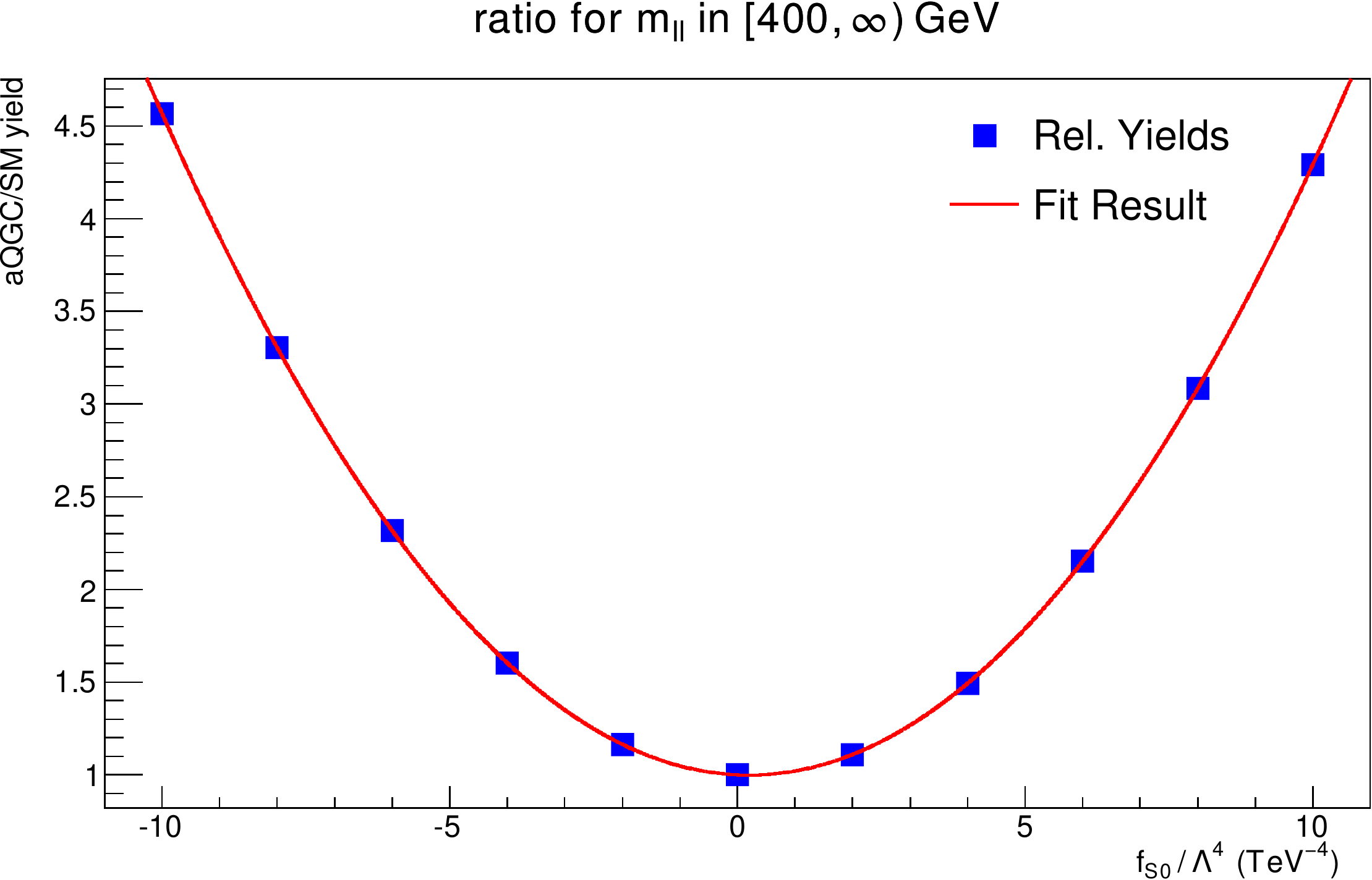}}
\caption[Reweighting example and signal scaling]
{
    Comparison of a MC sample generated at a nominal value of $f_\mathrm{S,0} = 8\, \mathrm{TeV}$reweighted to the SM scenario (left) and the yield ratios of discrete parameter values for $m_{\ell \ell} > 400\, \mathrm{GeV}$ in the ssWW channel (right). The reweighting results in discrepancies smaller than $5\,\%$ and the quadratic polynomial provides a good model to describe the yield ratios.
}
\label{fig:ssww_rwgt_matrix}
\end{figure}

\begin{figure}[h]
\centering
\includegraphics[width=.45\textwidth]{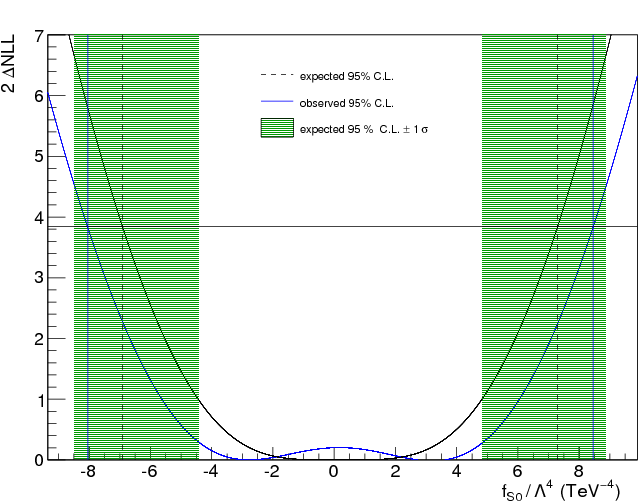}
\caption[$\mathrm{\Delta NLL}$ for $f_\mathrm{S,0}/ \mathrm{Lambda^4}$]{Scan of $2 \mathrm{\Delta NLL}$ for the coefficient $f_\mathrm{S,0} / \Lambda^4$ using the available information. The blue line shows the observed and the black line the expected distributions. A horizontal line at $2 \mathrm{\Delta NLL} = 3.84$ is drawn to derive the $95\,\%$ C.L. limits.}
\label{fig:deltaNLL_fs0}
\end{figure}

So far work was done only in the same sign WW channel~\cite{Sirunyan:2017ret}, since no more public data is available at the time of this workshop. Reweighting within the same parameter results in discrepancies smaller than $5\, \%$. 
This is examplarily shown in Fig.~\ref{fig:ssww_rwgt_matrix}, left, 
where the $m_{\ell \ell}$ distribution generated at $f_\mathrm{S,0} = 8\, \mathrm{TeV}$ is reweighted to the SM scenario. Eleven equally spaced parameter values are chosen along the $f_\mathrm{S,0}$-axis and the yields are fitted with a quadratic polynomial. The resulting function is then normalized to the standard model value. This provides a good model and describes the reweighted yields excellently as illustrated in Fig.~\ref{fig:ssww_rwgt_matrix}, right. 
The limit setting machinery is based on the Higgs combine package~\cite{Khachatryan:2016vau} and is running smoothly. A semi-analytic description of the signal contribution $S(\vec{\theta}, \vec{f})$ is used for constructing the likelihood function:
\begin{equation}
\label{eq:combination_likelihood}
L (\vec{\theta} \, \vert \, \vec{n}) = \prod _i \mathrm{Poisson} ( \,n_i \, \vert \, S_i (\vec{\theta}, \vec{f}) + \sum _j B_{i,j} (\vec{\theta}) \, ) \cdot \rho (\vec{\theta} \, \vert \, \hat{\vec{\theta}}) \, ,
\end{equation}
where $S_i$ and $B_{i,j}$ are the signal and background contributions in bin $i$, $\vec{f}$ the Wilson coefficients, and $\vec{\theta}$ the nuisance parameters with pdf $\rho$. A scan of the negative logarithm of the likelihood ratio, $\mathrm{\Delta NLL}$, under variation of $f_\mathrm{S,0}, \Lambda ^4$ with all other parameters set to zero is shown in Fig.~\ref{fig:deltaNLL_fs0}. Intersection with the horizontal line marks the resulting $95\, \%$ C.L. limits. In this case, they are given in terms of $\mathrm{TeV^{-4}}$ by: 
\begin{equation}
-8.04 < f_\mathrm{S,0} / \Lambda^4 < 8.45\, ,
\end{equation}
which is about $7\, \%$ looser when compared to the officially reported limits. Similar results are expected in the case of other parameters.\\

Currently, the the accuracy of the reweighting procedure is further studied in detail when different parameters are considered. This is needed to set two-dimensional limits and generalize the procedure to other parameters.\\
Future endeavors include the actual combination as more data becomes public, a closer look at the uncertainties and correlations, the clipping method as a unitarisation method, and cross-checks with other fitting frameworks. At the present time data is public for the same sign WW channel from CMS~\cite{Sirunyan:2017ret} and expected from \mbox{ATLAS}~\cite{ATLAS:2018ogo}. In addition public data is expected in the WZ channel from CMS and ATLAS~\cite{ATLAS:2018ucv}, as well as in the ZZ channel from the CMS collaboration~\cite{Sirunyan:2017fvv}.\\

\section{Machine learning for jets (reconstruction) \footnote{speaker: H. Kirschenmann}}

Jets are the experimental signatures of quarks and gluons produced
in high-energy processes such as hard scattering of partons in proton-proton
collisions. The process of hadronization leads to a collimated
spray of color neutral hadrons which is referred to as a jet.
As these particles propagate through the detector, they leave signals
in the detector components such as the tracker and the electromagnetic
and hadronic calorimeter. Within CMS, the {}``Particle Flow'' (PF) approach is used, which
attempts to reconstruct individually each particle in the event, prior
to the jet clustering, based on information from all relevant subdetectors.
Machine learning methods are explored for optimizing jet reconstruction at 
all levels, including the low-level reconstruction of, e.g., 
tracks, the calibration of the jet energy scale, and the identification of 
jet types. The wide availability of industry-supported frameworks for 
the training of deep neural networks (DNNs) such as Tensorflow 
\cite{Abadi2016} has boosted the usage of these techniques, recently.

A fast and reliable track reconstruction is vital to the overall physics 
performance of CMS and becomes increasingly difficult in high pileup conditions. 
In order to reduce the combinatorial complexity of the problem an iterative 
approach has been chosen in CMS\cite{Chatrchyan:2014fea}. The tracks that are 
easiest to find are searched in the early iterations and the signals associated 
to the found good quality tracks are masked from the later iterations to reduce 
the computational load. An accurate method for estimating the track
quality is needed both for masking the signals that are associated to reconstructed 
tracks and for rejecting fake tracks. For this purpose, a single DNN classifier is 
being developed to replace individually trained BDTs at each iterative step. 
First results promise a comparable efficiency at fake rates lowered by ca. 50\% for track \pt
in the range of 1-100\GeV \cite{Havukainen:2017}. 
In view of HL-LHC, there is also an influx of ideas from the data science community for
more fundamental changes to track reconstruction, e.g., with the TrackML particle 
tracking challenge ongoing until end of 2018.

The technique of b-jet energy regression has been extensively used in searches for and measurements of
the Higgs boson decay to a bottom quark-antiquark pair, e.g. \cite{Aaltonen:2011bp,Sirunyan:2017elk}, 
where the latest iterations use DNNs. The b-jet energy regression is a perfect use-case for showcasing 
regression techniques, because standard jet energy corrections only correct back to the particle level, 
excluding neutrinos. The regression is often used to correct the jet energy to the b quark energy, 
recovering energy lost to neutrinos in e.g. semileptonic b hadron decays, which translates to a 
significant improvement of the mass resolution. More generically applicable jet energy regressions are 
also being explored by LHC experiments.

\begin{figure}
  \centering
  \includegraphics[width=.45\linewidth]{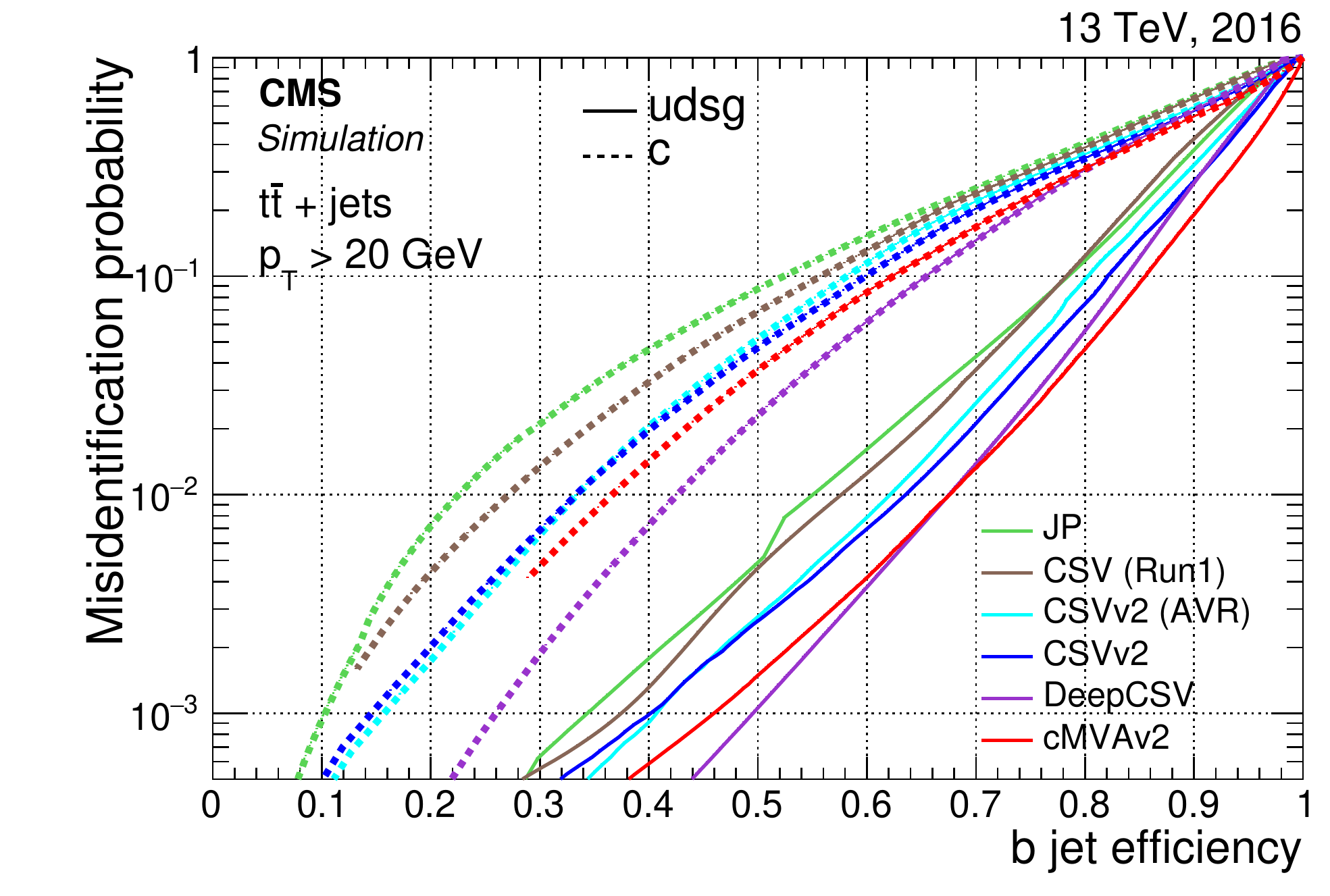}  
  \includegraphics[width=.45\linewidth]{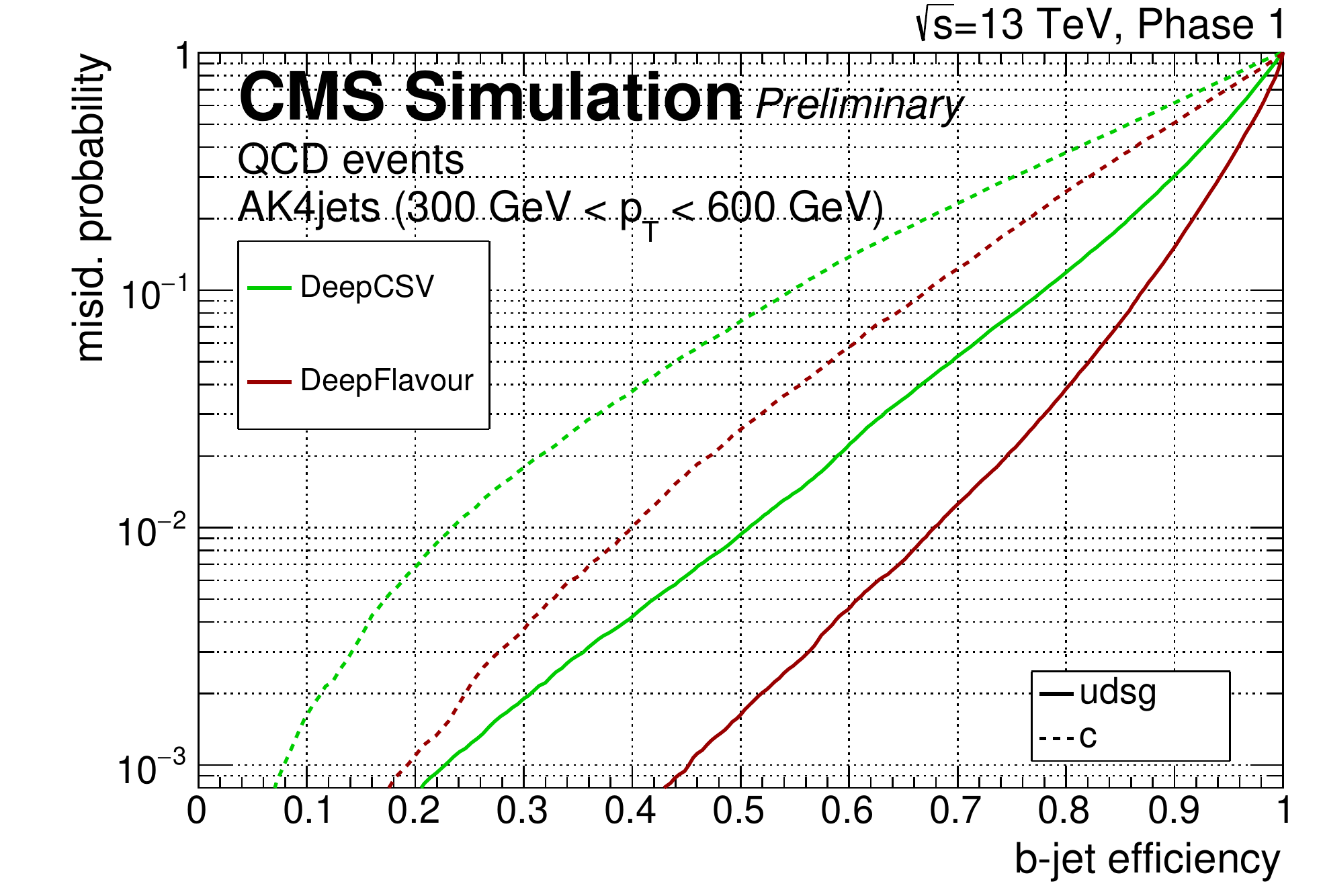}	
  \caption{(Left) Misidentification probability for c and light-flavour jets versus b jet identification efficiency for various b tagging algorithms applied to jets in $t\bar{t}$ events \cite{Sirunyan:2017ezt}. (Right)  Performance of the DeepCSV (retrained for the Phase 1 detector geometry) and DeepFlavour b jet identification algorithms demonstrating the probability for non-b jets to be misidentified as b jet, as a function of the efficiency to correctly identify b jets \cite{CMS-DP-2017-013}.}
  \label{fig:MLreco}
\end{figure}

Many measurements and searches for physics beyond the standard model at the LHC rely on the 
efficient identification of heavy-flavour jets, i.e. jets originating from bottom or charm quarks. 
During Run 2, the heavy-flavor identification has seen significant improvements. The current standard 
algorithm in online and offline reconstruction is based on a deep neural network, using similar inputs 
as the previous factorized approach, significantly improving the signal efficiency by ca. 15\% for the 
same misidentification probability (DeepCSV) \cite{Sirunyan:2017ezt}. An extended tagging algorithm using directly
PF candidates as inputs (DeepFlavour) enables further gains, reducing the misidentification by more than 50\% for the same signal efficiency at high \pt
\cite{CMS-DP-2017-013}. The performance of both new approaches is compared to previous methods in Fig.~\ref{fig:MLreco}.

More recently, these DNN techiques have also been applied to W/Z/H/t tagging \cite{CMS-DP-2017-049}, 
using low-level PF candidates as input, yielding superior performance to combination of just a few 
substructure observables using BDTs, reducing the misidentification rate by more than 50\% for the same signal efficiency. 

A likelihood discriminator using three input observables capable of distinguishing between jets 
originating from quarks and from gluons has been pioneered by CMS during Run 1 \cite{CMS:2013kfa}. 
It has been recommissioned at 13 \TeV and more sophisticated discriminators are being studied, such as 
BDTs using more input observables or extending the novel DNN heavy flavor tagging techniques towards a 
full multiclassification of light quark, gluon, and heavy flavor jets at the same time \cite{CMS-DP-2017-027}.

\section{Machine learning for jets (substructure) \footnote{speaker: S. Schramm}}

The usage of Machine Learning (ML) in hadronic physics is a rapidly growing field.  ML has been studied in the context of everything from jet reconstruction, to calibration, to identification.  The last topic is a particularly promising area of study, where ML algorithms are able to exploit additional non-trivial information to select hadronically-decaying bosons (W, Z, and H), quarks (charm, bottom, and top), and to discriminate between quark- and gluon-initiated jets.

In order to identify such different types of jets, it is important to understand the differences between each type of hadronic shower.  The different hadronic showers result in different energy profiles and angular correlations between the energy deposits within the jet, which can be used to infer the most probable initiating particle for the jet.  This practice of studying the energy and angular distributions within a single jet is referred to as jet substructure, and variables that have been designed to quantify such distributions are referred to as jet substructure variables.

\begin{figure}
  \centering
  \includegraphics[width=.45\linewidth]{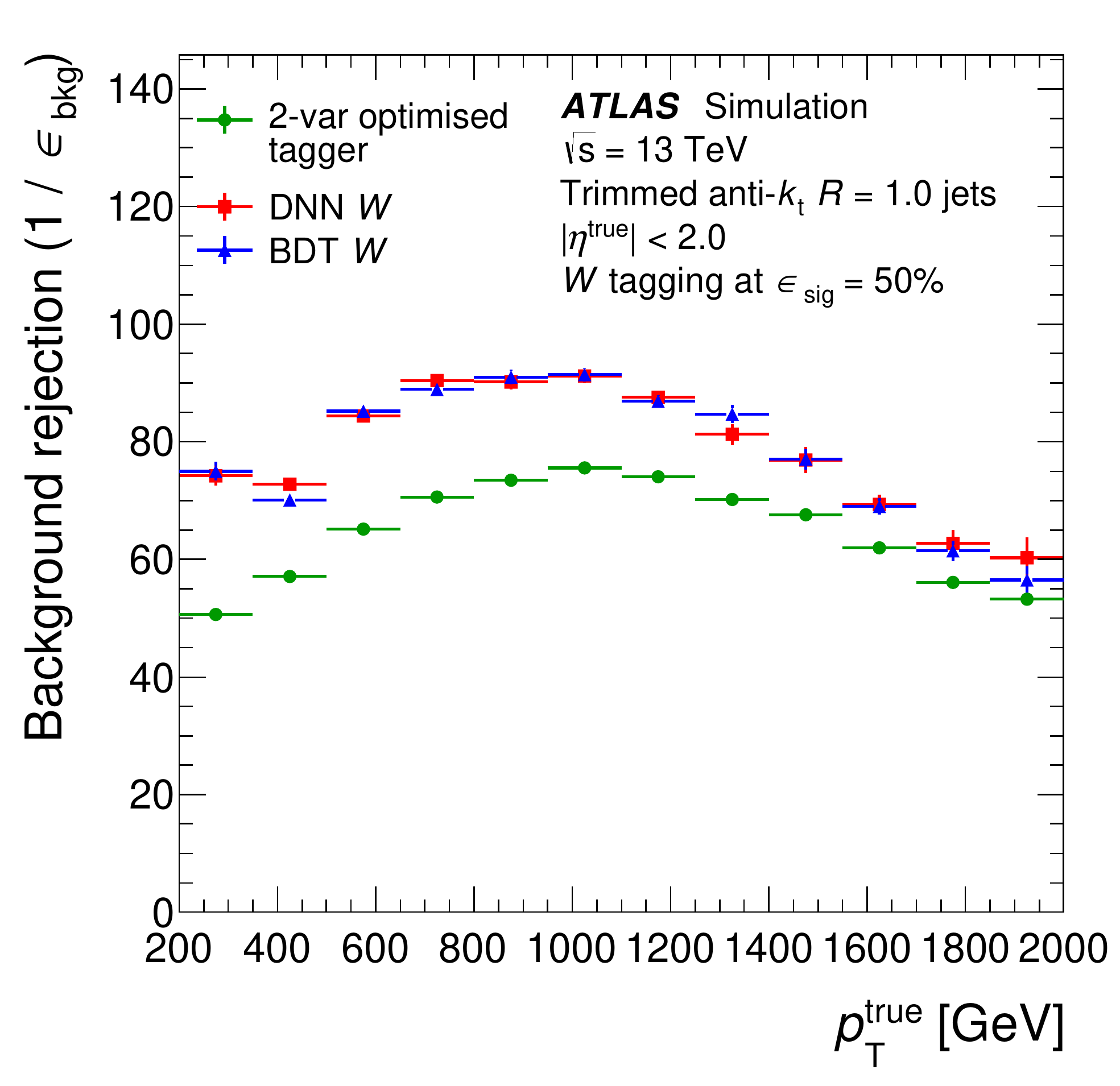}	
  \includegraphics[width=.45\linewidth]{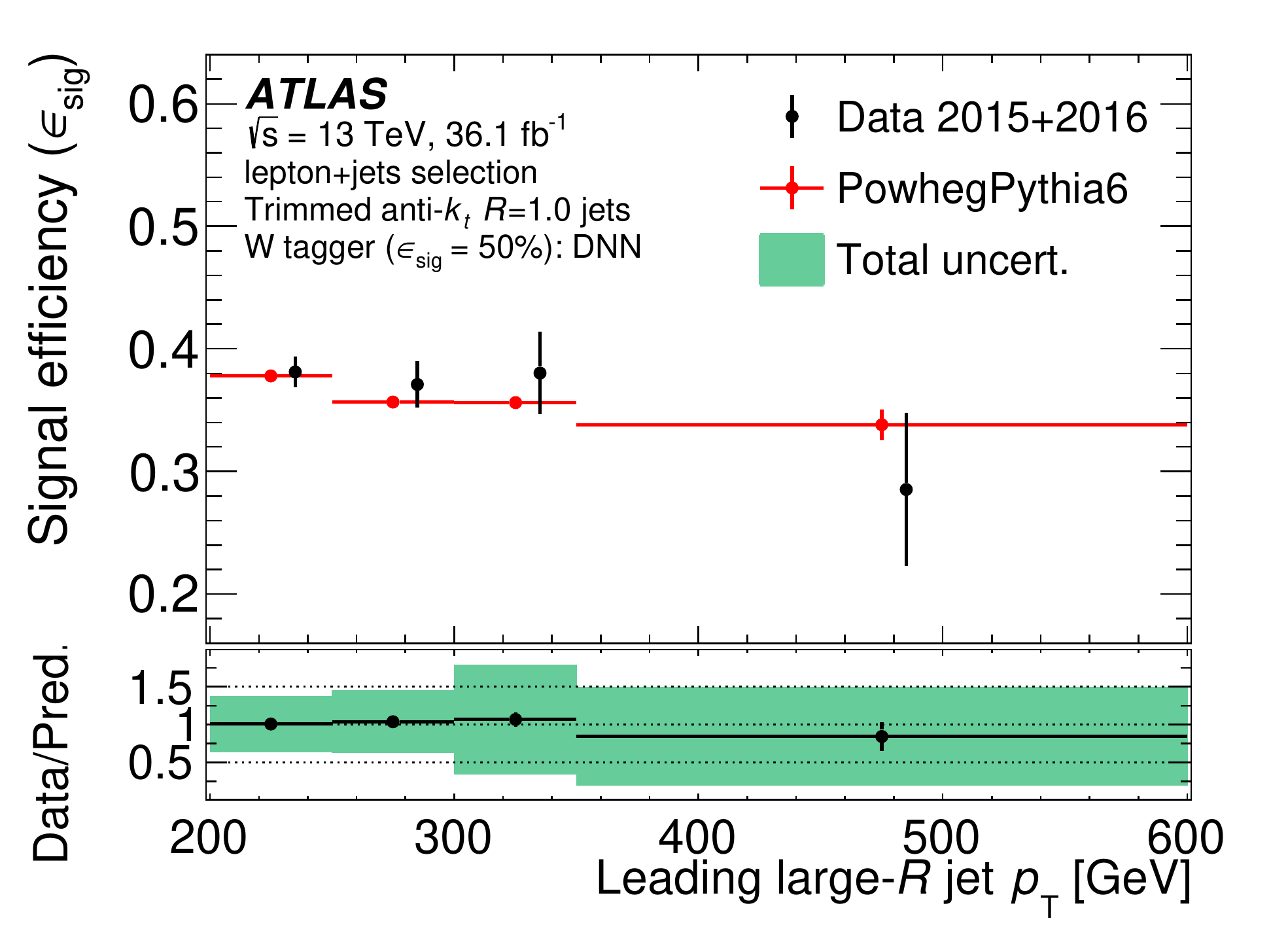}  
  \caption{(Left) a comparison of W-taggers based on simple two-variable cuts or ML classifiers trained using high-level jet substructure information, showing a moderate improvement when using either BDTs or DNNs~\cite{Aaboud:2018psm}.  (Right) the uncertainty associated with the W-tagging DNN ML classifier as evaluated in semi-leptonic $t\bar{t}$ events~\cite{Aaboud:2018psm}.  When using the same approach to derive uncertainties for the simple two-variable cut-based tagger, similar uncertainties are observed, indicating that the DNN tagger does not increase the tagging uncertainty with respect to non-ML taggers.}
  \label{fig:MLtag}
\end{figure}

ATLAS has combined a variety of ML techniques with the use of jet substructure, from simple extensions of cut-based taggers to much more advanced approaches.  The identification of W bosons is a useful example.  Previously, hadronically-decaying W bosons were identified with simple two-variable cuts~\cite{Aad:2015rpa}.  More recently, combinations of several jet substructure variables either using Boosted Decision Trees (BDTs) or Deep Neural Networks (DNNs) have been studied and have shown moderate improvements at the level of 20-30\%~\cite{Aaboud:2018psm} as shown in Fig.~\ref{fig:MLtag}.

While combinations of jet substructure variables are an easy first step, it limits the potential of ML techniques, as it restricts the network to exploiting correlations between existing variables rather than identifying entirely new representations of the data.  The use of "low-level" information such as the four-vectors of the jet constituents, rather than "high-level" information representing a given jet property, should provide additional benefits.  Indeed, this has been demonstrated in phenomenological studies in the context of Higgs-tagging~\cite{Baldi:2014kfa}, and has more recently been confirmed by ATLAS for top-tagging~\cite{Aaboud:2018psm}).  Such methods of providing additional "low-level" information are expected to continue to improve the ability of the LHC experiments to identify different types of jets.

ATLAS has additionally demonstrated the power of "low-level" information in two other contexts, namely b-tagging using Recurrent Neural Networks (RNNs)~\cite{ATL-PHYS-PUB-2017-003} and quark vs gluon discrimination using Convolutional Neural Networks (CNNs)~\cite{ATL-PHYS-PUB-2017-017}.  In the first case, substantial benefits of order 100\% are observed due to the ability to exploit correlations between individual tracks and the possibility of adding additional variables without dealing with the increased dimensionality required when calculating exact discriminants.  The second case saw smaller but still relevant benefits, indicating that different final states and different types of ML techniques may be more or less applicable.

When using ML for jet identification, it is important to understand how to evaluate realistic uncertainties on the resulting discriminants.  An improved identification power is not useful if uncertainties cannot be evaluated, or if the uncertainties increase dramatically.  It is therefore critical to identify control samples or standard candles which can be used to compare data and simulation in a well-understood topology.  ATLAS currently makes use of semi-leptonic $t\bar{t}$ events to evaluate the uncertainties associated with W boson and top quark tagging~\cite{ATLAS-CONF-2017-064} (since updated~\cite{Aaboud:2018psm}), with uncertainties currently at the level of 50\% for W tagging as shown in Fig.~\ref{fig:MLtag}, where this tagging uncertainty is at the same level as non-ML taggers evaluated using the same approach.  V+jets events can be used for the same purpose for W and Z boson tagging~\cite{ATLAS-CONF-2018-016}, and $g\to{}bb$ events are used to evaluate uncertainties for H$\to{}bb$ tagging~\cite{ATLAS-CONF-2016-039}.  As ML techniques are increasingly exploited, it will be important to identify new control samples that can be used for other processes and more extreme regions of parameter space.

\section{Neural net jet reconstruction \footnote{speaker: K. Kallonen}}


The identification of the origin of jets after they have been clustered requires dedicated classification algorithms. This final step of jet reconstruction attempts to map a hadron-level jet back to an initiating parton. The task of accurately classifying jets originating from gluons and light quarks is difficult, because they appear superficially similar at the observed hadron-level. Traditionally, in both the ATLAS \cite{Aad:2014gea} and the CMS \cite{CMS:2013kfa} experiments, likelihood-based discriminators for quark and gluon jets have been built upon a few theoretically motivated jet-level variables.

Since the "Particle Flow" approach of event reconstruction allows access to particle-level information, more detailed representations of jets can be used in the hope of improving the classification performance. In the past few years, multiple deep learning approaches to identifying different types of jets have been proposed. These approaches utilize specialized neural network models, which can exploit various structural features of the jets.

Some prominent approaches include the construction and analysis of jet images \cite{Komiske:2016rsd}, the application of natural language processing techniques \cite{Louppe:2017ipp} and organizing the jets as graphs in order to use message passing \cite{Henrion:2017nips}. Additionally, in the CMS experiment, a deep learning model capable of more general jet classification has been developed \cite{Stoye:2017nips}.

The goal is to perform a comparative study of the performance of these different approaches to the discrimination of gluon and light quark jets with the detector response taken into account. The CMS likelihood-based quark/gluon jet discriminator is used as a benchmark model. If no significant improvement in performance is acquired by choosing a particular model, other properties of the classifiers need to be considered, such as the model's complexity and its ease of use. The robustness of the classifiers needs to be tested as well, by comparing the performance in different jet transverse momentum and pseudorapidity regions. In the low transverse momentum region, a model's resilience to the effects of pile-up becomes especially important.

The neural network models also face another problem, as they are trained on simulated quark and gluon jets for which the class labels are known. Quark and gluon jets produced by different event generators look different \cite{Gras:2017jty}. Hence, it is reasonable to question whether this method of supervised machine learning results in classifiers learning idiosyncratic features of event generators and thus render them unsuitable for accurately classifying quark and gluon jets in recorded data. In another test of robustness, the neural network models should thus be validated by cross-evaluating their performance on quark and gluon jets produced by different event generators. Notably, the neural network approach based on jet images was shown to be remarkably indifferent to the choice of event generator used for the simulation of the jets \cite{Komiske:2016rsd}.

\phantomsection
\addcontentsline{toc}{part}{Acknowledgements} 

\chapter*{Acknowledgements\markboth{Acknowledgements}{Acknowledgements}}

\indent 
The authors acknowledge the finacial support of the COST Action CA16108,
and are grateful to the Thessaloniki University team for the great hospitality.
MP is supported by the European Research Council Consolidator Grant
NNLOforLHC2. AS acknowledges support from the National Science Centre, Poland Grant No.
2016/23/D/ST2/02605 and the grant 18-07846Y of the Czech Science Foundation (GACR). MZ 
is supported by the Netherlands National Organisation for Scientific Research (NWO).



\renewcommand\leftmark{References}
\renewcommand\rightmark{References}

\bibliographystyle{./StyleFilesMacros/atlasnote}
\cleardoublepage
\phantomsection
\addcontentsline{toc}{part}{References}
\bibliography{ThessalonikiReport}


\end{document}